\documentclass[]{elsart}

\usepackage{epsfig}
\usepackage{graphics}
\usepackage{graphicx}
\usepackage{amsmath}
\usepackage{url}
\usepackage{subfigure}

\newtheorem{proposition}{Proposition}

\pdfoutput=1

\begin{document}

\begin{frontmatter}

\title{On growing connected $\beta$-skeletons\\  \vspace{0.5cm} \footnotesize{Final version is published in \\
\textit{Computational Geometry},  46 (2013) 6, 805--816.\\     http://dx.doi.org/10.1016/j.comgeo.2012.11.009}}

\author{Andrew Adamatzky}

\address{University of the West of England, Bristol, United Kingdom\\ andrew.adamatzky@uwe.ac.uk}

\maketitle

\begin{abstract}

\noindent
A $\beta$-skeleton, $\beta \geq 1$, is a planar proximity undirected graph of an Euclidean points set, 
where nodes are connected by an edge if their lune-based neighbourhood contains no 
other points of the given set. Parameter $\beta$ determines the size and shape of the lune-based neighbourhood.
A $\beta$-skeleton of a random planar set  is usually a disconnected graph for $\beta>2$. With the increase of $\beta$,
the number of edges in the $\beta$-skeleton of a random graph decreases. We show how to grow stable $\beta$-skeletons, which are connected for any given value of $\beta$ and characterise morphological 
transformations  of the skeletons governed by $\beta$ and a degree of approximation. We speculate how the results obtained can be applied in biology and chemistry.  

\vspace{0.5cm}

\noindent
\emph{Keywords: proximity graphs, $\beta$-skeletons, pattern formation, morphogenesis} 
\end{abstract}

\end{frontmatter}

\markboth{xxx}{xxx}

\section{Introduction}

A planar graph consists of nodes which are points of the Euclidean plane and edges which are straight segments connecting the points. A planar proximity graph is the planar graph where two points are connected by an edge if they are close in some sense. Usually a pair of points is assigned a certain neighbourhood, and points of the pair are connected by an edge if their neighbourhood is empty (does not contain any points of the given set).  Delaunay triangulation~\cite{delauanay}, relative neighbourhood graph~\cite{jaromczyk}, Gabriel graph~\cite{matula_1980}, and spanning tree, are the classical examples of the proximity graphs.

The $\beta$-skeletons~\cite{kirkpatrick} make a unique family of the proximity graphs monotonously parameterised by 
$\beta$. Two neighbouring points of a planar set are connected by an edge in a $\beta$-skeleton if a lune-shaped domain between the points contains no other points of the planar set. The size and shape of the lune is governed by $\beta$.   The $\beta$-skeletons are worth to study because they are amongst the key representatives of the family of proximity graphs. Proximity graphs are applied in many fields of science and engineerings: from image processing (eps. 
reconstructing the shape of a two-dimensional object, given a set of sample points on the boundary of the object), visualisation  and physical modelling to analysis and design of communication and transport 
networks~\cite{adamatzky_ppl_2008,adamatzky_bioevaluation,Amenta_1998,billiot_2010,dale_2000,dale_2002,gabriel_1969,jombart_2008,legendre_1989,li_2004,magwene_2008,matula_1980,muhammad_2007,runions_2005,santi_2005,sokal_2008,song_2004,sridharan_2010,toroczkai_2008,wan_2007,watanabe_2005, watanabe_2008}

A $\beta$-skeleton is the Gabriel graph~\cite{matula_1980} for $\beta=1$ and it is the relative neighbourhood graph for $\beta=2$~\cite{jaromczyk,kirkpatrick}.  A $\beta$-skeleton, in general case, becomes disconnected for $\beta>2$ and continues losing its edges with further increase of $\beta$. In our previous paper~\cite{adamatzky_betaskeletons} we demonstrated that $\beta$-skeletons of random planar sets lose edges by a power low with the rate of edge disappearance proportional to a number of points in the sets. Some $\beta$-skeletons conserve their edges for 
any $\beta$ as large as it could be. These are usually skeletons built on a regularly arranged sets of planar points, however even minuscule impurity in the regular arrangement of points leads to propagation of an edge loss wave across the otherwise stable skeleton. Can we produce connected $\beta$-skeletons for arbitrarily large values of $\beta$?  How do these skeletons look like? What are properties of these skeletons? How topological features of the connected $\beta$-skeletons are changed with the increase of $\beta$? We answer these questions in the paper.

\section{$\beta$-skeletons}
\label{themodel}

\begin{figure}[!tbp]
\centering
\subfigure[$\beta=1$]{\includegraphics[scale=0.3]{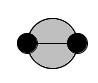}\hspace{0.9cm}} 
\subfigure[$\beta=2$]{\includegraphics[scale=0.3]{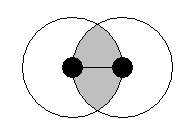}\hspace{0.9cm}} 
\subfigure[$\beta=10$]{\includegraphics[scale=0.3]{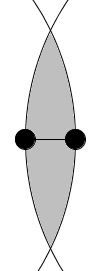}\hspace{0.9cm} } 
\subfigure[$\beta=100$]{\includegraphics[scale=0.3]{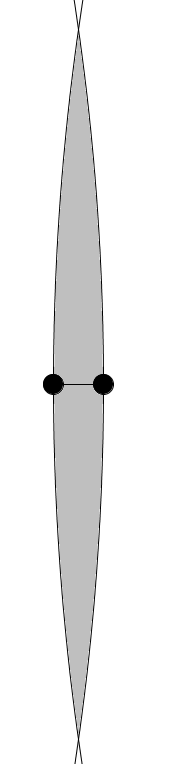}\hspace{0.9cm}} 
\caption{Examples of lunes, $\beta$-neighbourhoods, of two planar points, black disks, for various values of $\beta$.
The $\beta$-neighbourhoods are shaded in grey.}
\label{lunes}
\end{figure}

\begin{figure}[!tbp]
\centering
\subfigure[$\mathbf{V}$]{\includegraphics[width=0.24\textwidth]{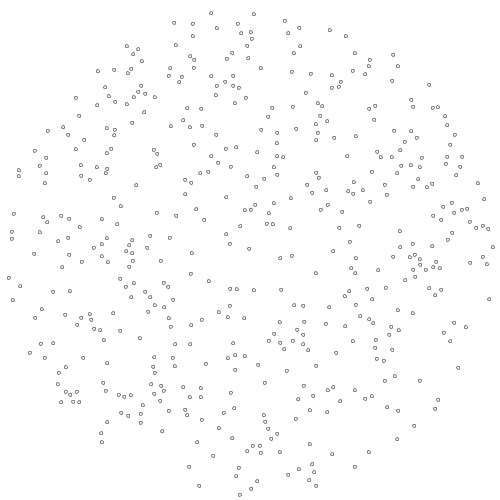}}
\subfigure[$\beta=0.7$]{\includegraphics[width=0.24\textwidth]{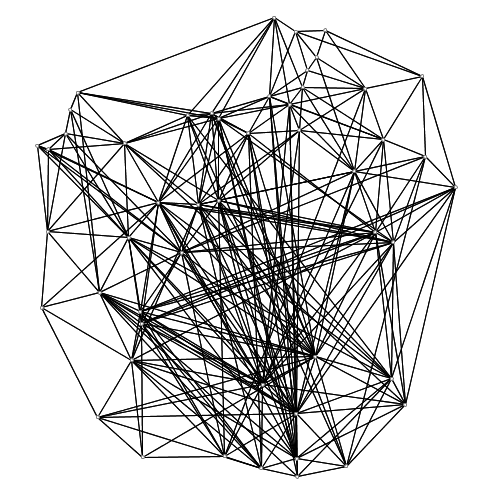}}
\subfigure[$\beta=1$]{\includegraphics[width=0.24\textwidth]{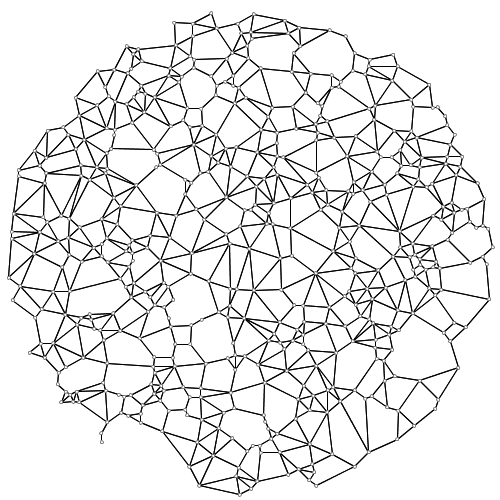}}
\subfigure[$\beta=2$]{\includegraphics[width=0.24\textwidth]{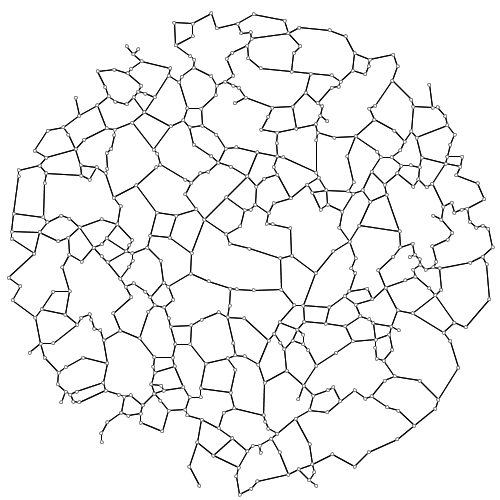}}
\subfigure[$\beta=3$]{\includegraphics[width=0.24\textwidth]{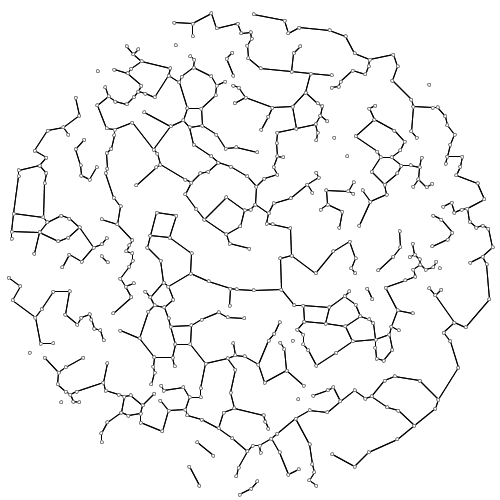}}
\subfigure[$\beta=4$]{\includegraphics[width=0.24\textwidth]{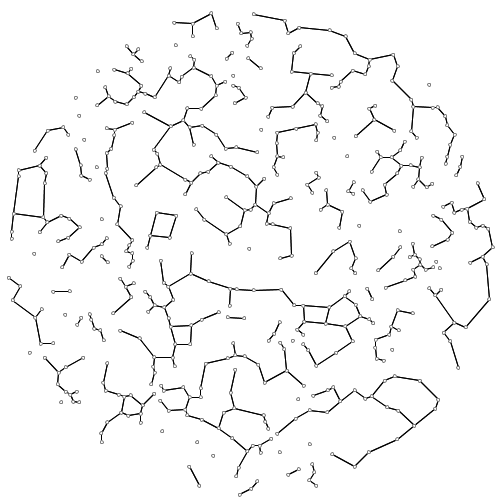}}
\subfigure[$\beta=7$]{\includegraphics[width=0.24\textwidth]{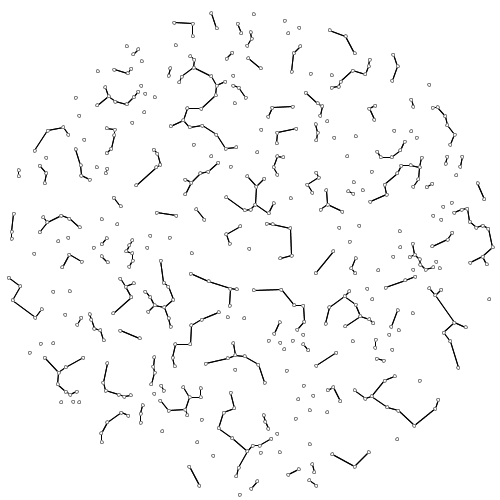}}
\subfigure[$\beta=20$]{\includegraphics[width=0.24\textwidth]{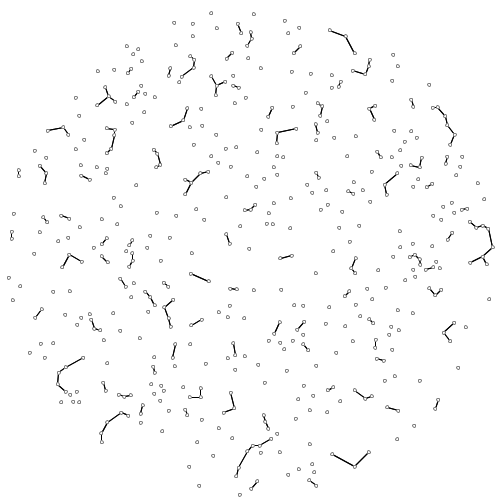}}
\subfigure[$e(n,\beta)$]{\includegraphics[width=0.85\textwidth]{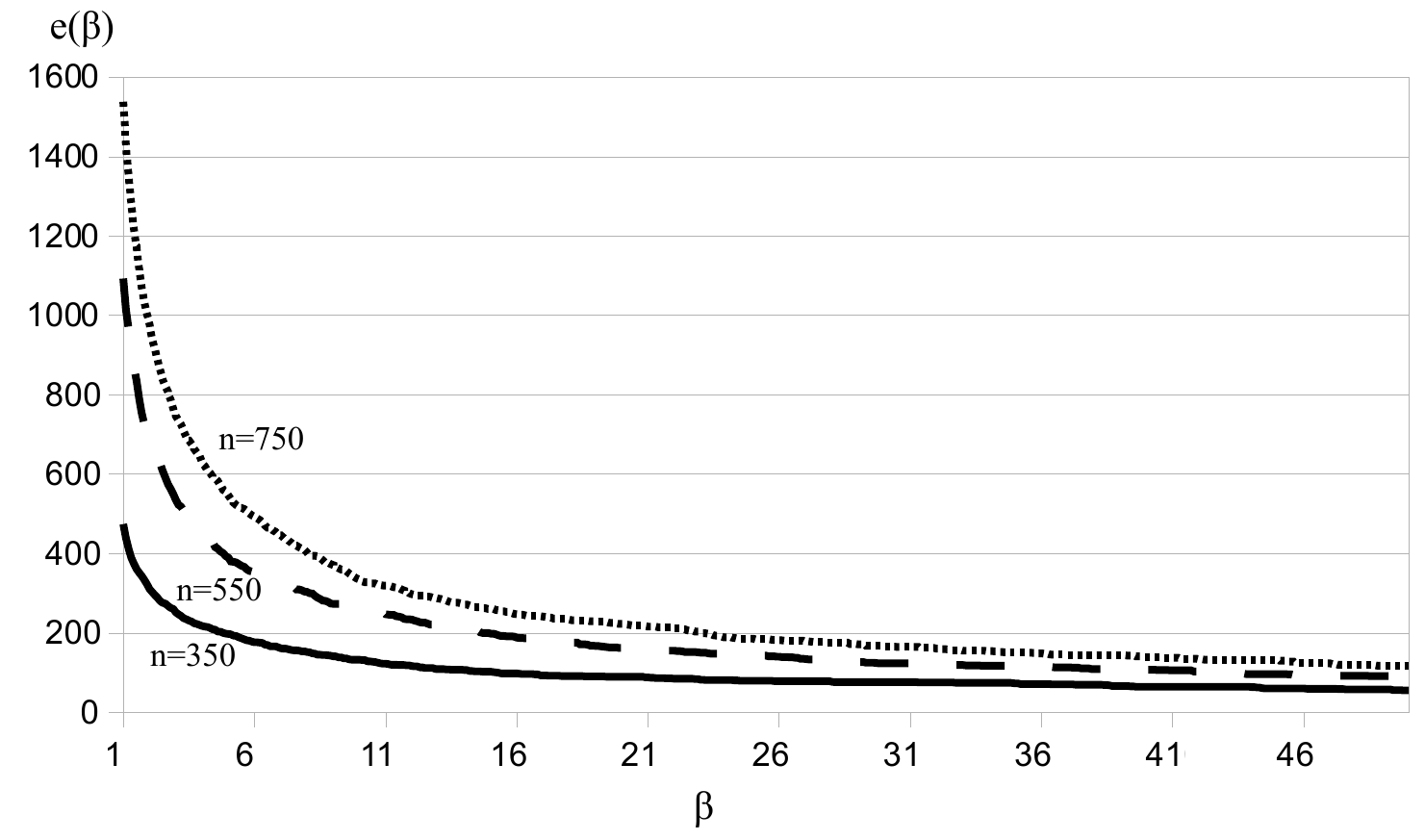}}
\caption{Example of $\beta$-skeletons of a random planar set. 
(a)~A random planar set $\mathbf{V}$ of 500 discs, radius 2.5 each,  randomly distributed in a disc radius 250.
(b--h)~Examples of $\beta$-skeletons on planar set $\mathbf{V}$.
(i)~Example power curves of edge loss, $e(n,\beta)$ is a number of edges in $\beta$-skeleton 
constructed on a random set with $n$ nodes, $1 \leq \beta \leq 100$, $n=350$ (solid line),
$n=550$ (dashed line) and $n=750$ (fine dashed line), values of $\beta$ are incremented by 0.1.}
\label{randomdiscexamples}
\end{figure}

Given a set $\mathbf V$ of planar points, for any two points $p$ and $q$ we define a
$\beta$-neighbourhood $U_\beta(p,q)$ as the intersection of two discs 
with radius $\beta |p-q| / 2$ centered at points $((1-\frac{\beta}{2})p,\frac{\beta}{2}q)$ and 
$(\frac{\beta}{2}p, (1-\frac{\beta}{2})q)$,  $\beta \geq 1$~\cite{kirkpatrick,jaromczyk}, see examples 
of the lunes in Fig.~\ref{lunes}.  Points $p$ and $q$ are connected by an edge in $\beta$-skeleton if the pair's $\beta$-neighbourhood contains no  other points from $\mathbf V$.

A $\beta$-skeleton is a graph $\mathbf{B}_\beta({\mathbf V})= \langle {\mathbf V}, {\mathbf E}, \beta \rangle$, 
where nodes ${\mathbf V} \subset {\mathbf R}^2$, edges $\mathbf E$, and for $p, q \in {\mathbf V}$ 
edge $(pq) \in \mathbf E$ if $U_\beta(p,q) \cap {\mathbf V}/\{p, q\} = \emptyset$. Parameterisation 
$\beta$ is monotonous: if $\beta_1 > \beta_2$ then  $\mathbf{B}_{\beta_1}({\mathbf V}) \subset  \mathbf{B}_{\beta_2}({\mathbf V})$~\cite{jaromczyk,kirkpatrick}.  

A $\beta$-skeleton is a non-planar graph for $\beta<1$, see example in Fig.~\ref{randomdiscexamples}b. Therefore 
we consider only skeletons with $\beta>1$.

\section{On stability of $\beta$-skeletons}

 $\beta$-skeletons of random planar sets lose their edges by a power law when $\beta$ increases linearly,  see details in~\cite{adamatzky_betaskeletons}. See examples of $\beta$-skeletons, $\beta=$1, 2, 3, 4, 7, 20, 100 and edge loss curve in Fig.~\ref{randomdiscexamples}. Most $\beta$-skeleton lose their edges with increase of $\beta$ however some $\beta$-skeleton do not. A stable $\beta$-skeleton retains its edges for any value of $\beta>1$. A most obvious example of a stable $\beta$-skeleton is a skeleton built on a set of planar points arranged in a rectangular array.

\begin{figure}[!tbp]
\centering
\includegraphics[width=0.95\textwidth]{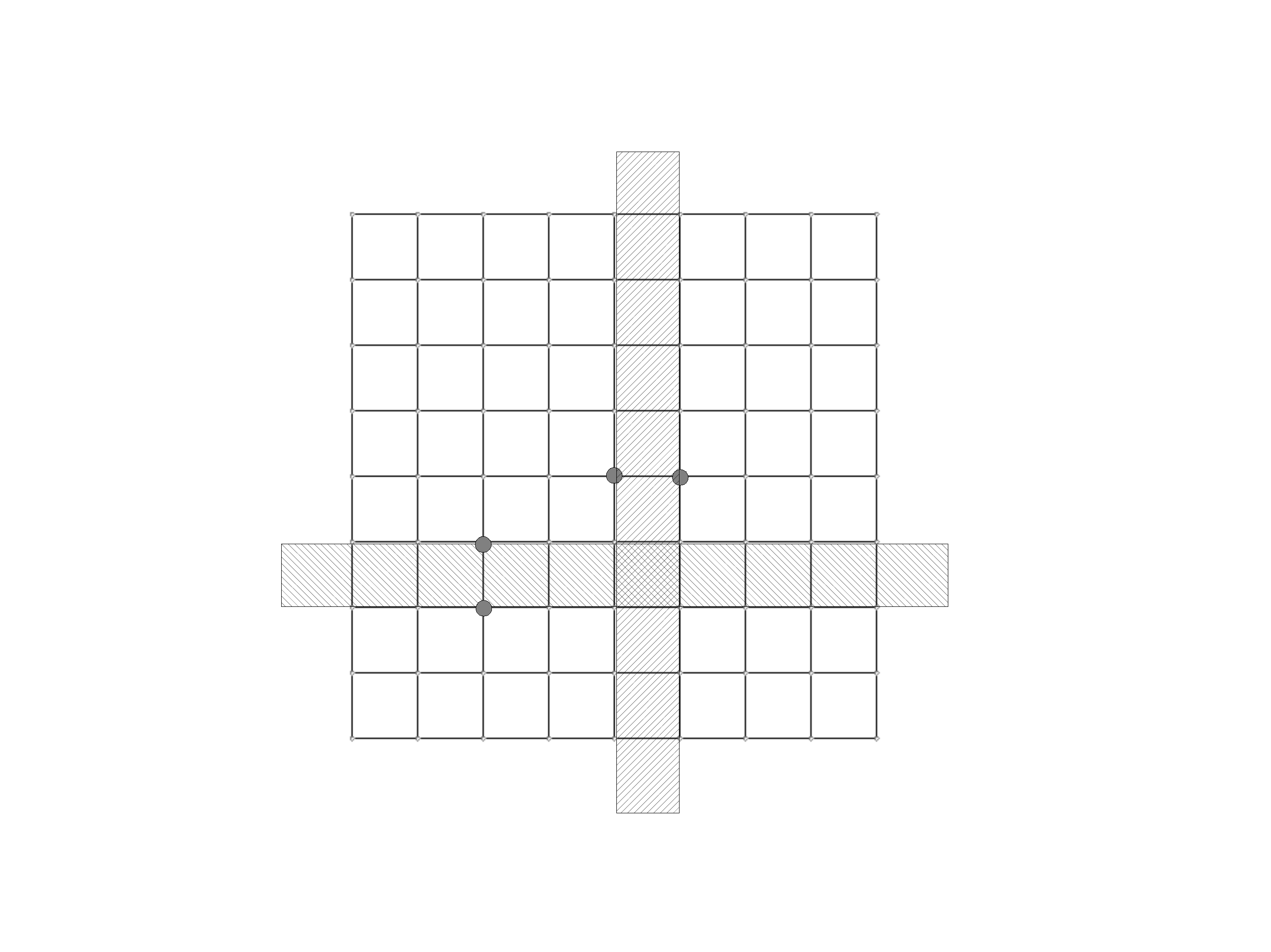}
\caption{Rectangular lattice is a stable $\beta$-skeleton. $\beta$-neighbourhoods, $\beta \rightarrow \infty$, of two pairs of nodes (marked by grey discs) are shown by hatched areas. }
\label{rectangularlatticescheme}
\end{figure}

\begin{proposition}
Rectangular lattice is a stable $\beta$-skeleton.
\end{proposition}

This is because when $\beta$ tends to infinity the $\beta$-neighbourhood tends to a rectangular shape 
and becomes an intersection of two half-planes. In details, let $\mathbf{H}_{ab}$ be an open half-plane bounded by an infinite straight line $l_a$ passing through $a$, perpendicular to segment $(a,b)$ and containing $b$; and  $\mathbf{H}_{ba}$ be an open half-plane bounded by an infinite straight line $l_b$ perpendicular to segment $(a,b)$, passing through $b$ and containing $a$. Let $\mathbf{M}_{ab} =  \mathbf{H}_{ab} \cap \mathbf{H}_{ba}$. When $\beta$ becomes extremely large, tends to infinity, a $\beta$-neighbourhood of any two neighbouring points $a$ and $b$ tends to $\mathbf{M}_{ab}$.  A $\beta$-skeleton of planar set $\mathbf{V}$ is stable if for any $a, b \in \mathbf{V}$  $\mathbf{M}_{ab}$ does not contain any points from $\mathbf{V}$ apart of $a$ and $b$.  The rectangular $\beta$-skeleton conserves its edges for any value of $\beta$ (Fig.~\ref{rectangularlatticescheme}). The rectangular lattice is stable because for any two neighbouring nodes $a$ and $b$ intersection $\mathbf{M}_{ab}$ of their half-planes  fits between rows or columns of nodes without covering any other nodes.  

Given $\beta$, is it possible to generate a planar set which $\beta$-skeleton is a connected graph? A method of growing such sets, and their $\beta$-skeletons, is presented further.

\section{Growing $\beta$-skeletons}
\label{procedure}

A graph is connected if there is a path along edges of the graph between any two nodes of the graph. An indirected graph is connected if there are no isolated nodes. To grow a connected $\beta$-skeleton for a given value of $\beta$ we start with a single planar point $p_0=(x_0,y_0)$ and then introduce
additional points one by one. When a new candidate point is introduced to a planar set we check if 
\begin{itemize}
\item the candidate point does not fell into $\beta$-neighbourhoods of existing nodes, and
\item the skeleton of the planar set  with the candidate point  retains its connectivity, i.e. there are no isolated nodes. 
\end{itemize}
Points can be added to the planar set either in a random fashion or in a regular manner. We adopt a regular addition of nodes by the following procedure.  

\vspace{0.2cm}

\textbf{Node Addition Procedure}
\begin{enumerate}
\item $r=5, \theta=0,  \delta = 10 $
\item $p = (x_0+r\cos \theta, y_0+r \sin \theta) $ 
\item if $\mathbf{B}_{\beta}(\mathbf{V} \cup \{ p \} )$ is connected graph and $(\forall q \in \mathbf{V}: |p-q|>\delta)$\\ 
then $\mathbf{V} \leftarrow \mathbf{V} \cup \{ p \}$
\item $\theta \leftarrow \theta + \Delta\theta$ 
\item if $\theta>360$ then $\theta \leftarrow 0$ and $r \leftarrow r + \Delta r$
\item go to step 2
\end{enumerate}

\vspace{0.2cm}

We use polar coordinates $p = (x_0+r\cos \theta, y_0+r \sin \theta)$ and assume that no two points can lie closer than $\delta$ to each other, in all experiments $\delta=2.5$.  Position of initial point is fixed, $p_0=(x_0,y_0)$. 
First candidate point is placed at distance $r$ from $p_0$ with angle $\theta=0$. The angle $\theta$ is incremented by 
$\Delta\theta$. When angle $\theta$ reaches 360 degrees radius $r$ is incremented by $\Delta r$ and $\theta$ is assigned value 0. The iterations may continue indefinitely but in experiments illustrated here we stop growing skeletons
when $r$ reached 90. Further in the paper we sometimes address $\beta$-skeletons grown by above procedure as simply $\beta$-skeleton.

\begin{proposition}
Let $\Delta \theta \rightarrow 0$ then grown $\beta$-skeleton is transformed from a hexagonal lattice, for $\beta=1$, to an orthogonal lattice, for $\beta \rightarrow \infty$.
\end{proposition}  

For $\beta=1$ lune  $U_\beta(p,q)$ is disc with diameter $pq$ (Fig.~\ref{lunes}a). Thus, in principle, we can arrange any number of points around an initial point $p_0$. However,  due to imposed minimal distance $\delta = 2.5$ between any two points, the points can be considered as discs.  A hexagonal packing is a densest arrangement of identical discs. 
Rectangular lattice is a stable $\beta$-skeleton, it remains connected for arbitrarily large $\beta$.

\begin{figure}[!tbp]
\centering
\subfigure[$\beta=1$]{\includegraphics[width=0.24\textwidth]{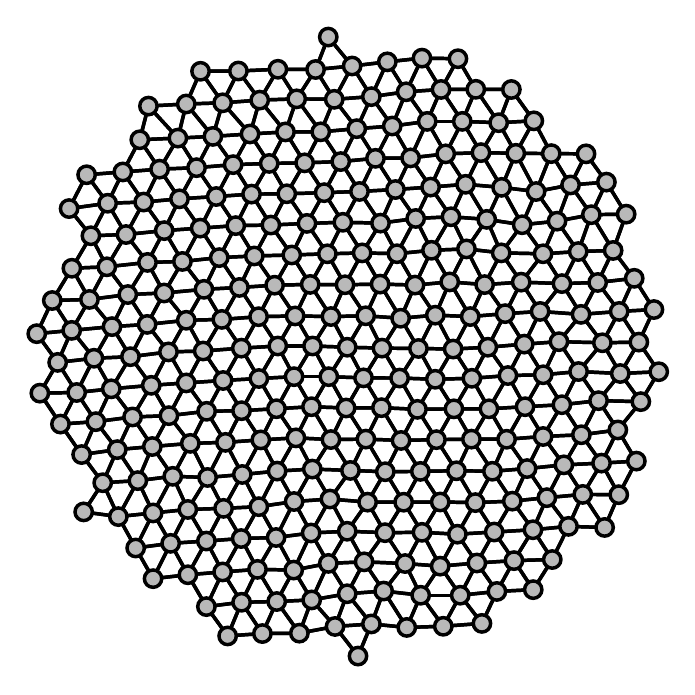}}
\subfigure[$\beta=1.5$]{\includegraphics[width=0.24\textwidth]{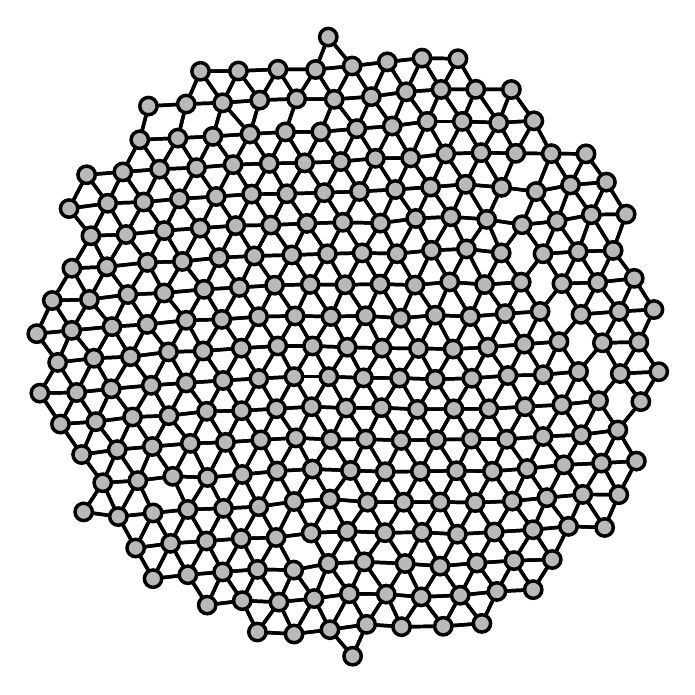}}
\subfigure[$\beta=2$]{\includegraphics[width=0.24\textwidth]{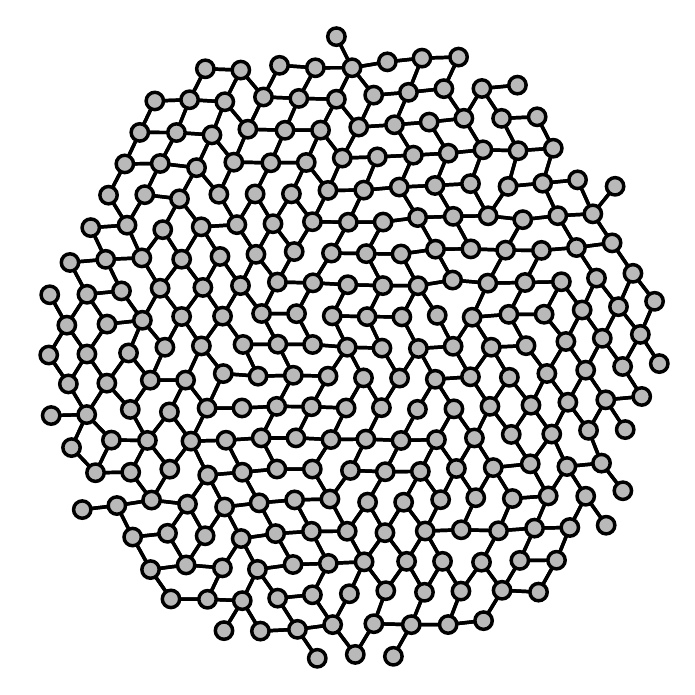}}
\subfigure[$\beta=2.5$]{\includegraphics[width=0.24\textwidth]{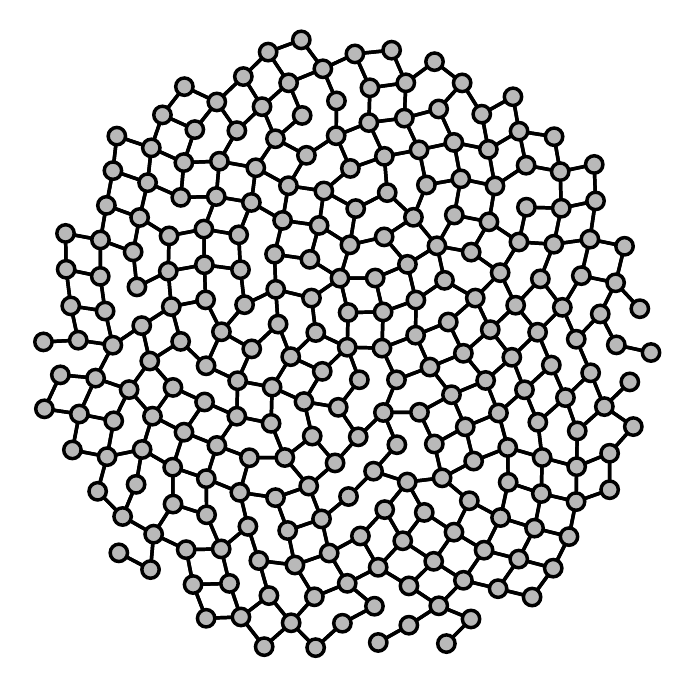}}
\subfigure[$\beta=3$]{\includegraphics[width=0.24\textwidth]{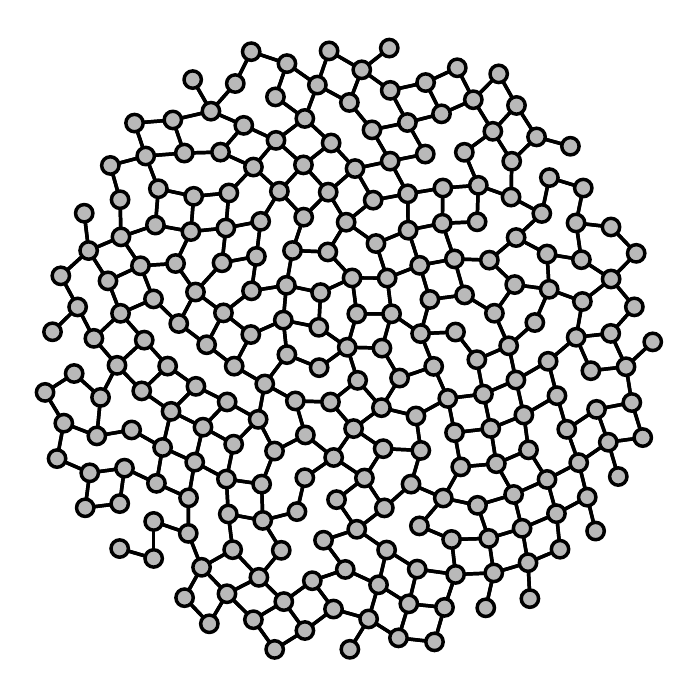}}
\subfigure[$\beta=3.5$]{\includegraphics[width=0.24\textwidth]{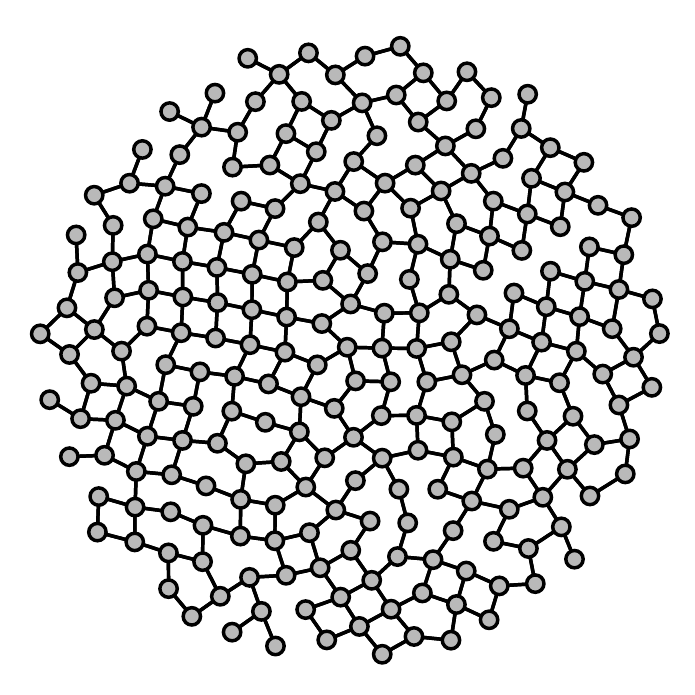}}
\subfigure[$\beta=4$]{\includegraphics[width=0.24\textwidth]{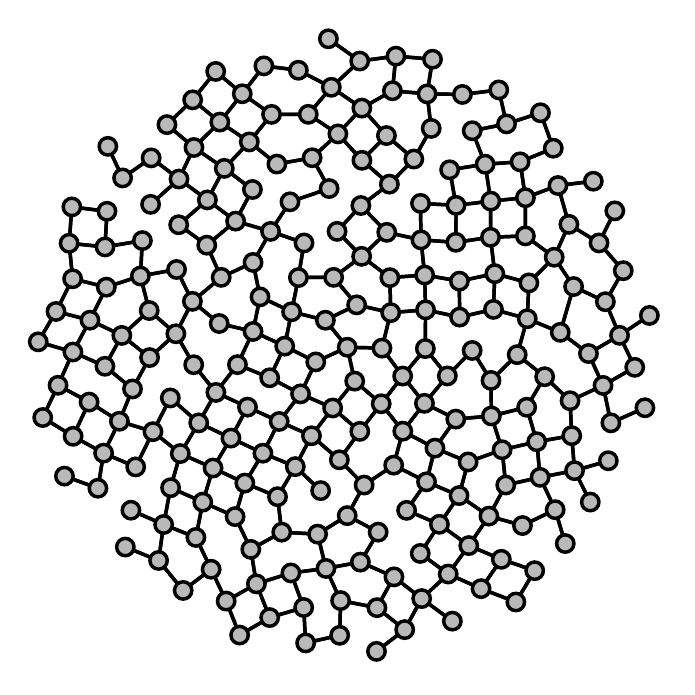}}
\subfigure[$\beta=4.5$]{\includegraphics[width=0.24\textwidth]{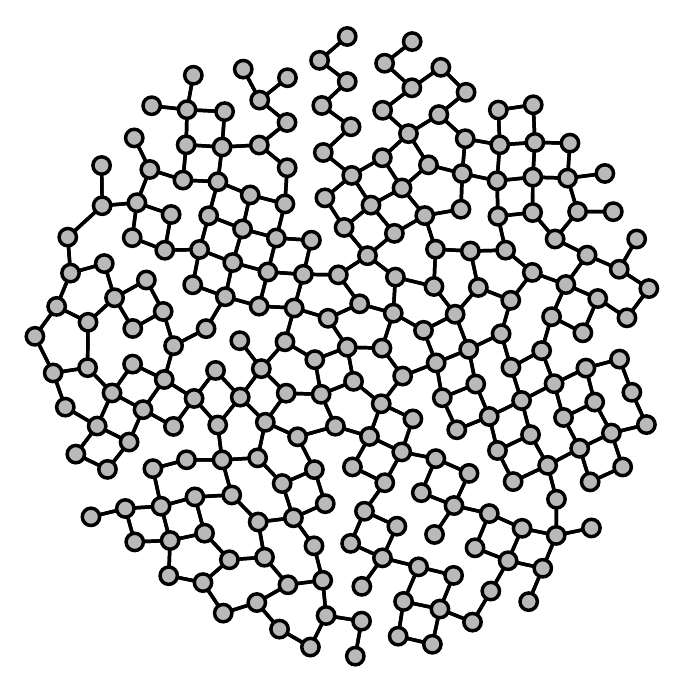}}
\subfigure[$\beta=10$]{\includegraphics[width=0.24\textwidth]{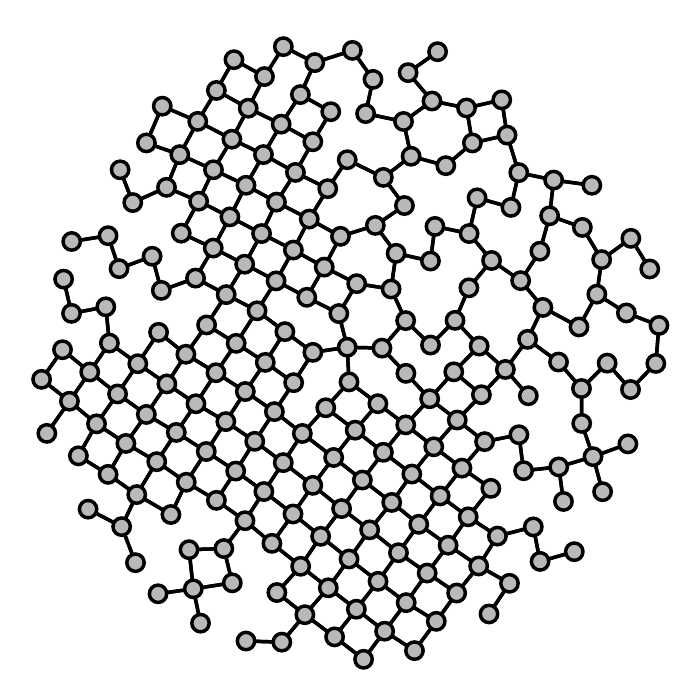}}
\subfigure[$\beta=30$]{\includegraphics[width=0.24\textwidth]{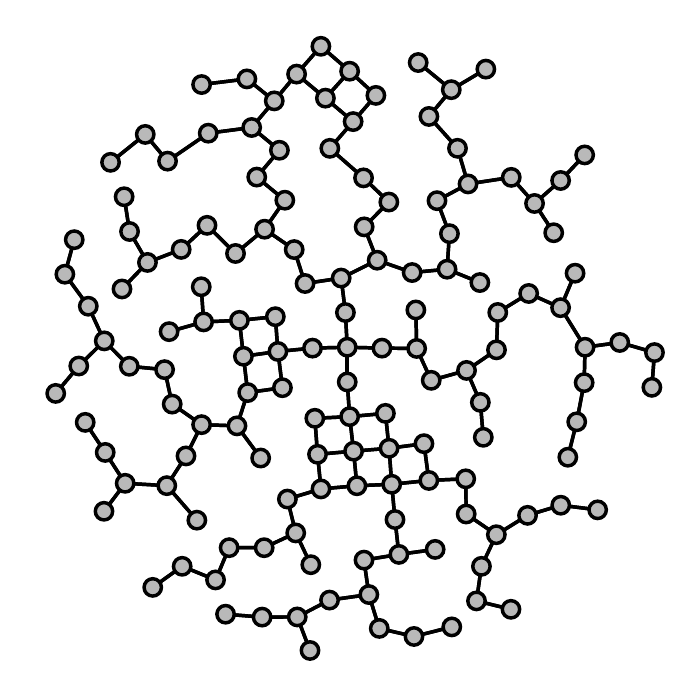}}
\subfigure[$\beta=40$]{\includegraphics[width=0.24\textwidth]{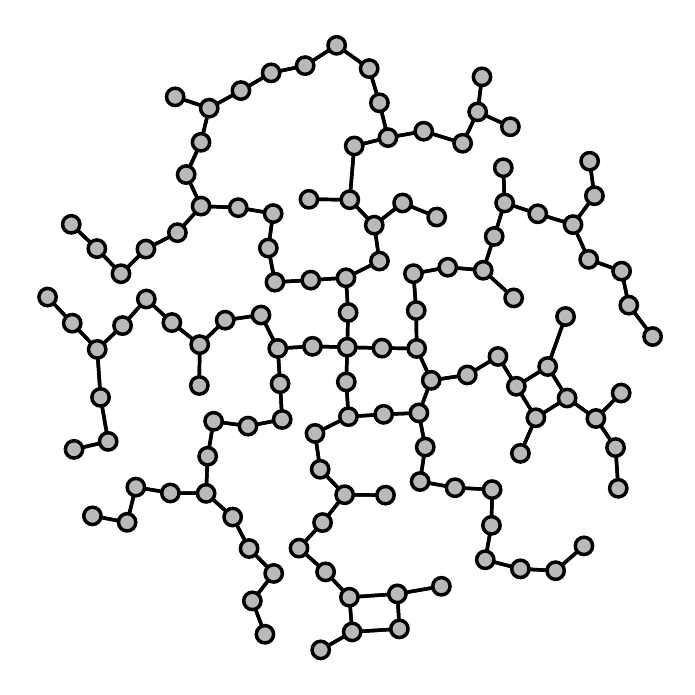}}
\subfigure[$\beta=50$]{\includegraphics[width=0.24\textwidth]{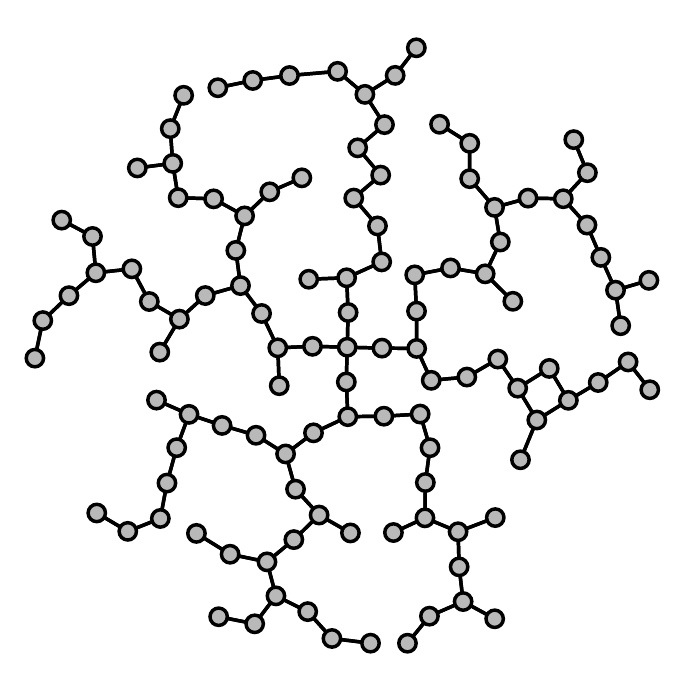}}
\subfigure[$\beta=100$]{\includegraphics[width=0.24\textwidth]{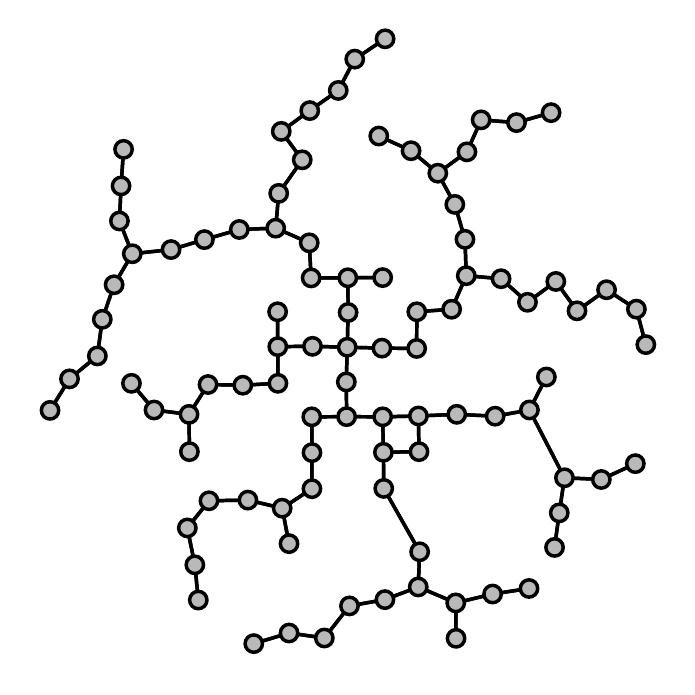}}
\subfigure[$\beta=200$]{\includegraphics[width=0.24\textwidth]{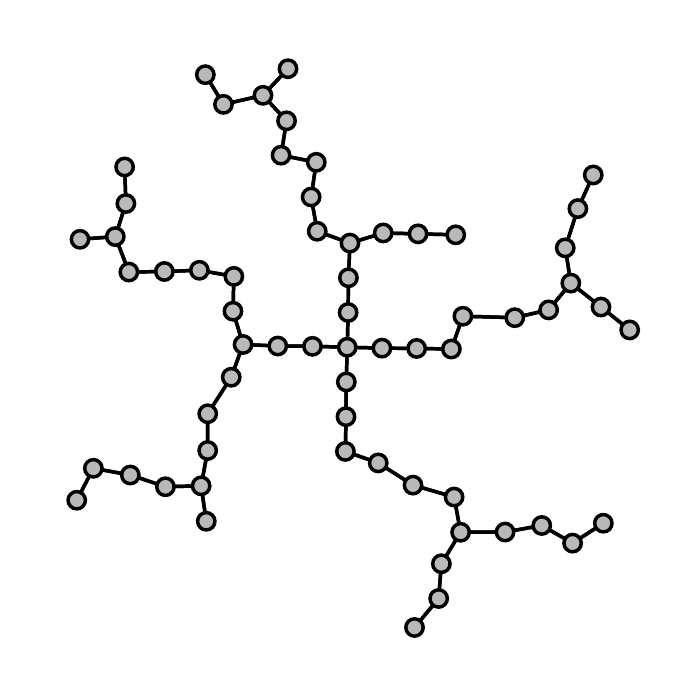}}
\subfigure[$\beta=300$]{\includegraphics[width=0.24\textwidth]{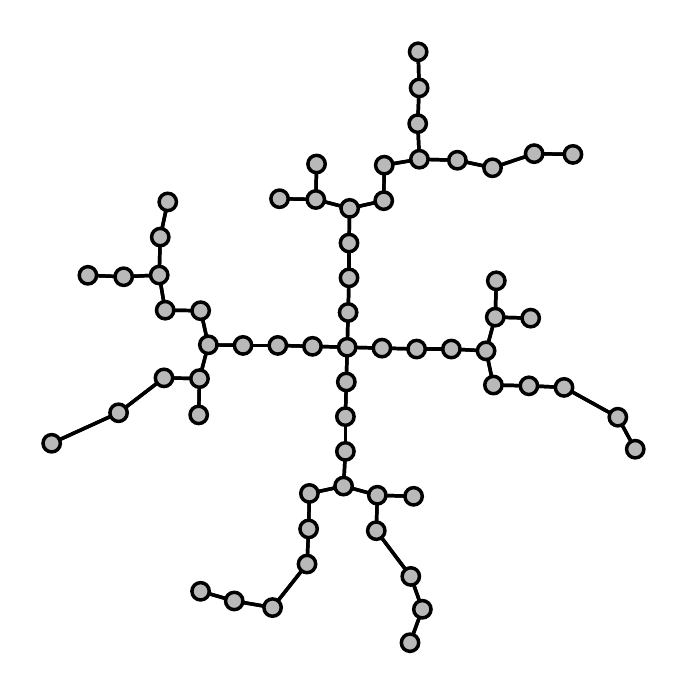}}
\subfigure[$\beta=400$]{\includegraphics[width=0.24\textwidth]{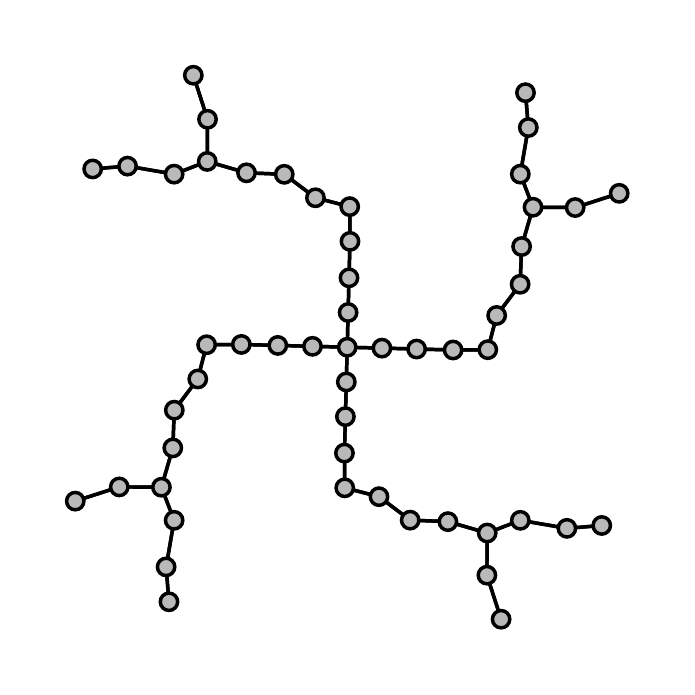}}
\subfigure[$\beta=500$]{\includegraphics[width=0.24\textwidth]{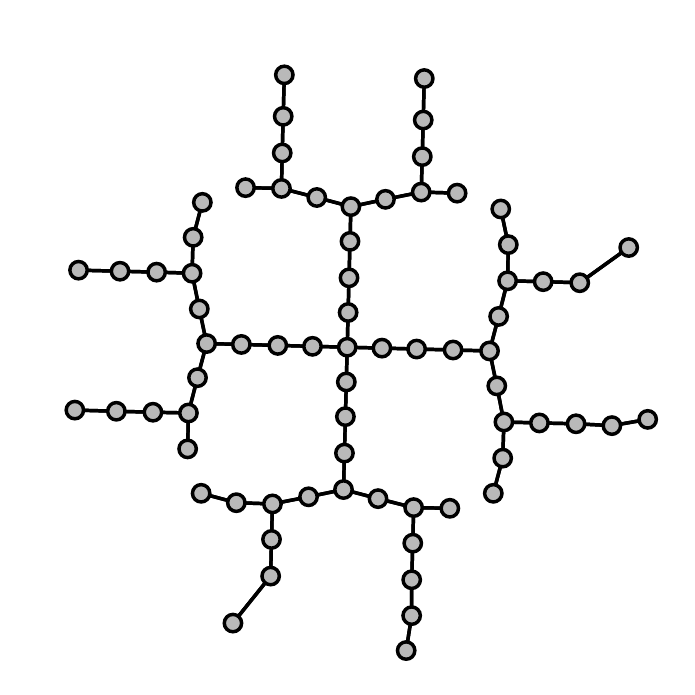}}
\subfigure[$\beta=600$]{\includegraphics[width=0.24\textwidth]{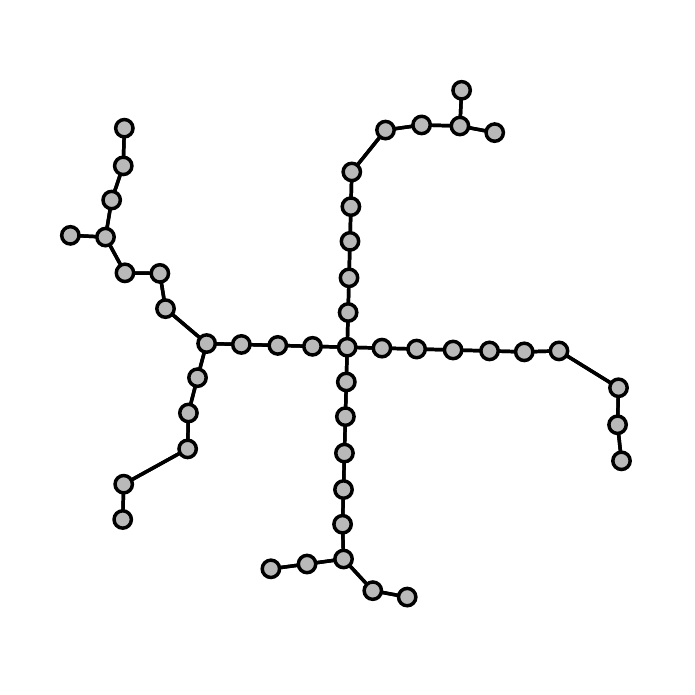}}
\subfigure[$\beta=700$]{\includegraphics[width=0.24\textwidth]{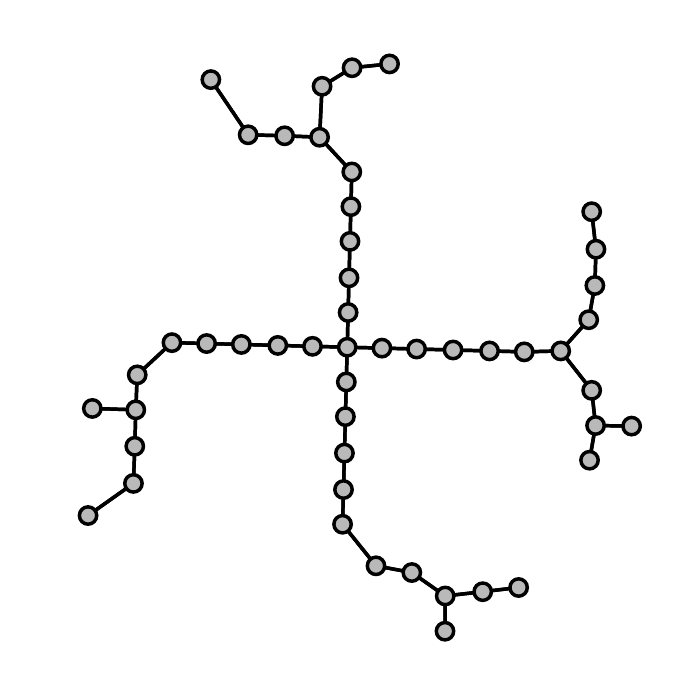}}
\subfigure[$\beta=800$]{\includegraphics[width=0.24\textwidth]{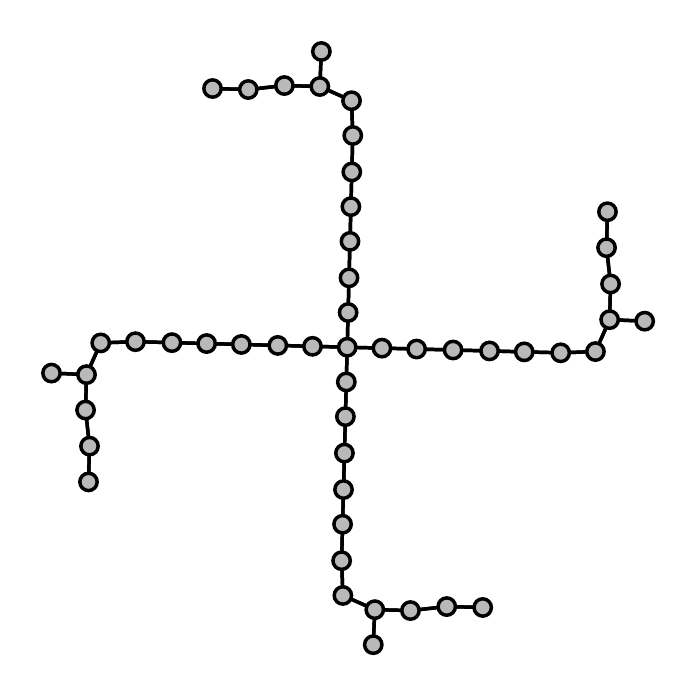}}
\caption{Examples of  $\beta$-skeletons grown from a single point, $r=5, \Delta r = 0.5, 
\Delta \theta = 0.5, \delta = 2.5 $}
\label{examplesofgrown}
\end{figure}

\begin{figure}[!tbp]
\centering
\subfigure[]{\includegraphics[width=0.65\textwidth]{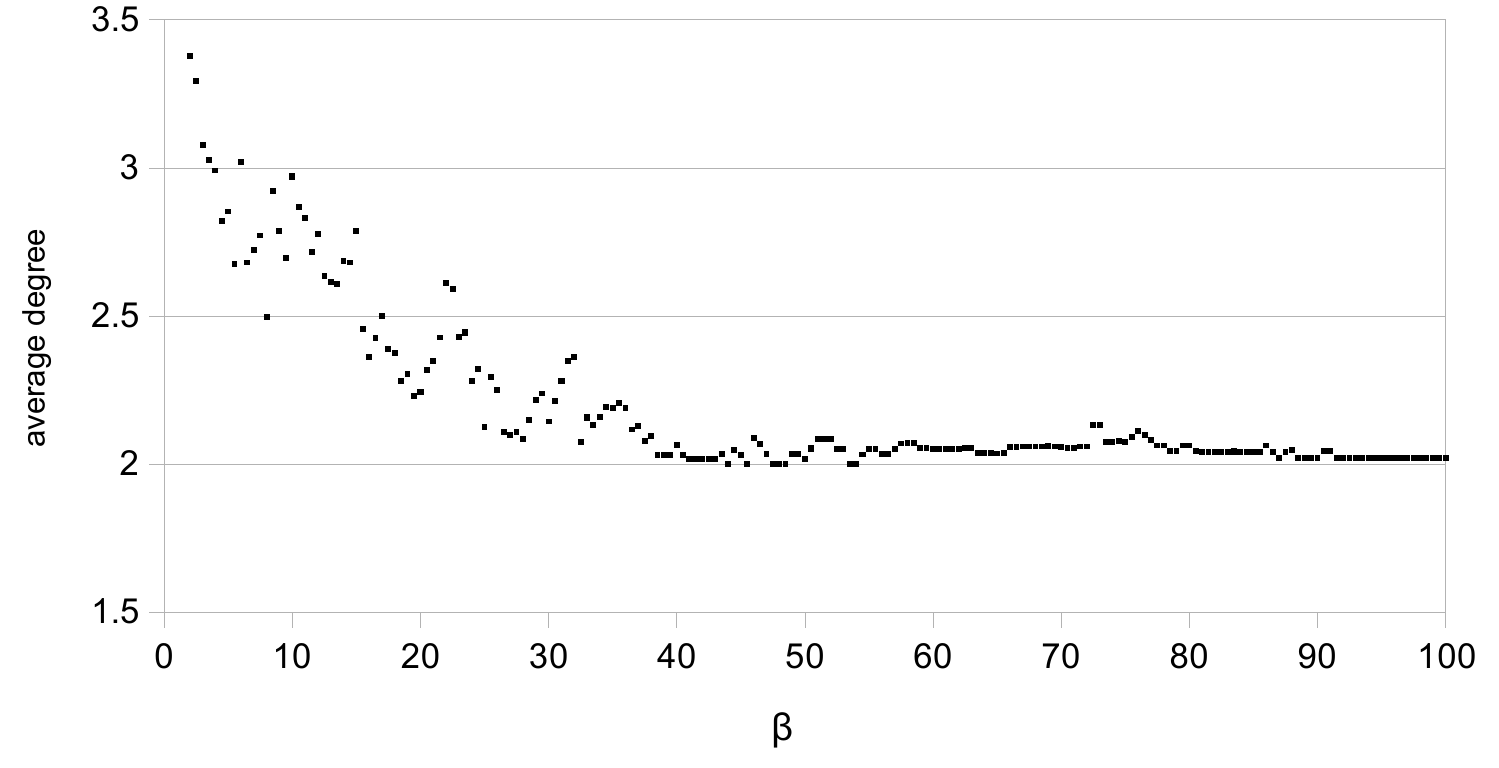}}
\subfigure[]{\includegraphics[width=0.65\textwidth]{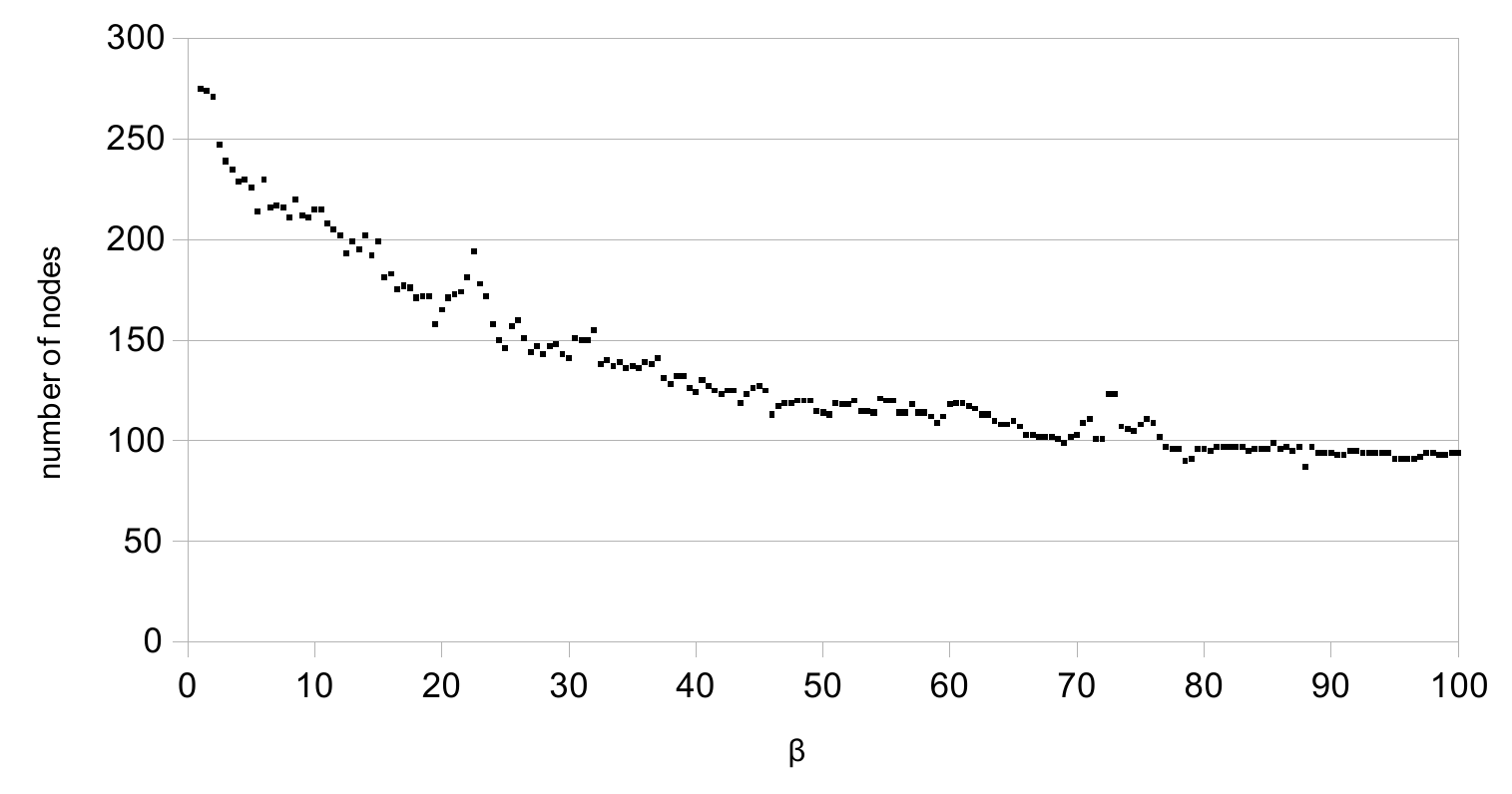}}
\subfigure[]{\includegraphics[width=0.65\textwidth]{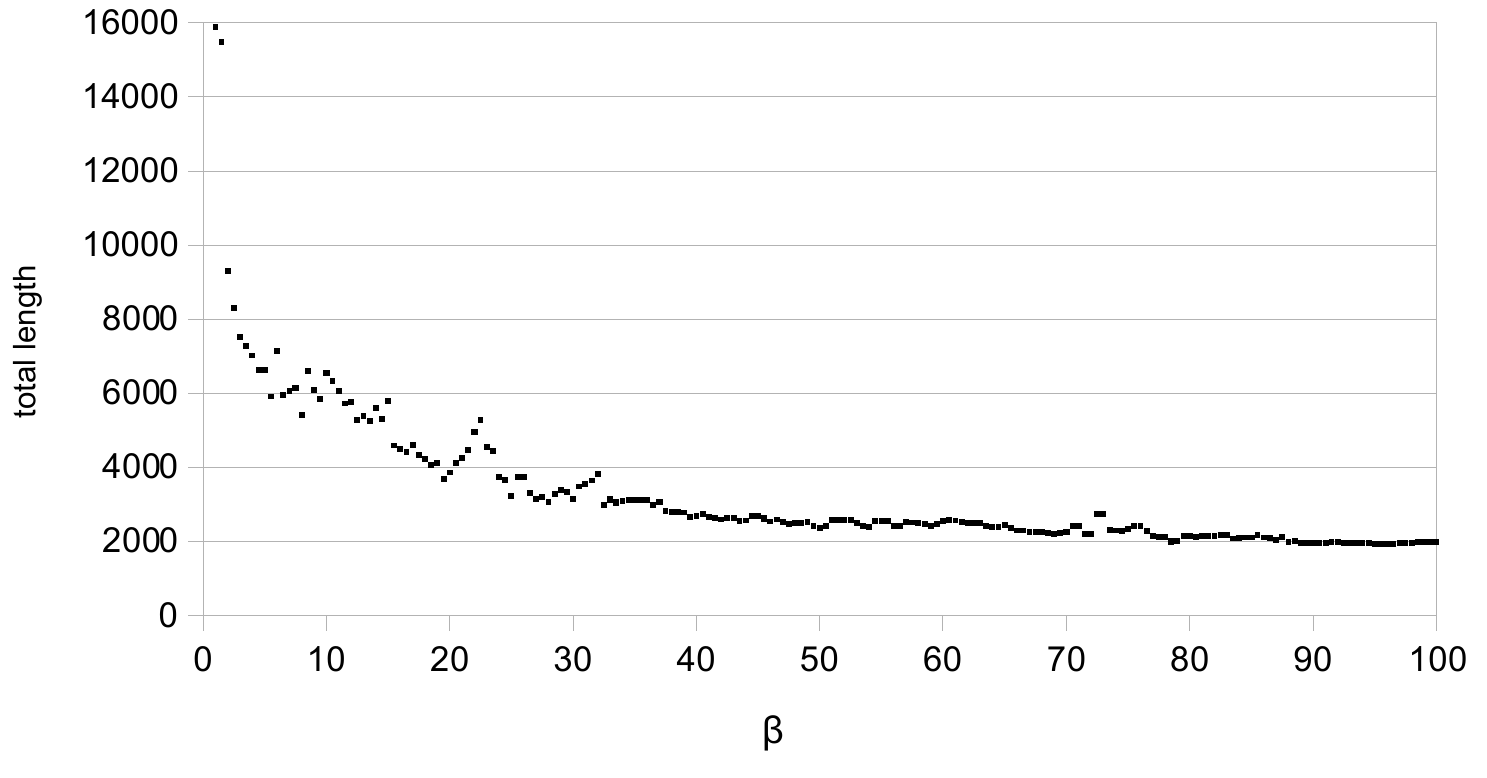}}
\subfigure[]{\includegraphics[width=0.65\textwidth]{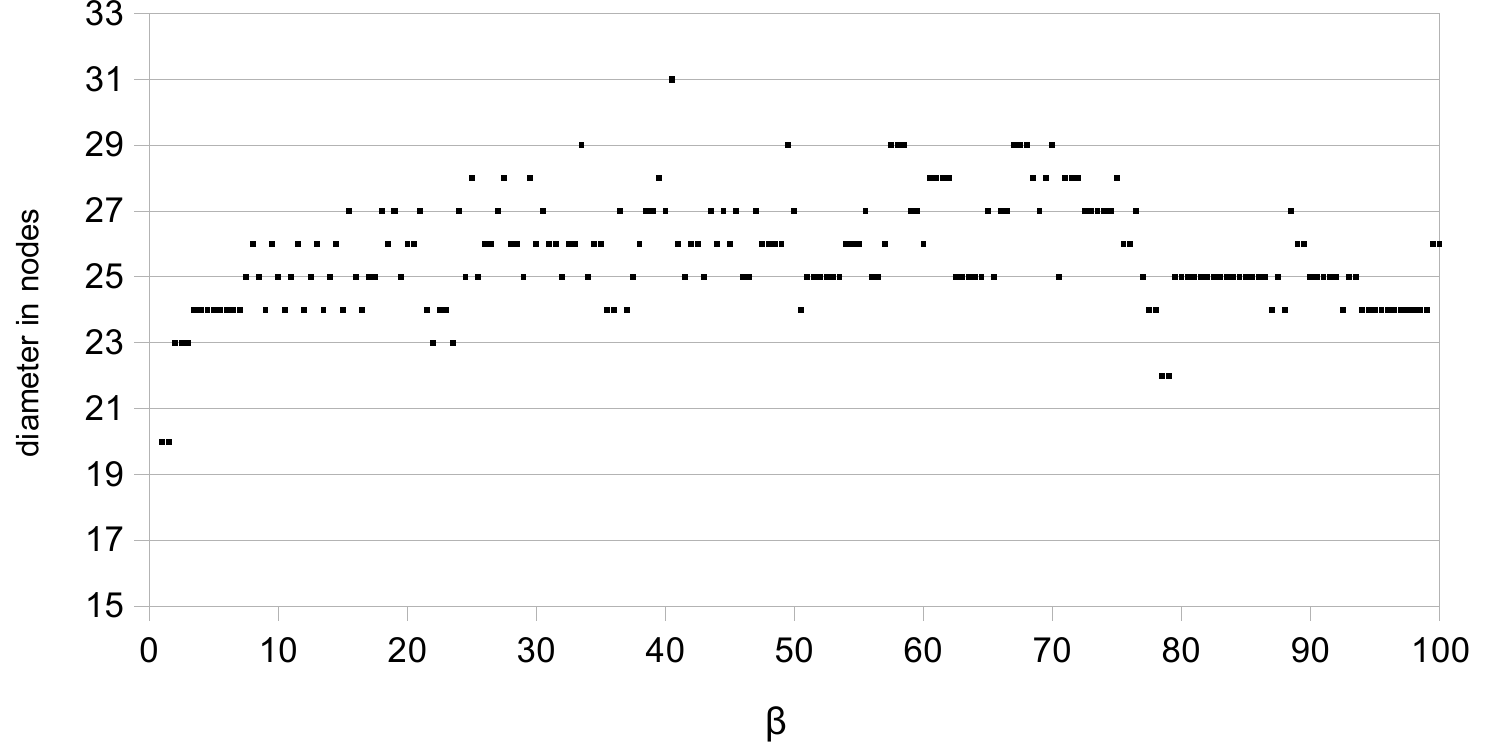}}
\caption{Average degree~(a),  number of nodes~(b), total lengths of edges~(c), and
diameter in nodes~(d) of $\beta$-skeletons grown with parameters 
$r=5, \Delta r = 0.5, \Delta \theta = 0.5, \delta = 2.5$.}
\label{statistics}
\end{figure}

\section{Dynamics of skeletons controlled by $\beta$}

Skeletons grown on computers are never ideal, and never become rectangular lattices, due to
impurities in their topologies introduced by increments of $\theta$. Examples of $\beta$-skeletons 
grown from a single point with angular increment $\Delta \theta=0.5$ are shown in 
Fig.~\ref{examplesofgrown}. A hexagonal arrangement of nodes in $\beta$-skeleton for $\beta=1$ 
is well seen  (Fig.~\ref{examplesofgrown}a).  The hexagonal arrangement is gradually  destroyed when 
$\beta$ increases from 1 to 2 (Fig.~\ref{examplesofgrown}abc) with majority of nodes having three or 
four neighbours (Fig.~\ref{statistics}a). 

Further increase of $\beta$ leads to dissociation of cycles and 
formation of tree-like skeletons with domains of lattice-like arrangements (Fig.~\ref{examplesofgrown}i--t). 
Sizeable domains of rectangular lattices are still observed at $\beta=10$, e.g. domains located in 
southern, western and north-westerns parts of the graph in  Fig.~\ref{examplesofgrown}i. Branching of the tree
is reduced with increase of $\beta$ till skeleton is transformed to a cross-like structures with a single binary branching 
at each of four main branches (Fig.~\ref{examplesofgrown}t). This $\beta$-induced transformation is reflected in 
decrease in average degree of the graphs' nodes, which almost stabilises around value 2 when $\beta$ exceeds 50  (Fig.~\ref{statistics}a).

These structural transformations are reflected in decrease of a total number of nodes (Fig.~\ref{statistics}b) and 
total length of edges (Fig.~\ref{statistics}c). Increase of $\beta$ does not affect diameters of 
the $\beta$-skeletons, which vary between 23 and 29 nodes for $1 \leq \beta \leq 100$ (Fig.~\ref{statistics}d).

\begin{figure}[!tbp]
\centering
\subfigure[$\Delta \theta=5$, $\beta=1$]{\includegraphics[width=0.24\textwidth]{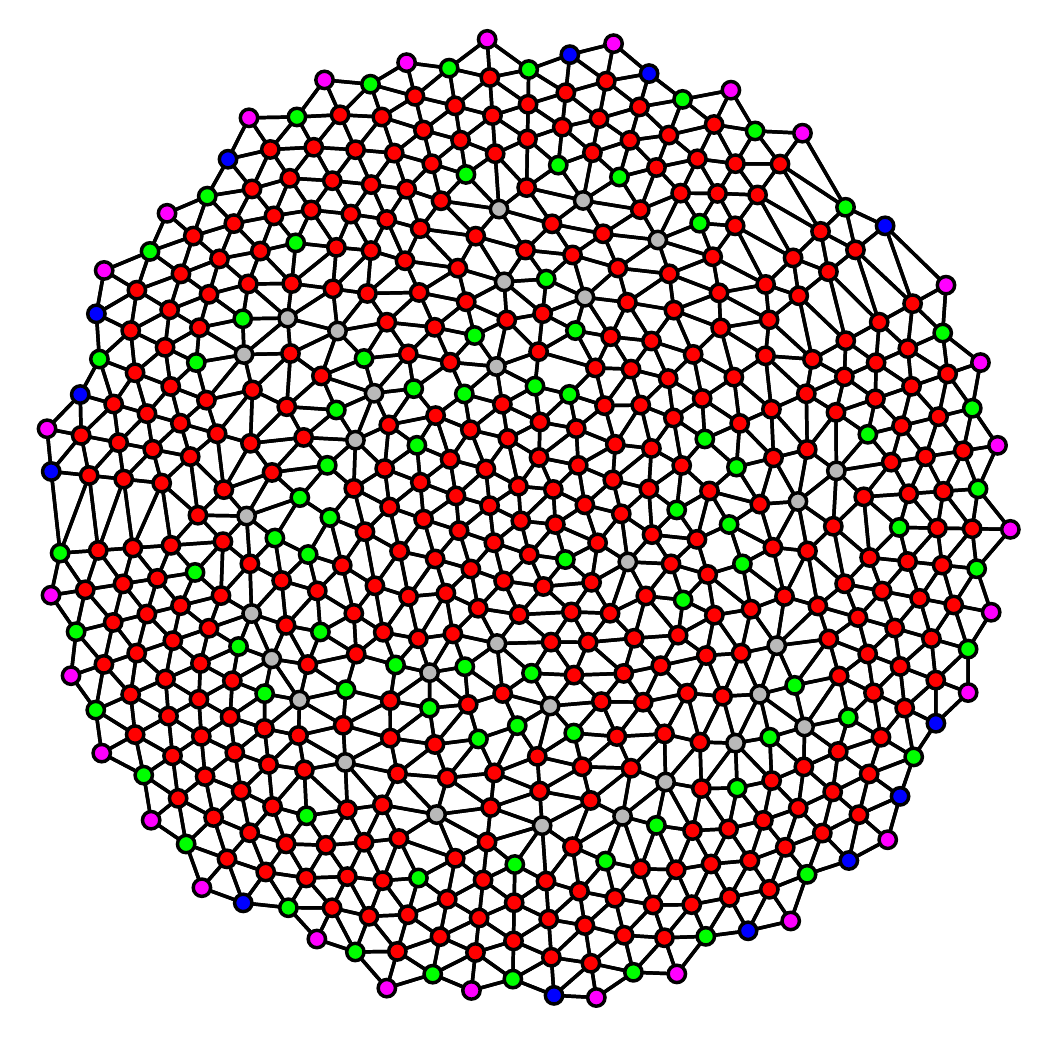}}
\subfigure[$\Delta \theta=5$, $\beta=2$]{\includegraphics[width=0.24\textwidth]{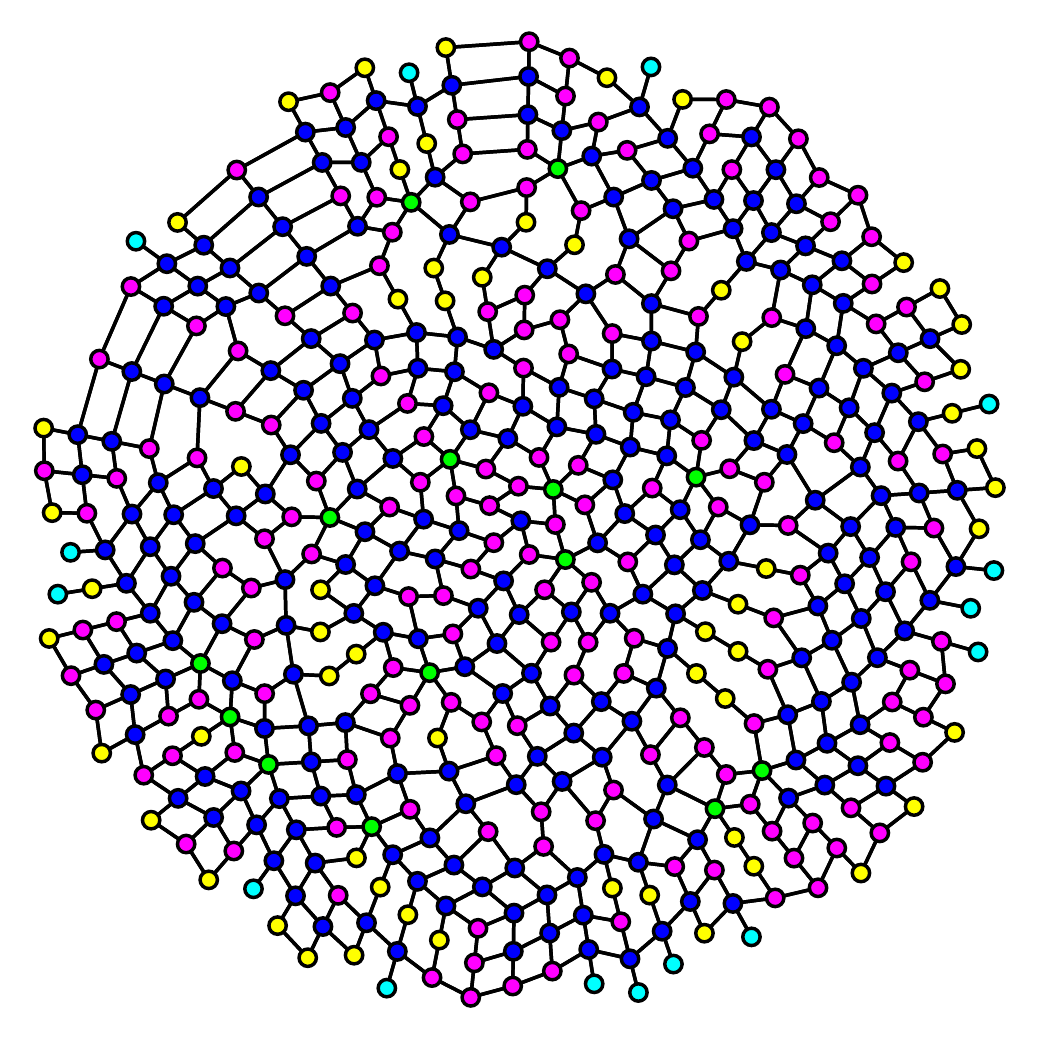}}
\subfigure[$\Delta \theta=5$, $\beta=3$]{\includegraphics[width=0.24\textwidth]{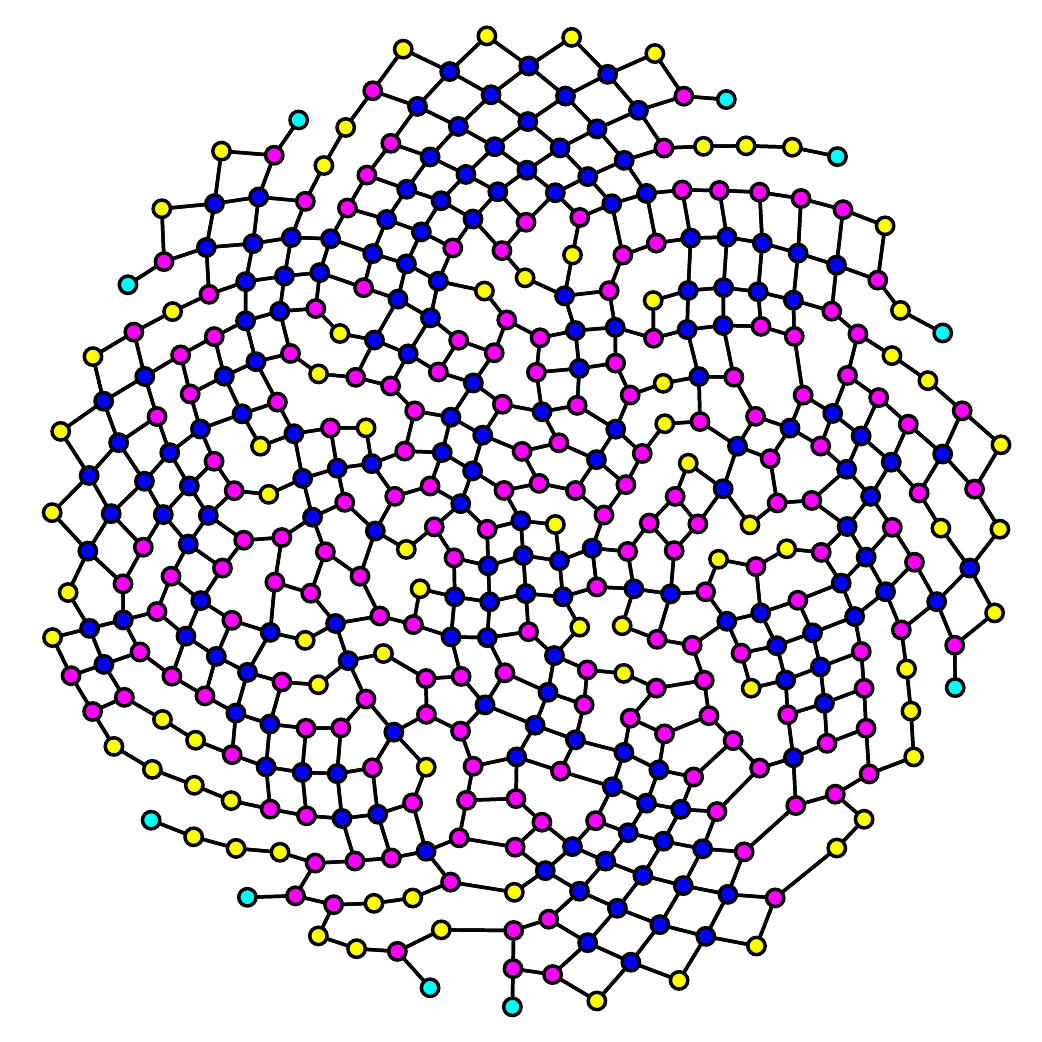}}
\subfigure[$\Delta \theta=5$, $\beta=10$]{\includegraphics[width=0.24\textwidth]{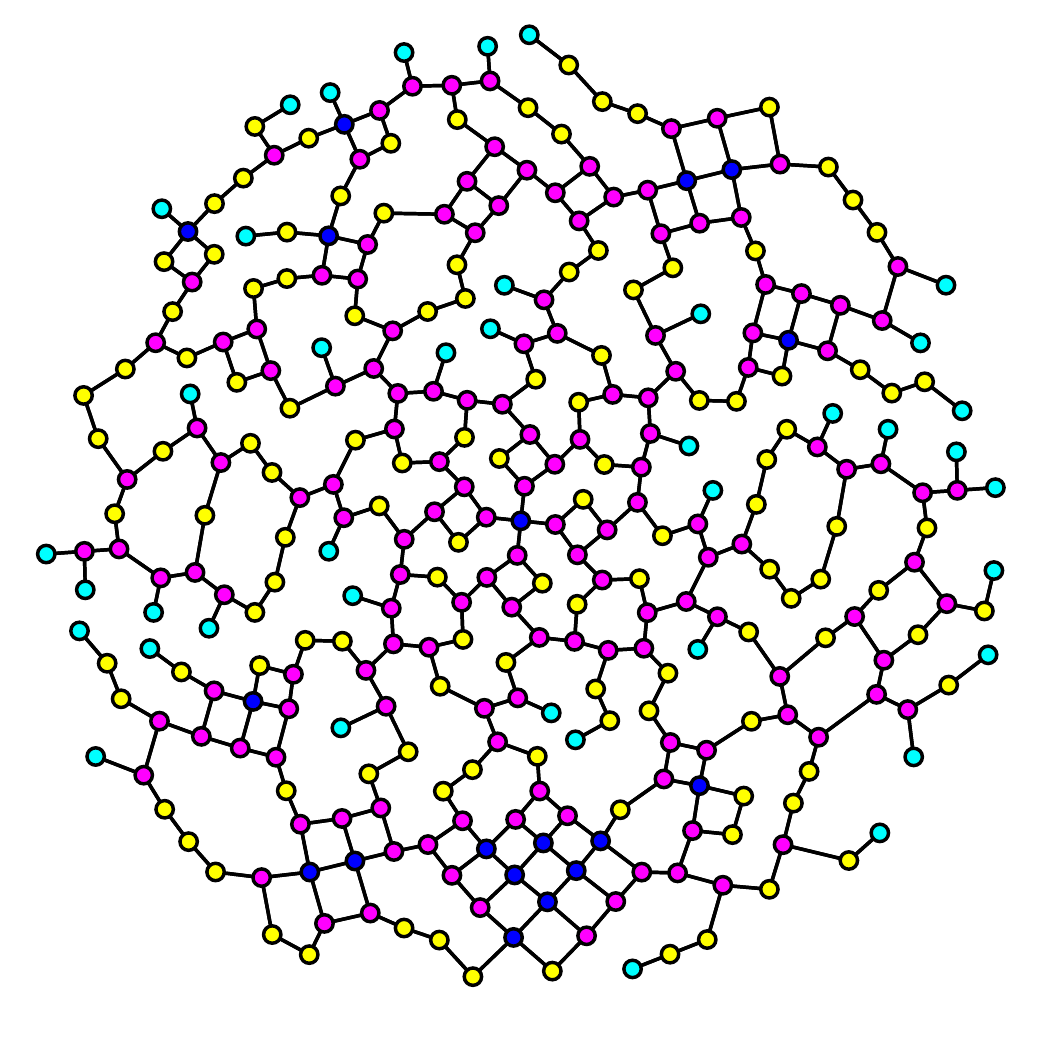}}
\subfigure[$\Delta \theta=5$, $\beta=15$]{\includegraphics[width=0.24\textwidth]{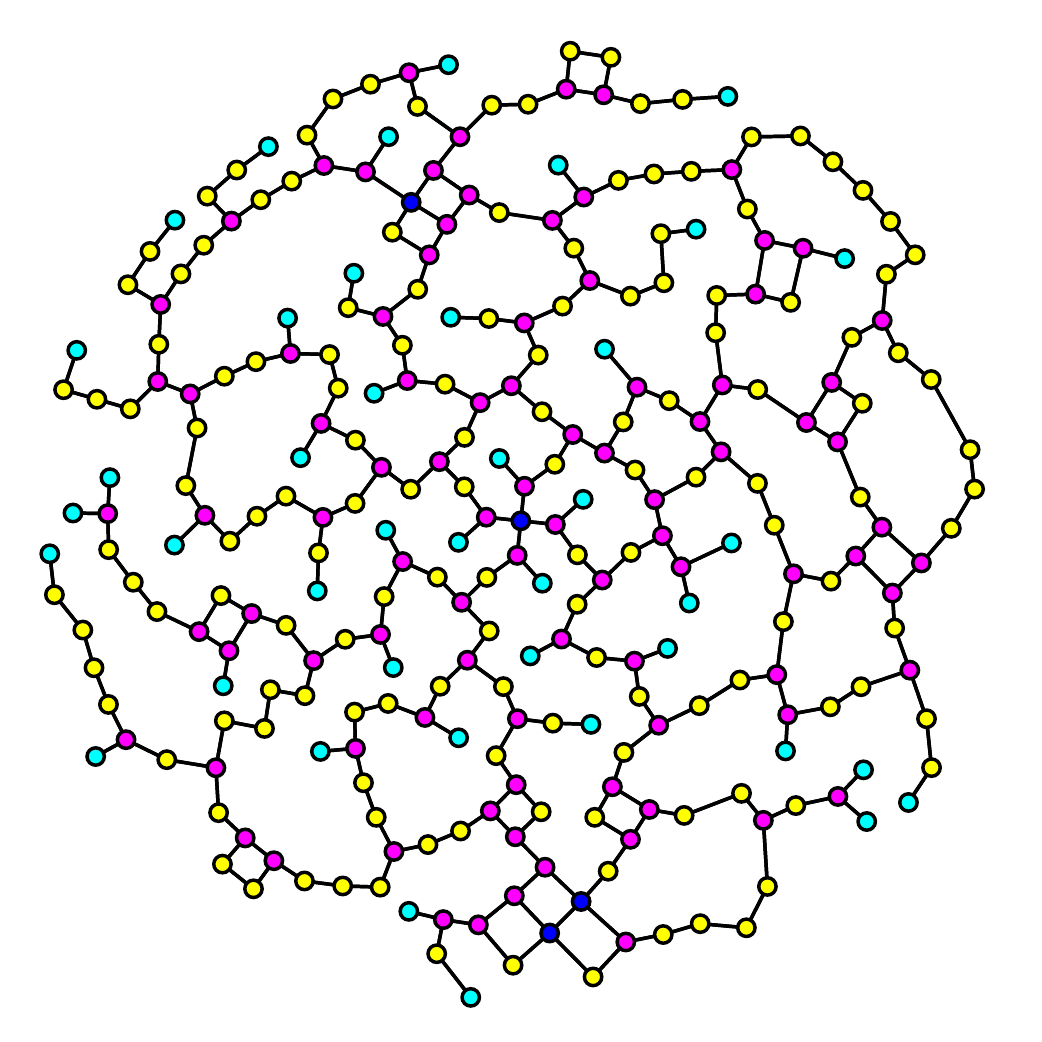}}
\subfigure[$\Delta \theta=5$, $\beta=20$]{\includegraphics[width=0.24\textwidth]{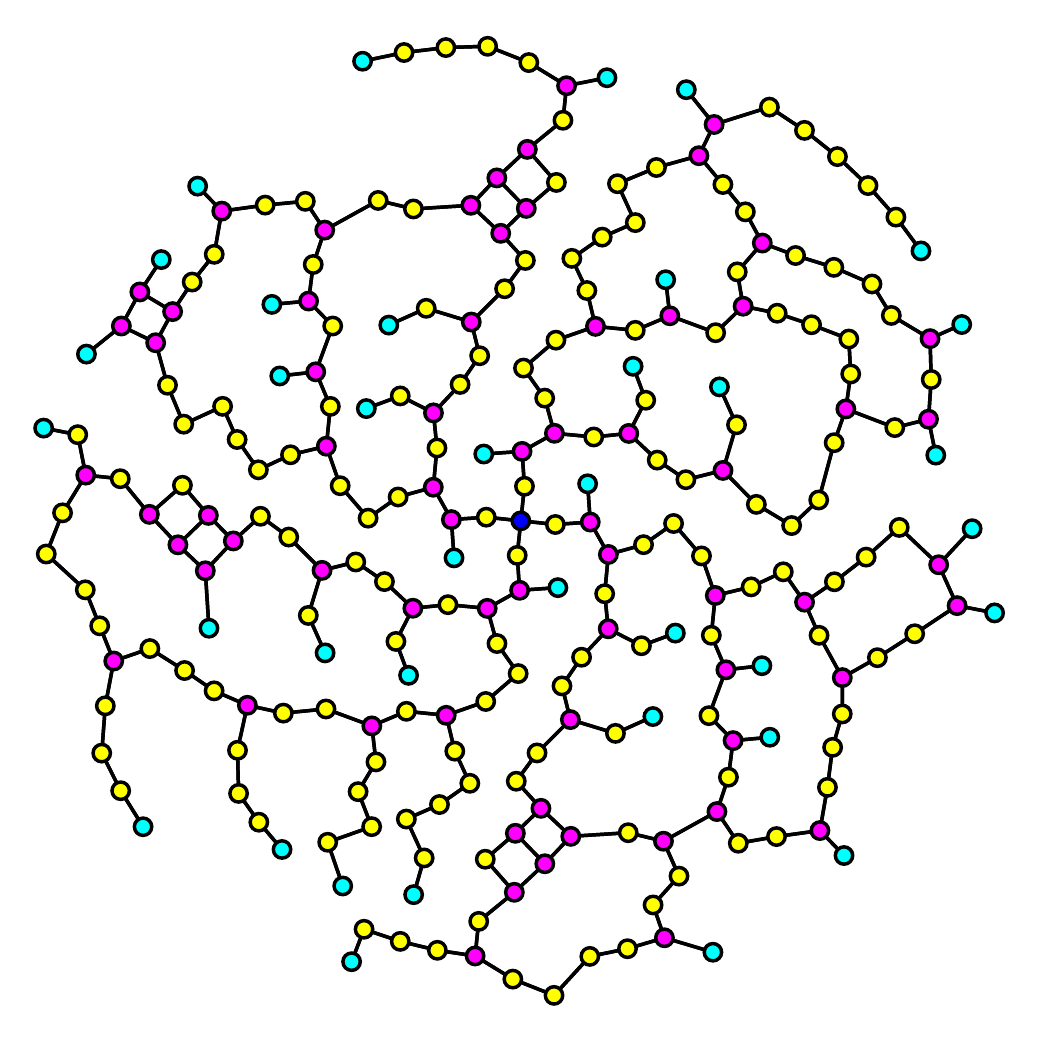}}
\subfigure[$\Delta \theta=5$, $\beta=25$]{\includegraphics[width=0.24\textwidth]{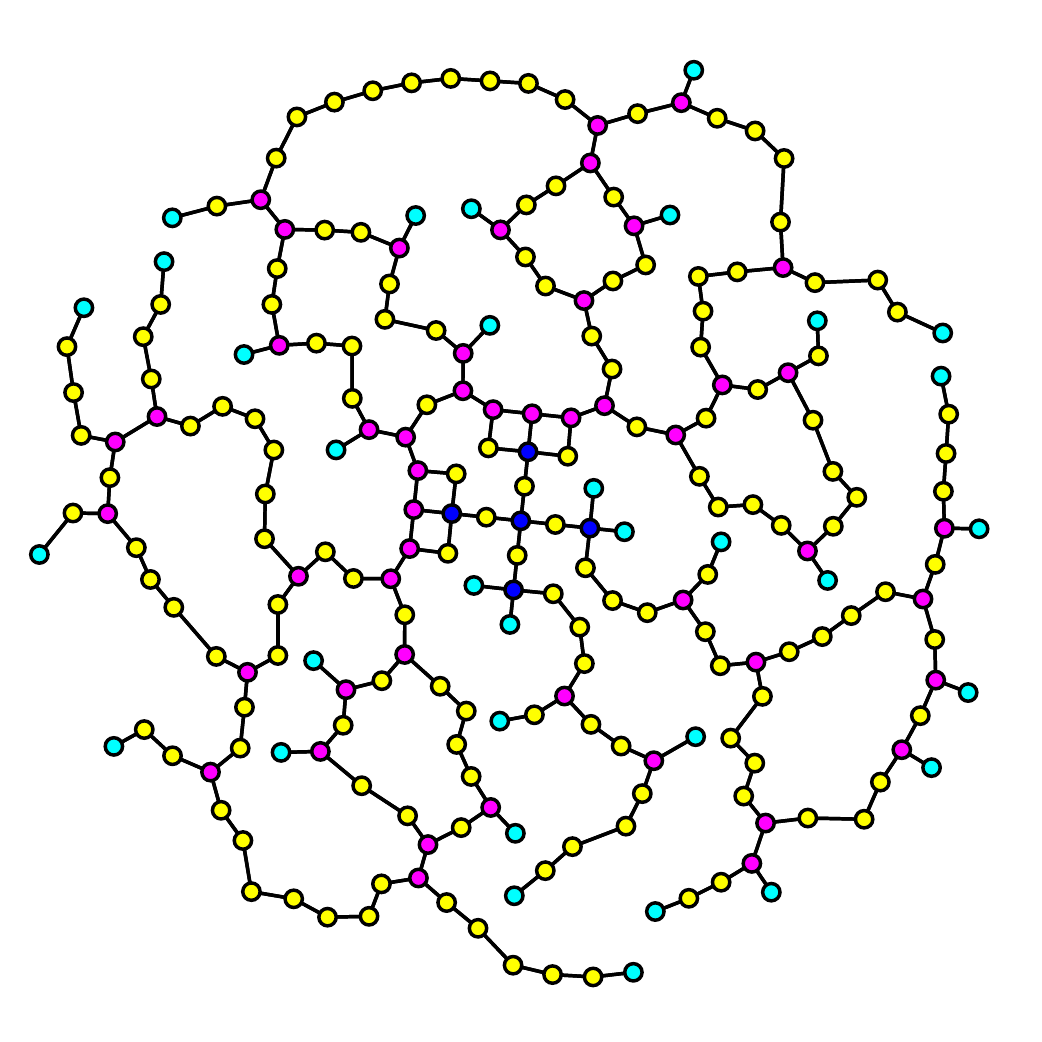}}
\subfigure[$\Delta \theta=5$, $\beta=30$]{\includegraphics[width=0.24\textwidth]{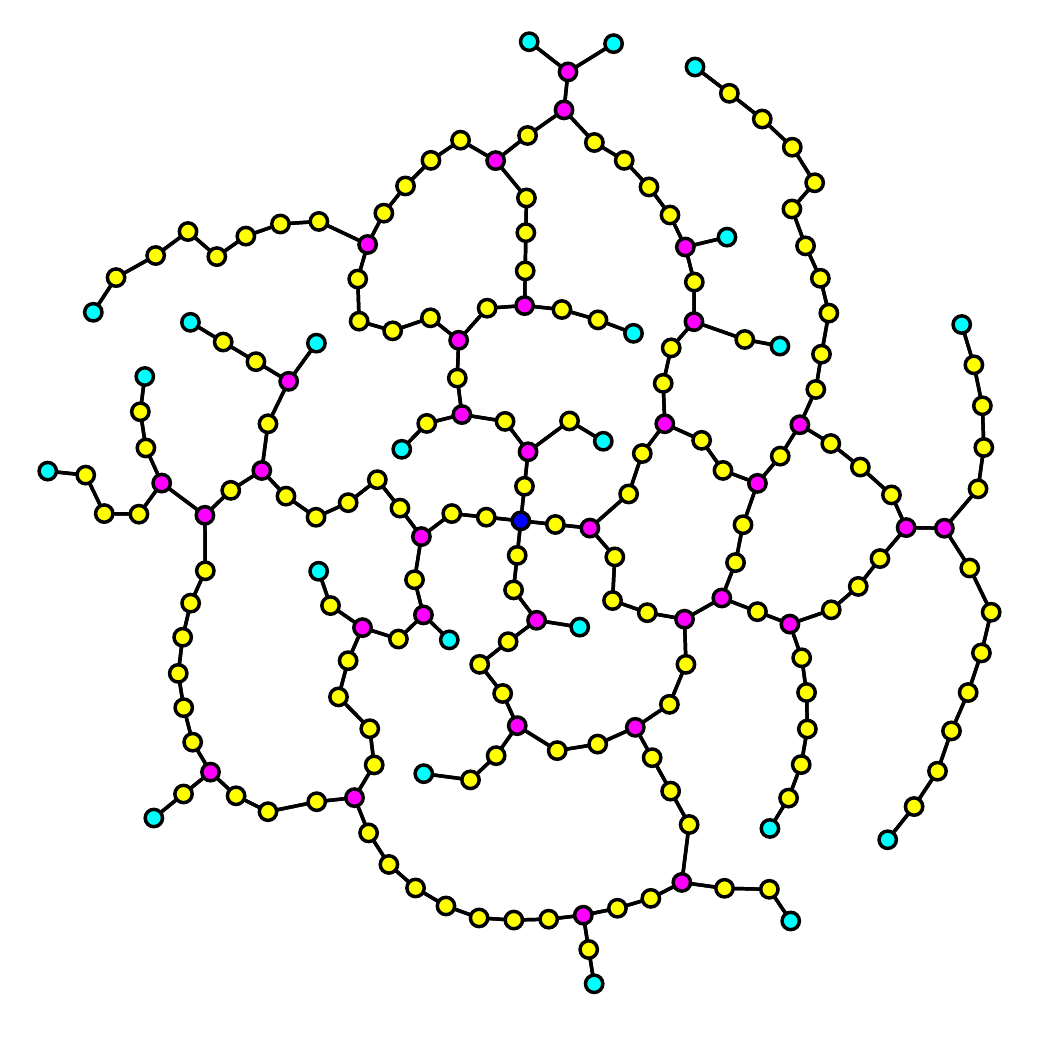}}
\subfigure[$\Delta \theta=5$, $\beta=35$]{\includegraphics[width=0.24\textwidth]{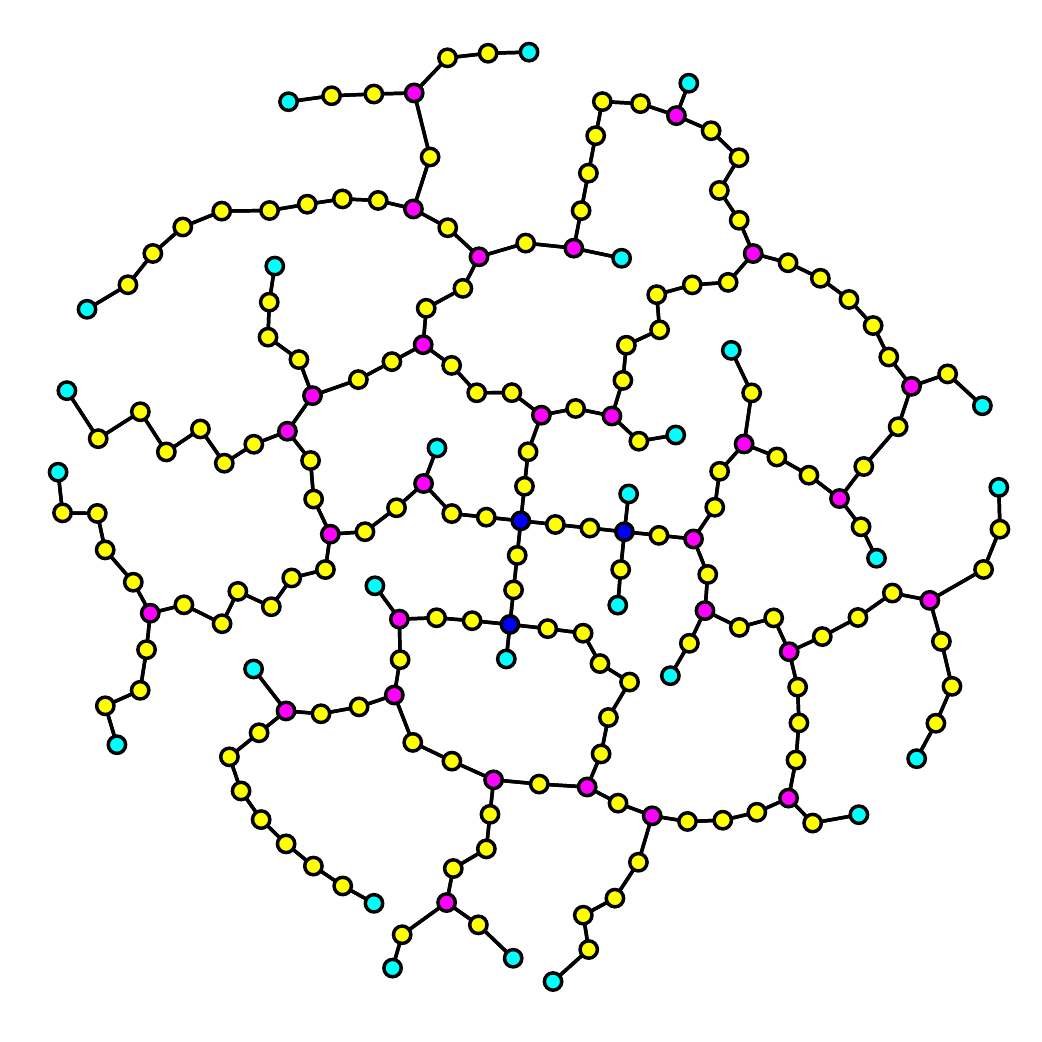}}
\subfigure[$\Delta \theta=5$, $\beta=40$]{\includegraphics[width=0.24\textwidth]{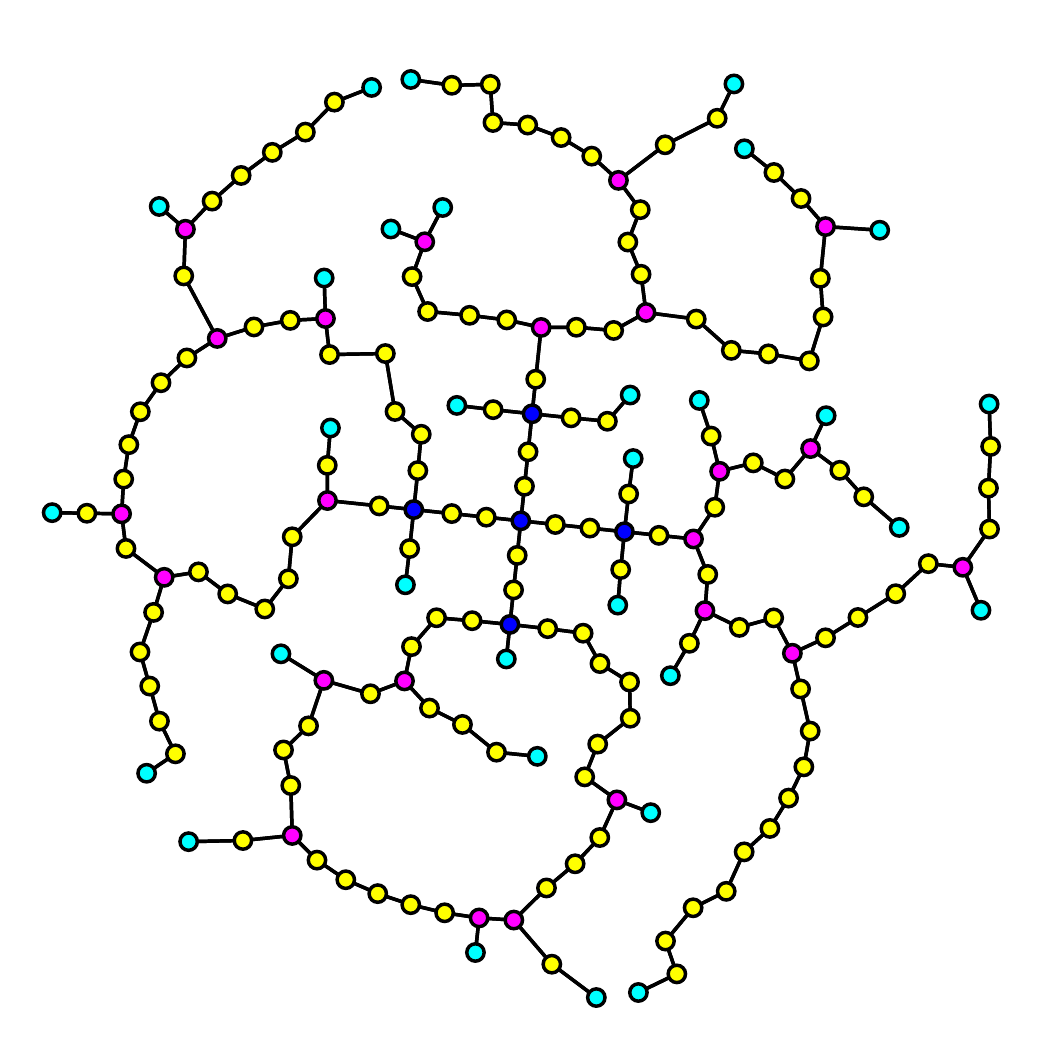}}
\subfigure[$\Delta \theta=5$, $\beta=45$]{\includegraphics[width=0.24\textwidth]{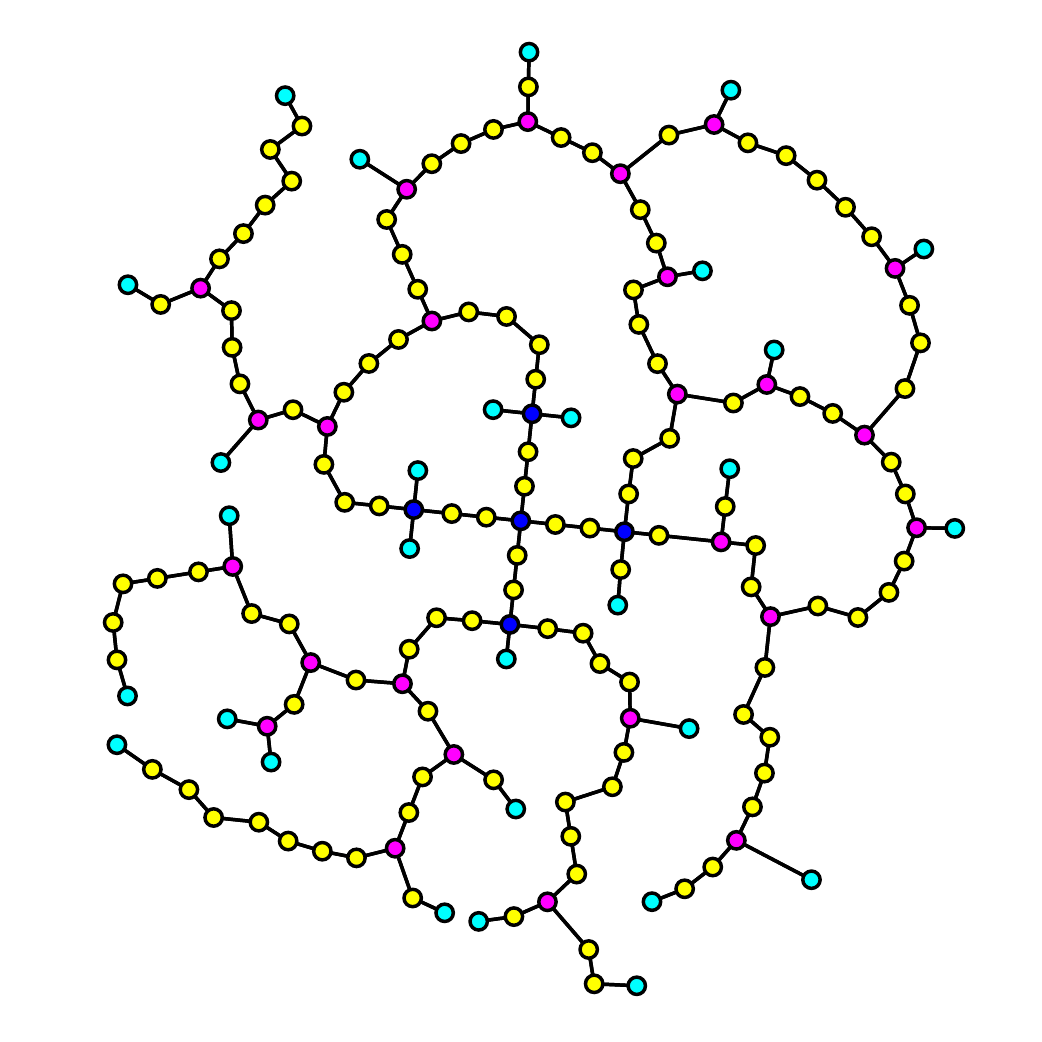}}
\subfigure[$\Delta \theta=5$, $\beta=50$]{\includegraphics[width=0.24\textwidth]{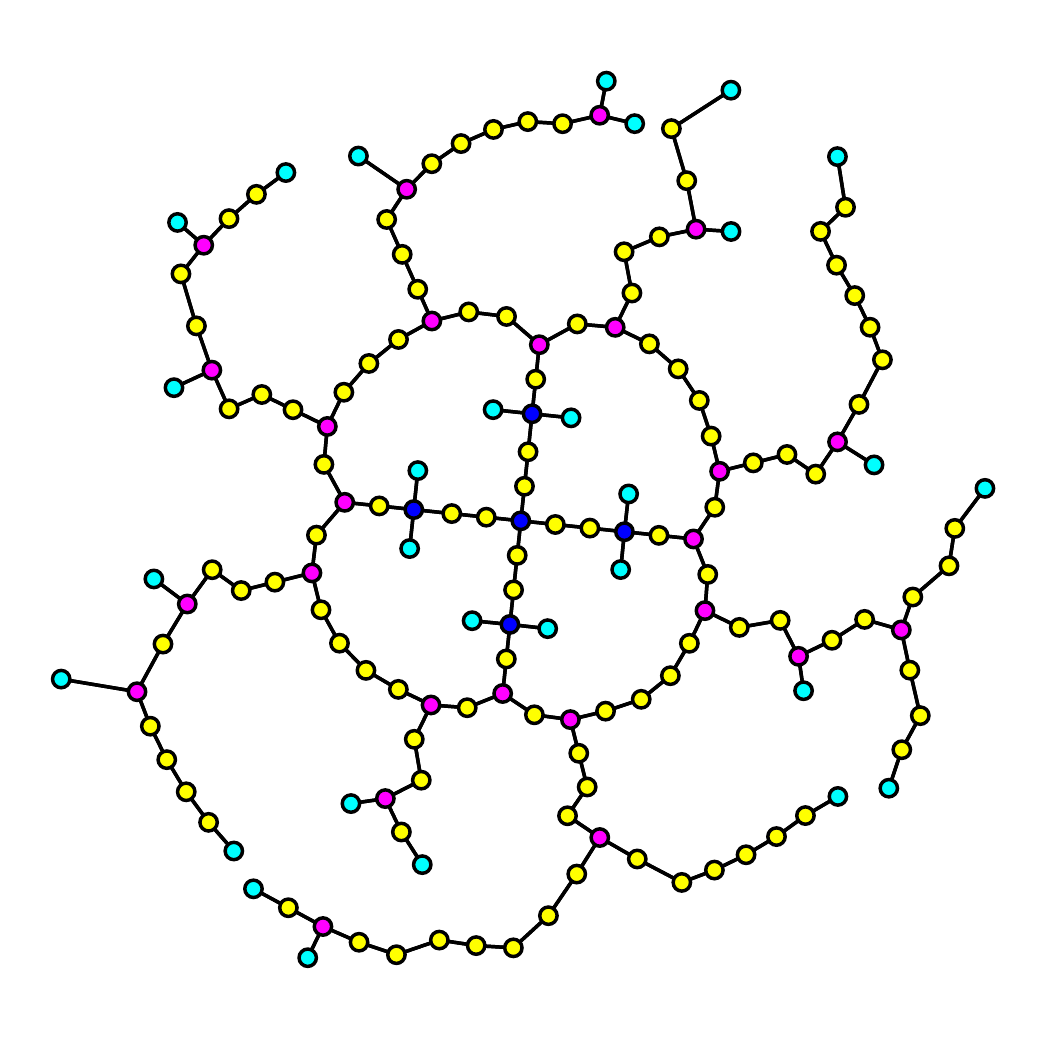}}
\subfigure[$\Delta \theta=10$, $\beta=1$]{\includegraphics[width=0.24\textwidth]{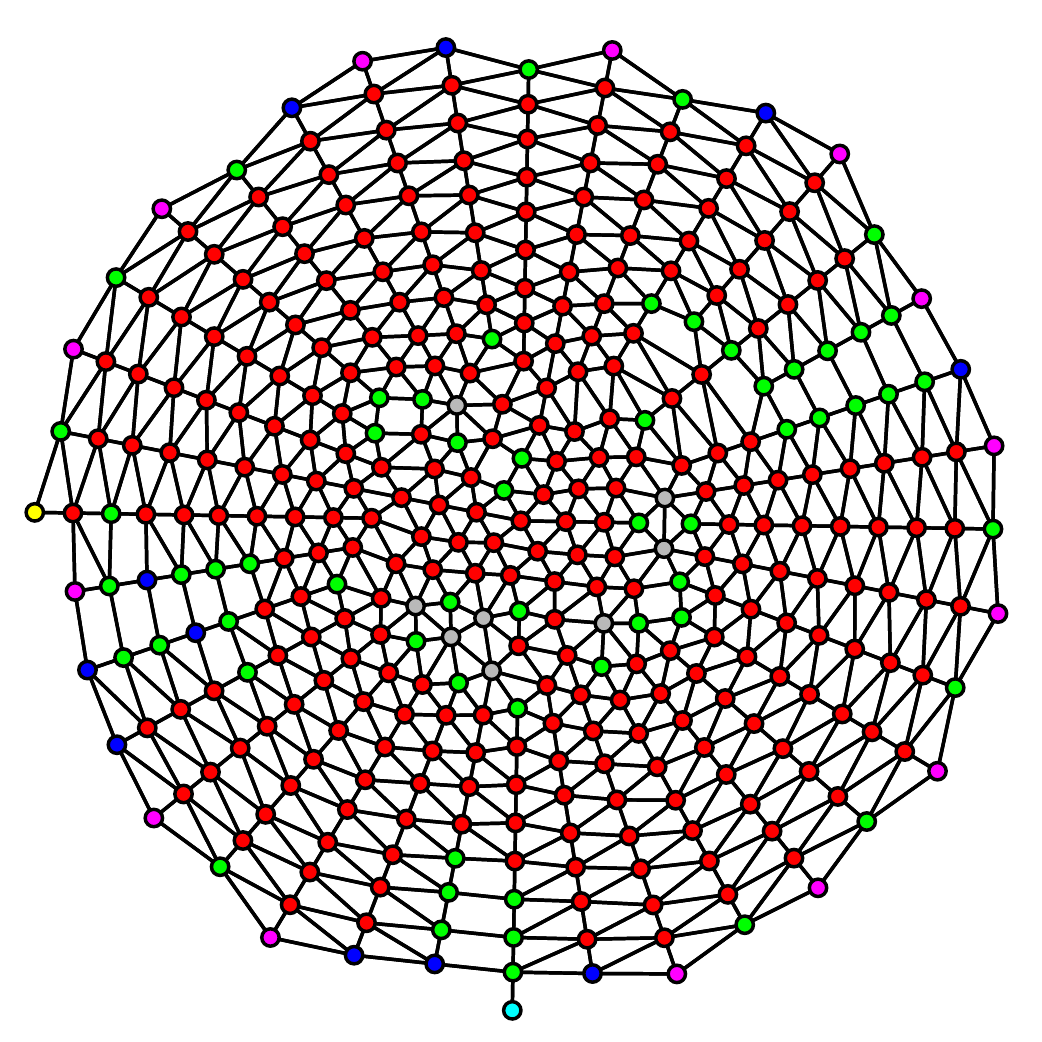}}
\subfigure[$\Delta \theta=10$, $\beta=2$]{\includegraphics[width=0.24\textwidth]{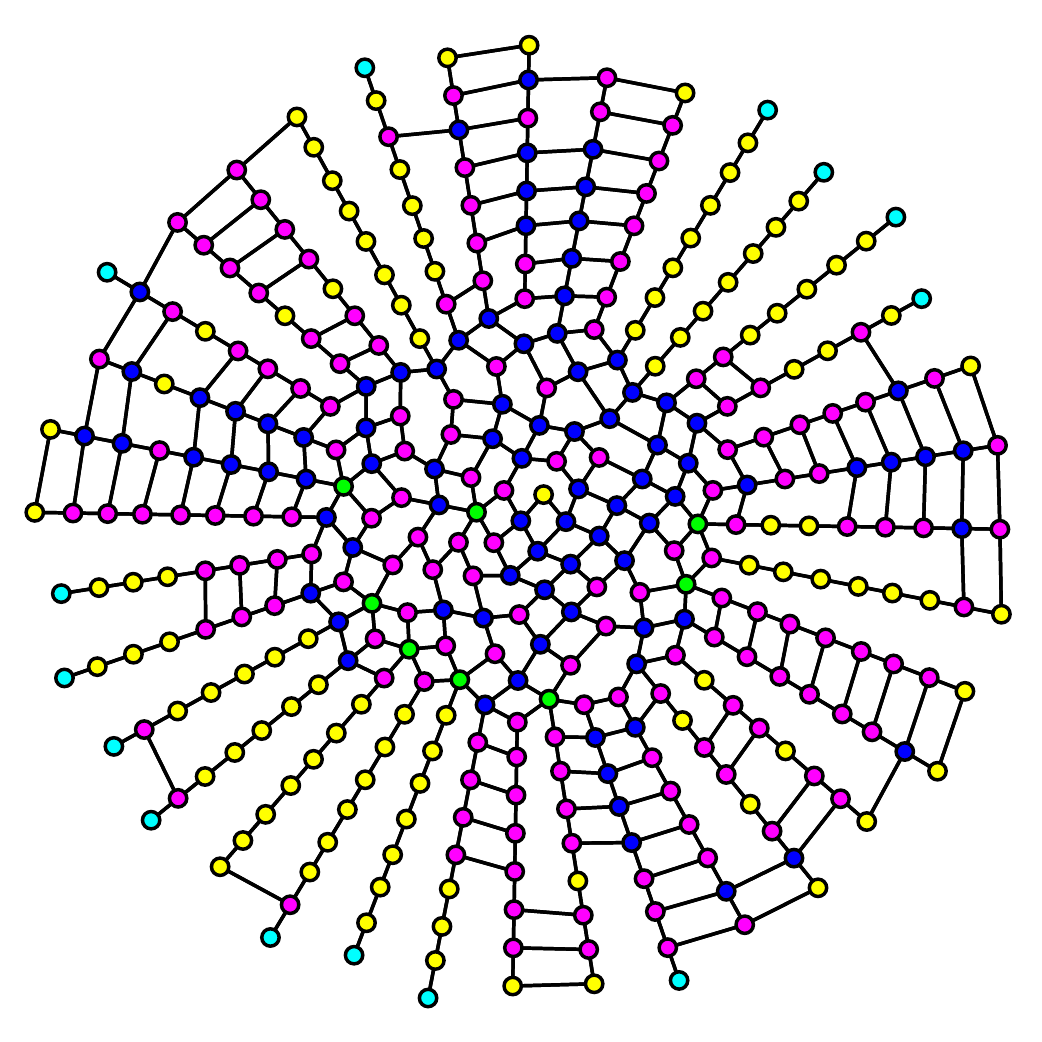}}
\subfigure[$\Delta \theta=10$, $\beta=3$]{\includegraphics[width=0.24\textwidth]{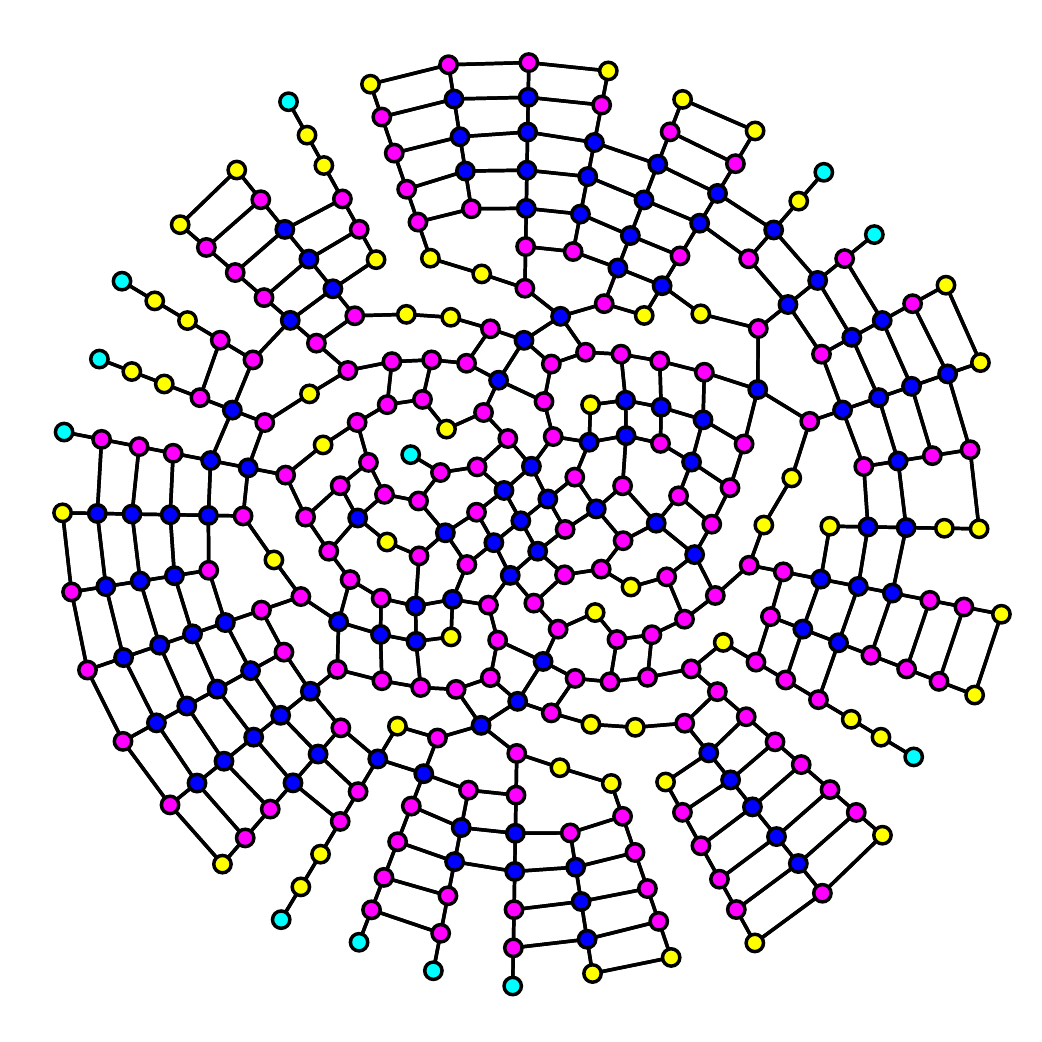}}
\subfigure[$\Delta \theta=10$,$\beta=10$]{\includegraphics[width=0.24\textwidth]{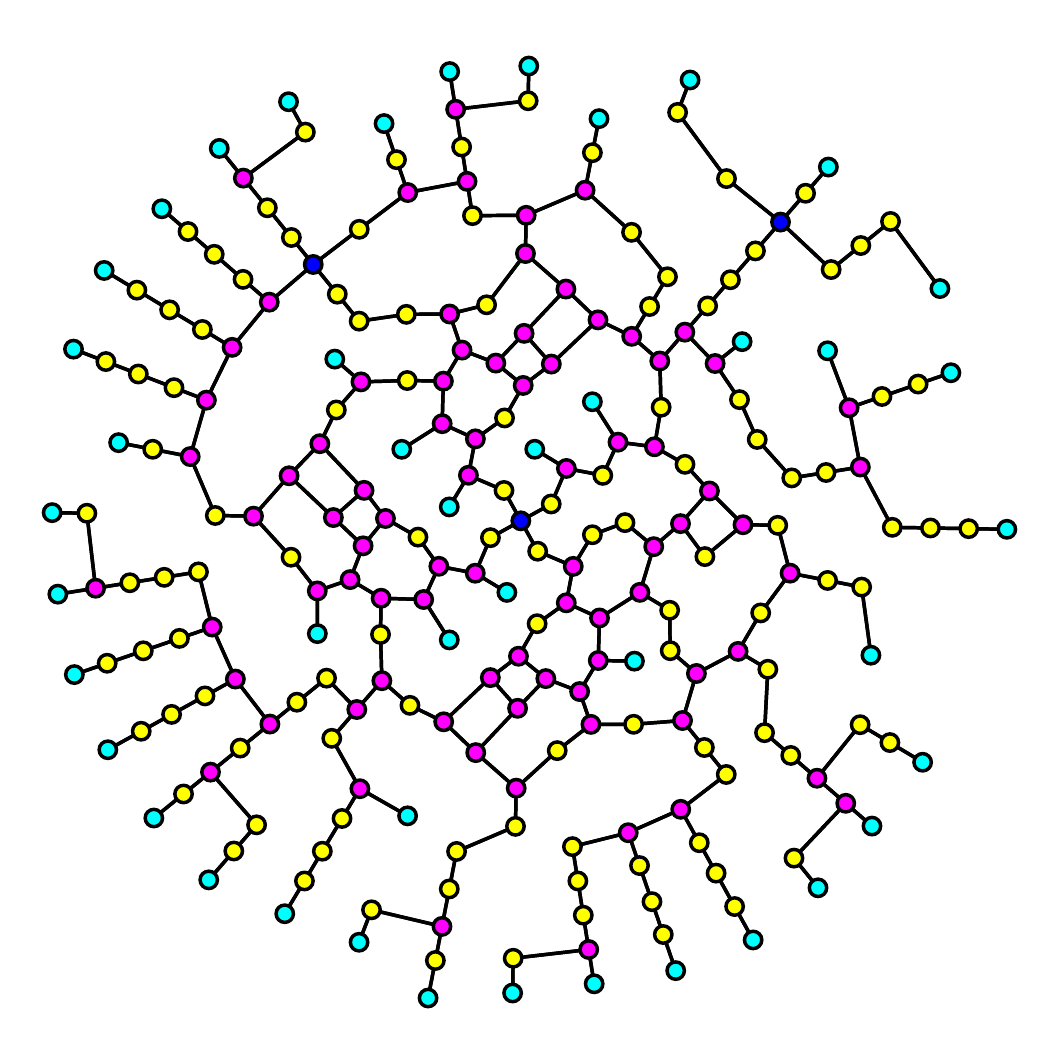}}
\subfigure[$\Delta \theta=10$,$\beta=20$]{\includegraphics[width=0.24\textwidth]{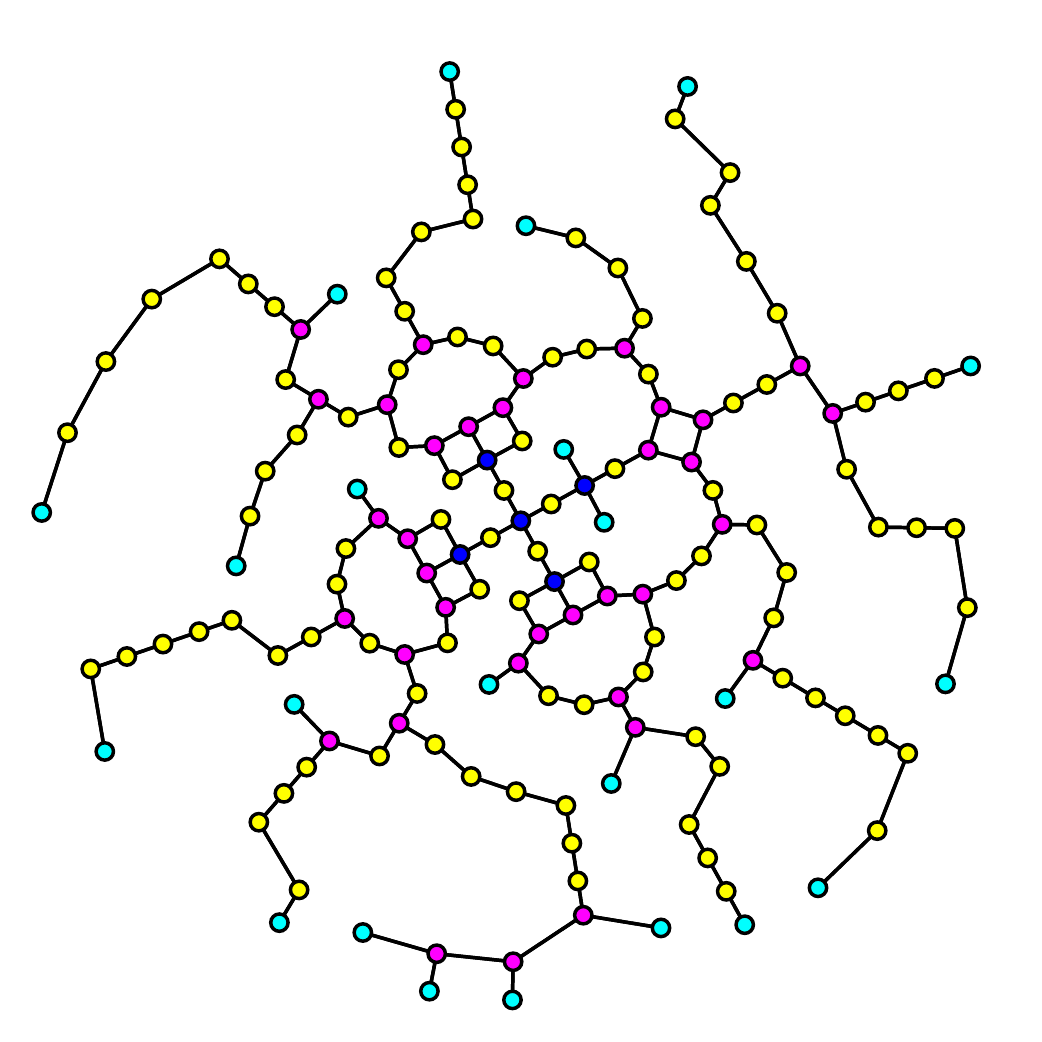}}
\subfigure[$\Delta \theta=10$, $\beta=30$]{\includegraphics[width=0.24\textwidth]{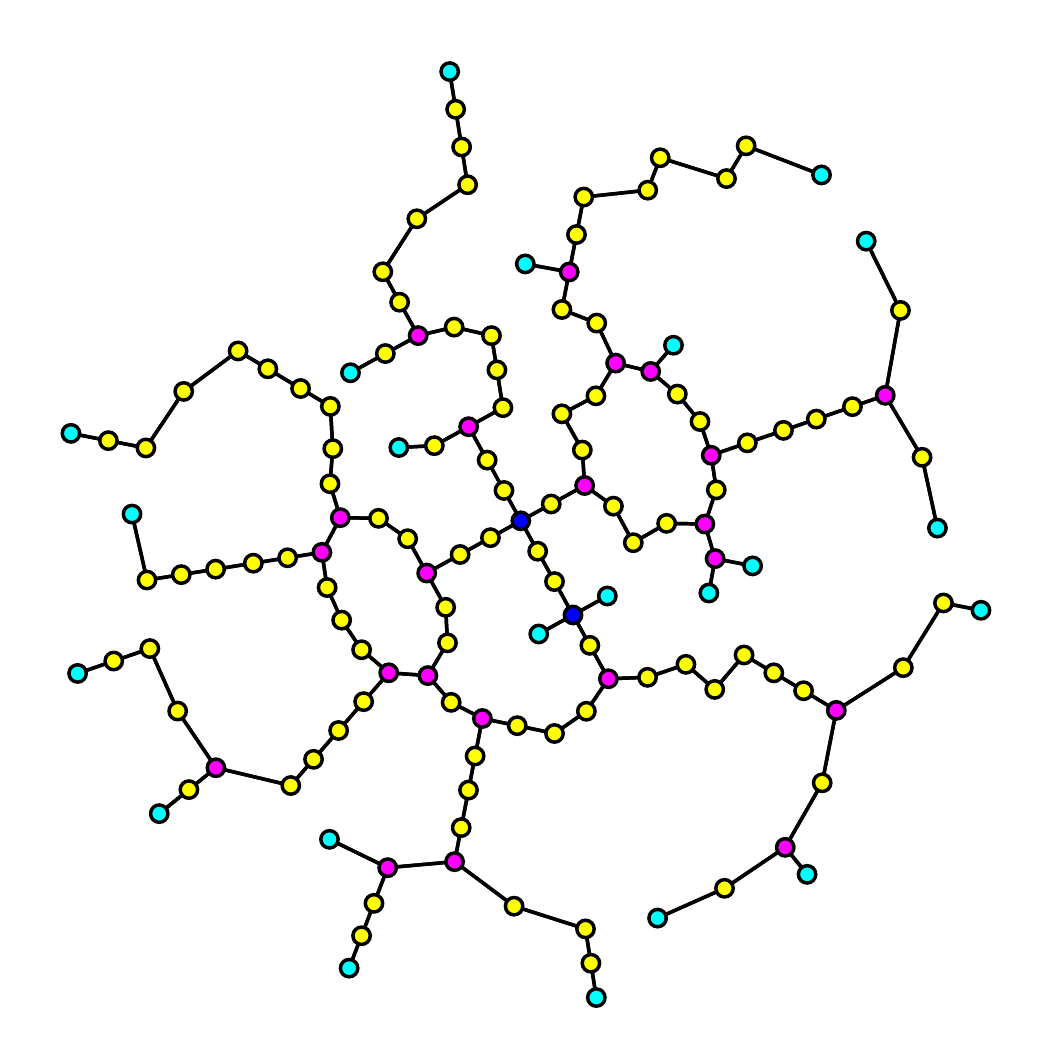}}
\subfigure[$\Delta \theta=10$, $\beta=40$]{\includegraphics[width=0.24\textwidth]{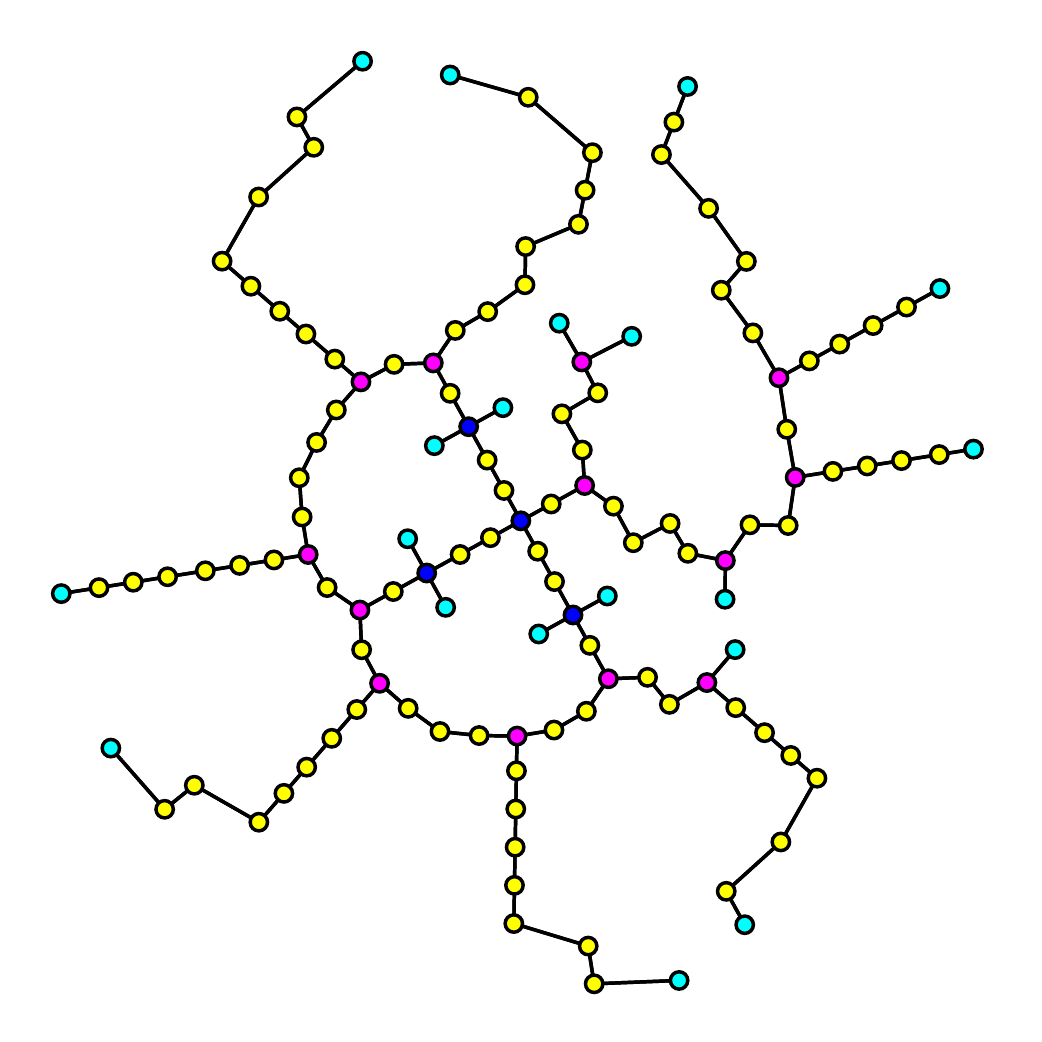}}
\subfigure[$\Delta \theta=10$, $\beta=50$]{\includegraphics[width=0.24\textwidth]{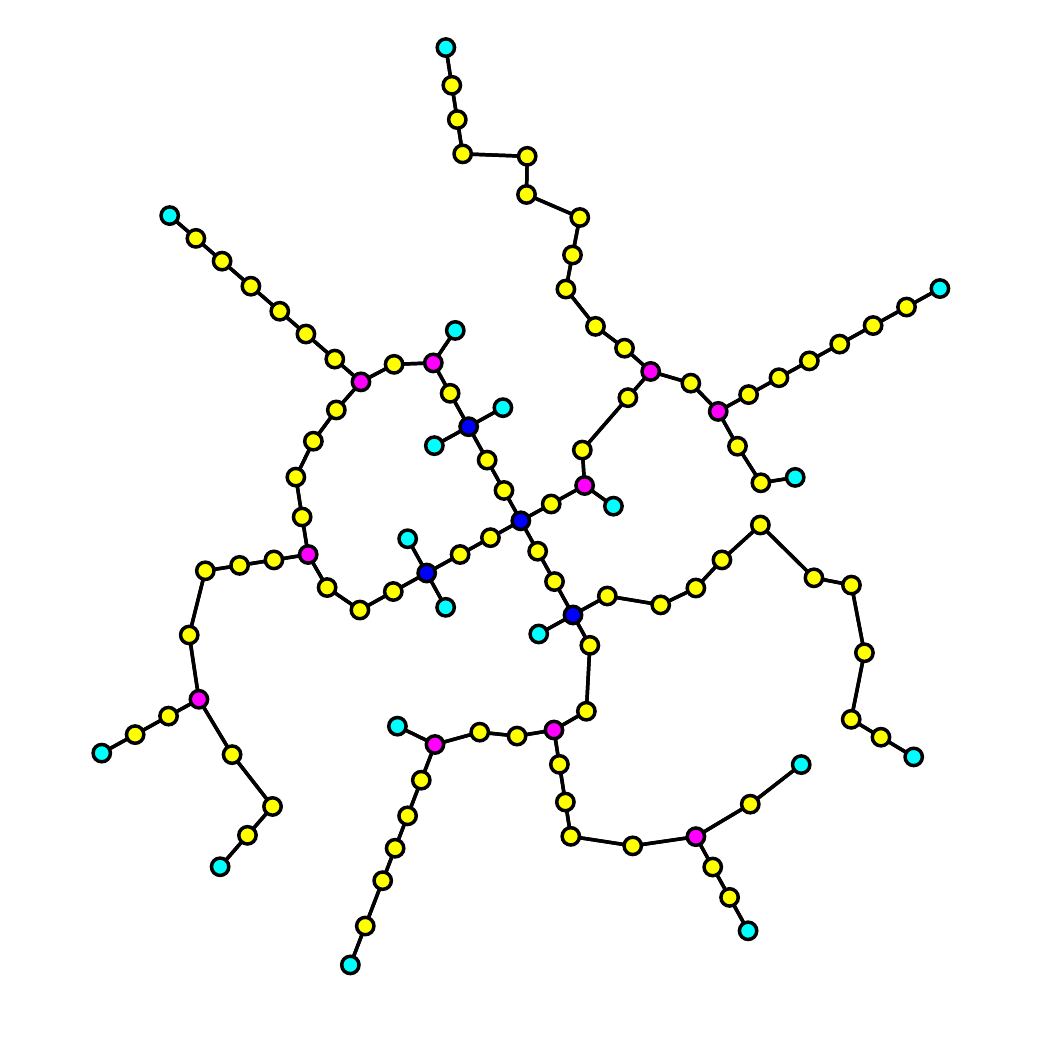}}
\caption{ $\beta$-skeletons grown with $\Delta\theta=5$ (a--l) and $\Delta\theta=10$ (m--t).}
\label{theta05}
\end{figure}

\begin{figure}[!tbp]
\centering
\subfigure[$\Delta \theta=15$, $\beta=1$]{\includegraphics[width=0.24\textwidth]{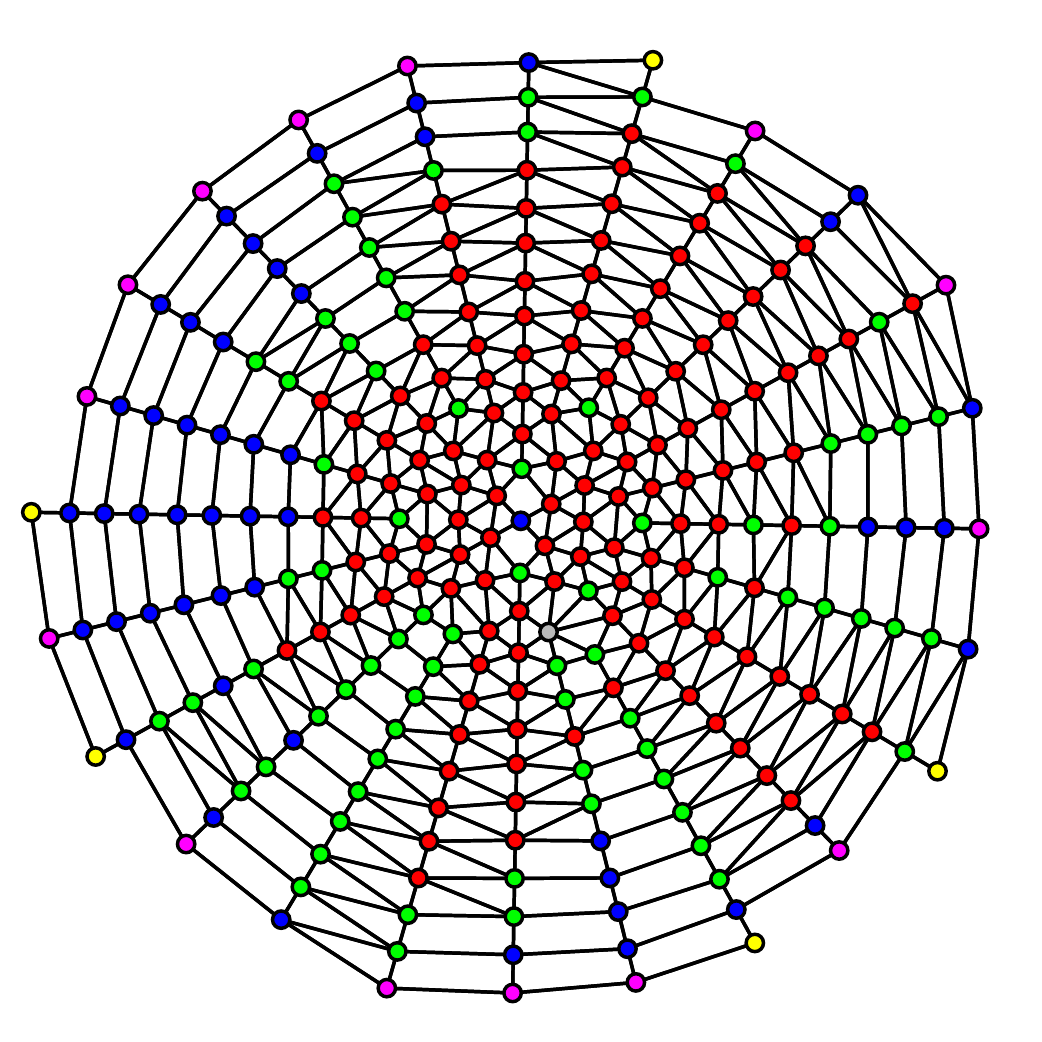}}
\subfigure[$\Delta \theta=15$, $\beta=2$]{\includegraphics[width=0.24\textwidth]{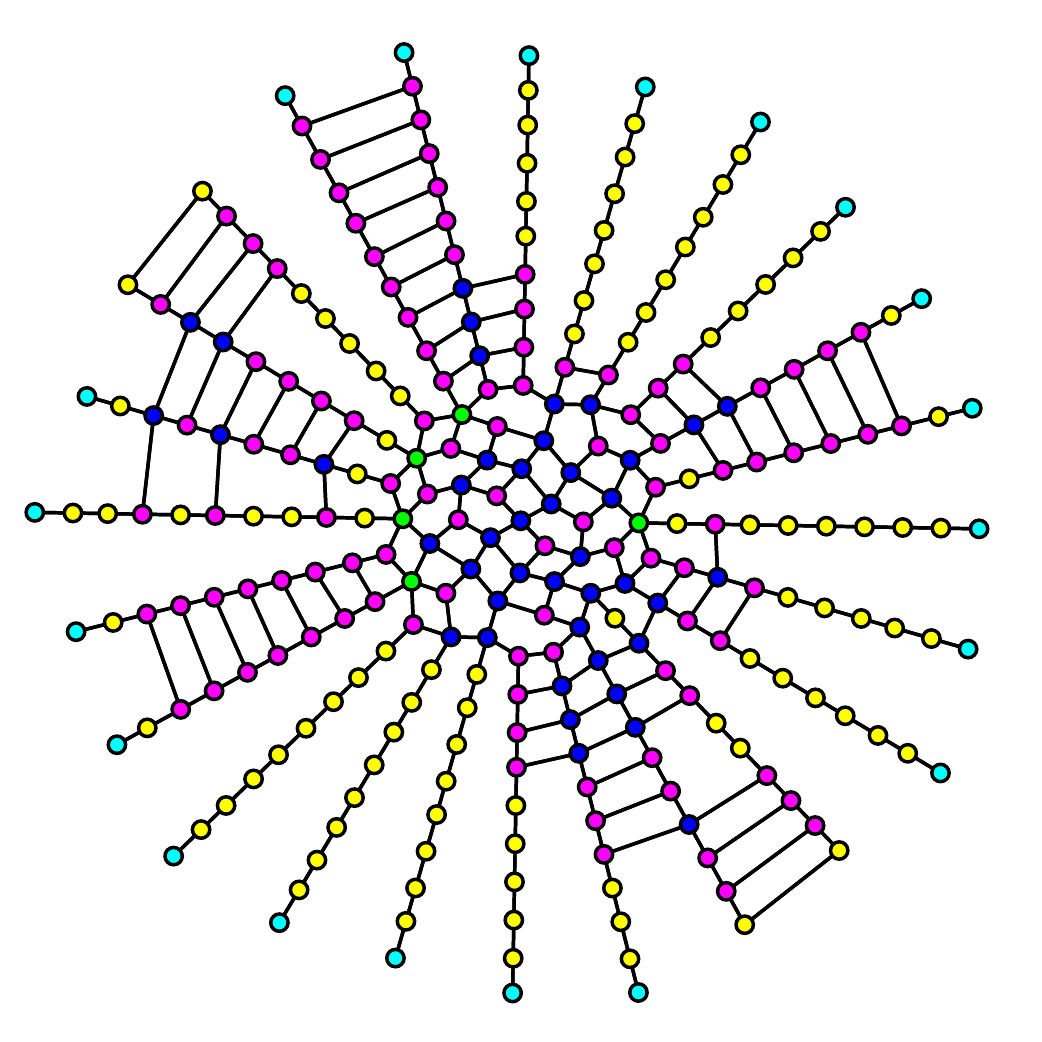}}
\subfigure[$\Delta \theta=15$, $\beta=3$]{\includegraphics[width=0.24\textwidth]{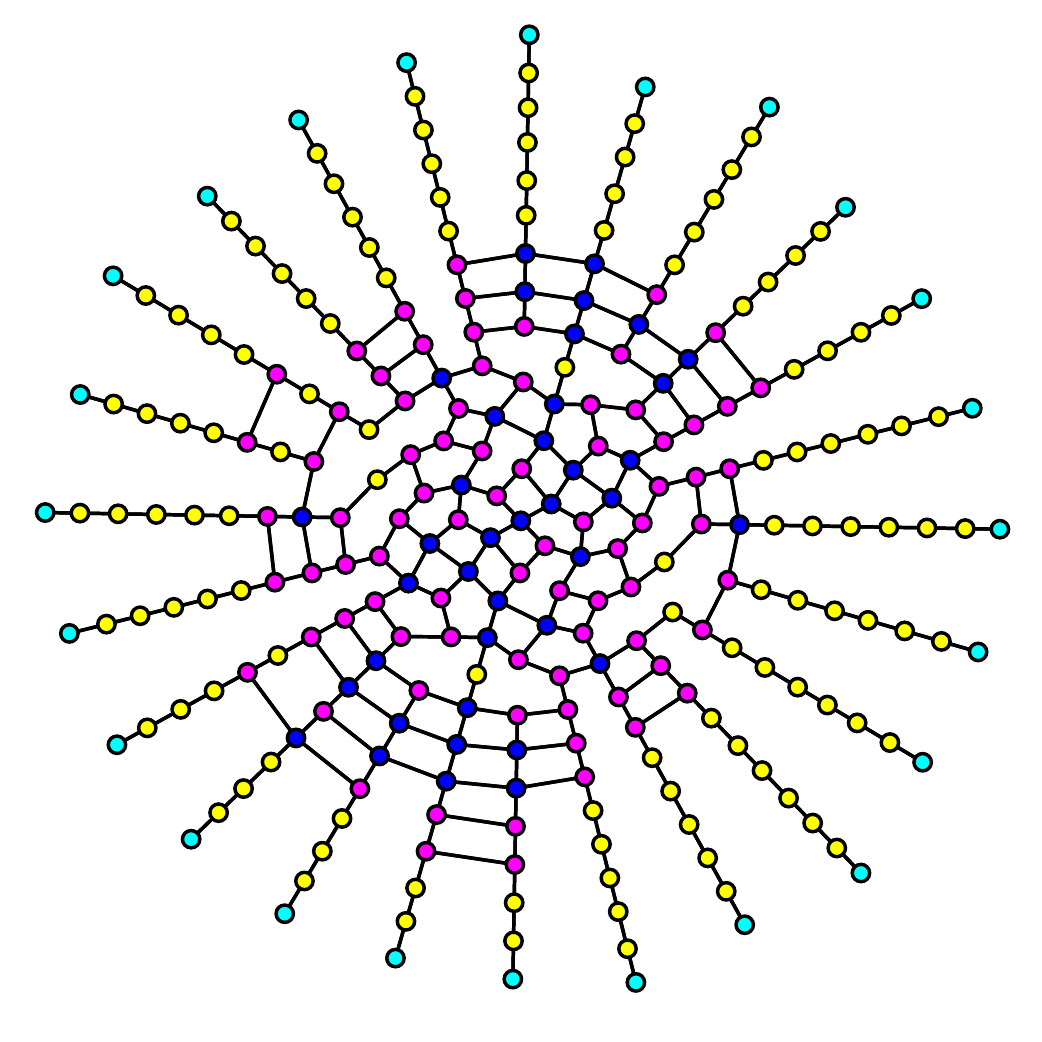}}
\subfigure[$\Delta \theta=15$, $\beta=10$]{\includegraphics[width=0.25\textwidth]{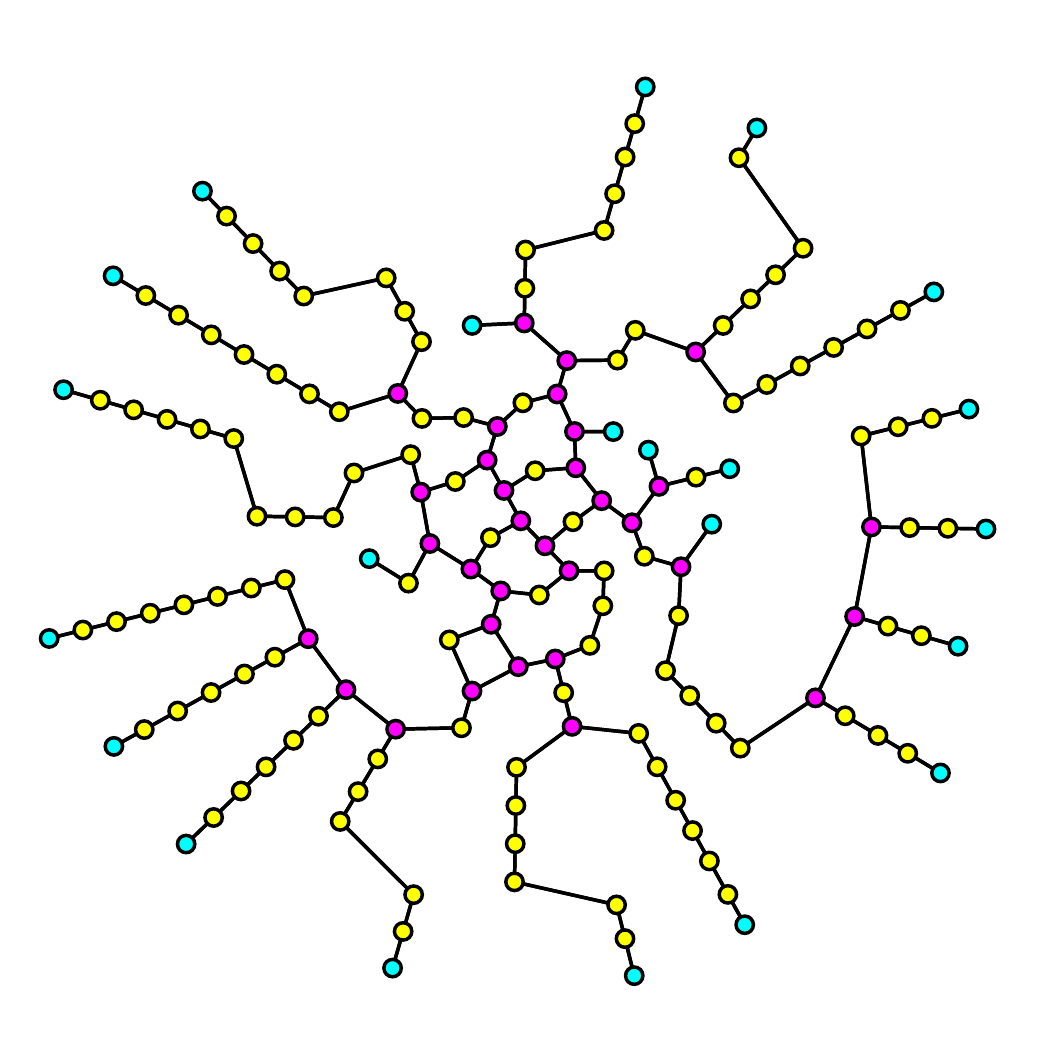}}
\subfigure[$\Delta \theta=15$, $\beta=20$]{\includegraphics[width=0.24\textwidth]{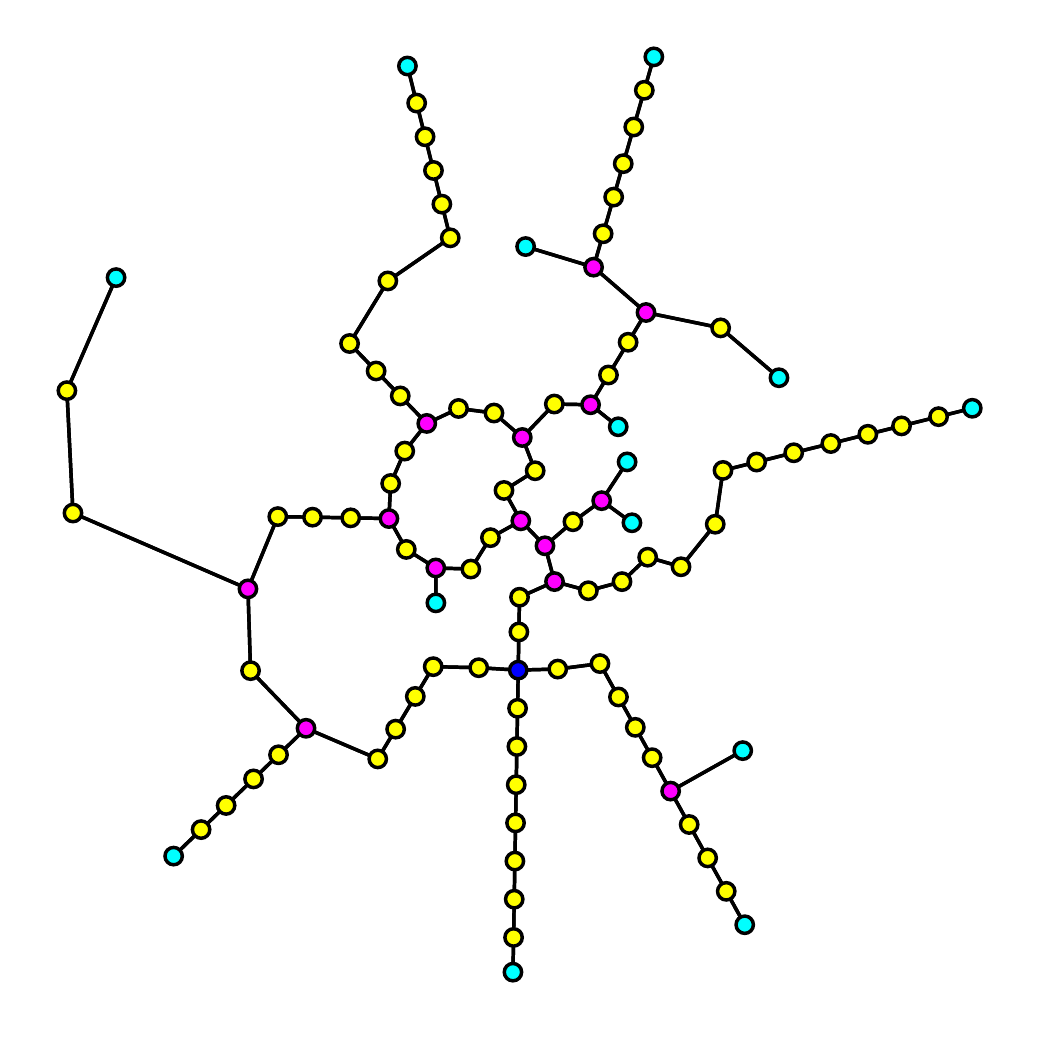}}
\subfigure[$\Delta \theta=15$, $\beta=30$]{\includegraphics[width=0.24\textwidth]{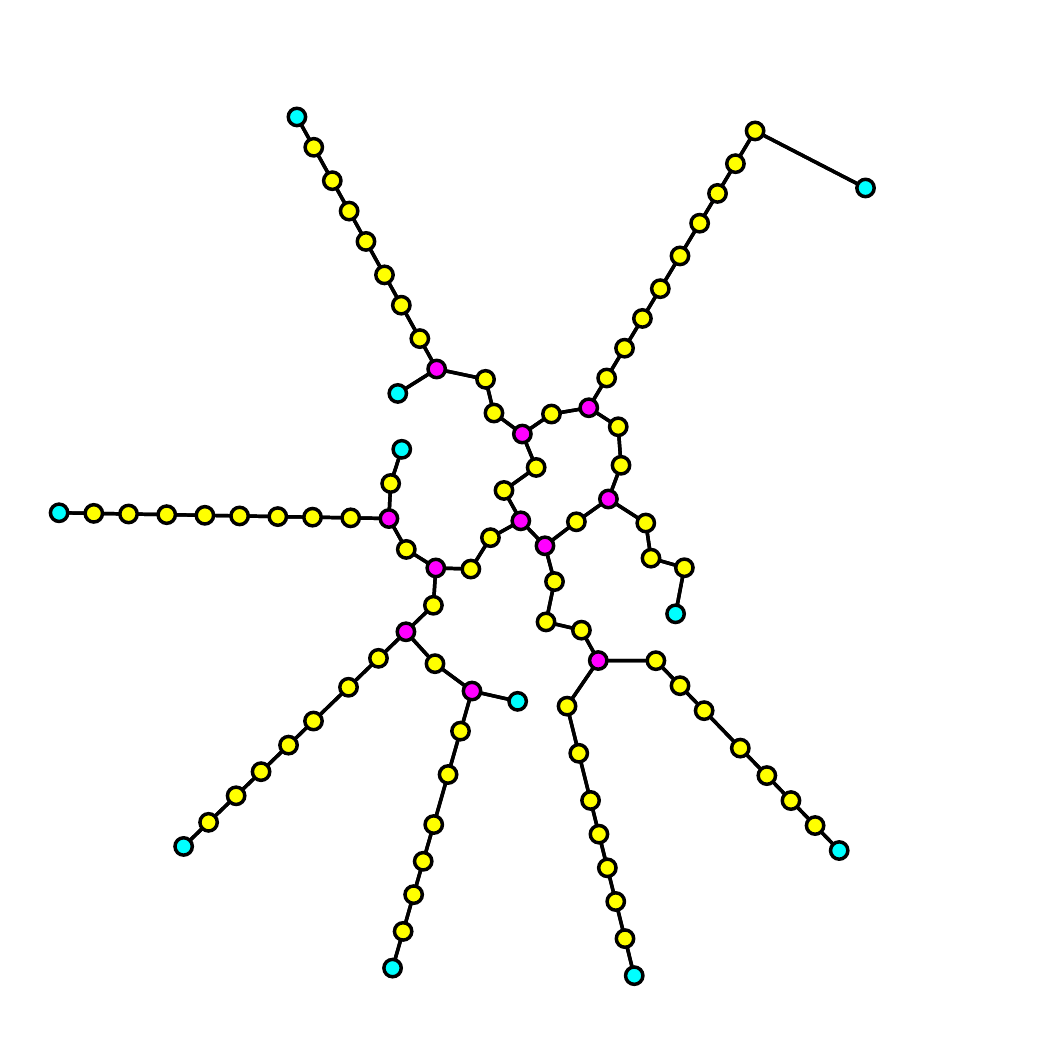}}
\subfigure[$\Delta \theta=15$, $\beta=40$]{\includegraphics[width=0.25\textwidth]{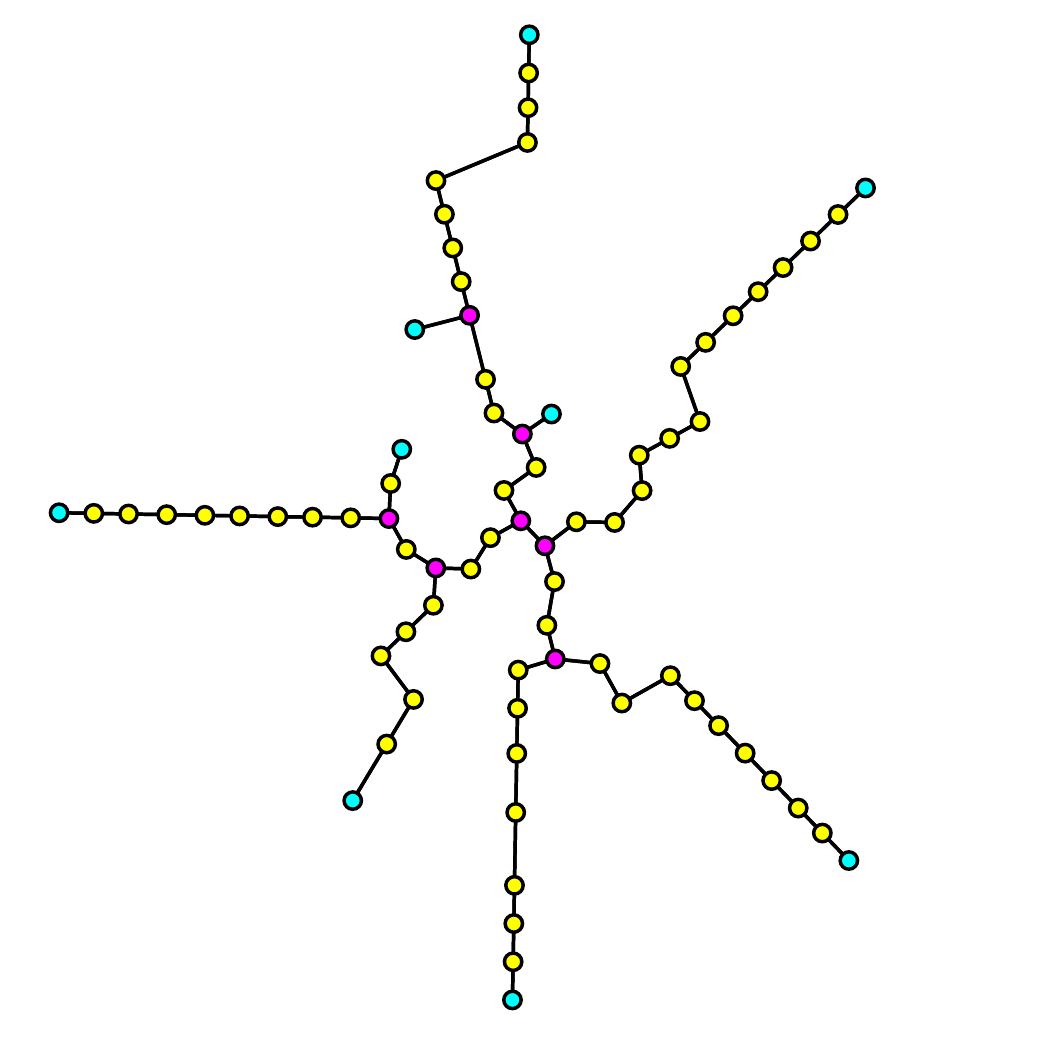}}
\subfigure[$\Delta \theta=15$, $\beta=50$]{\includegraphics[width=0.25\textwidth]{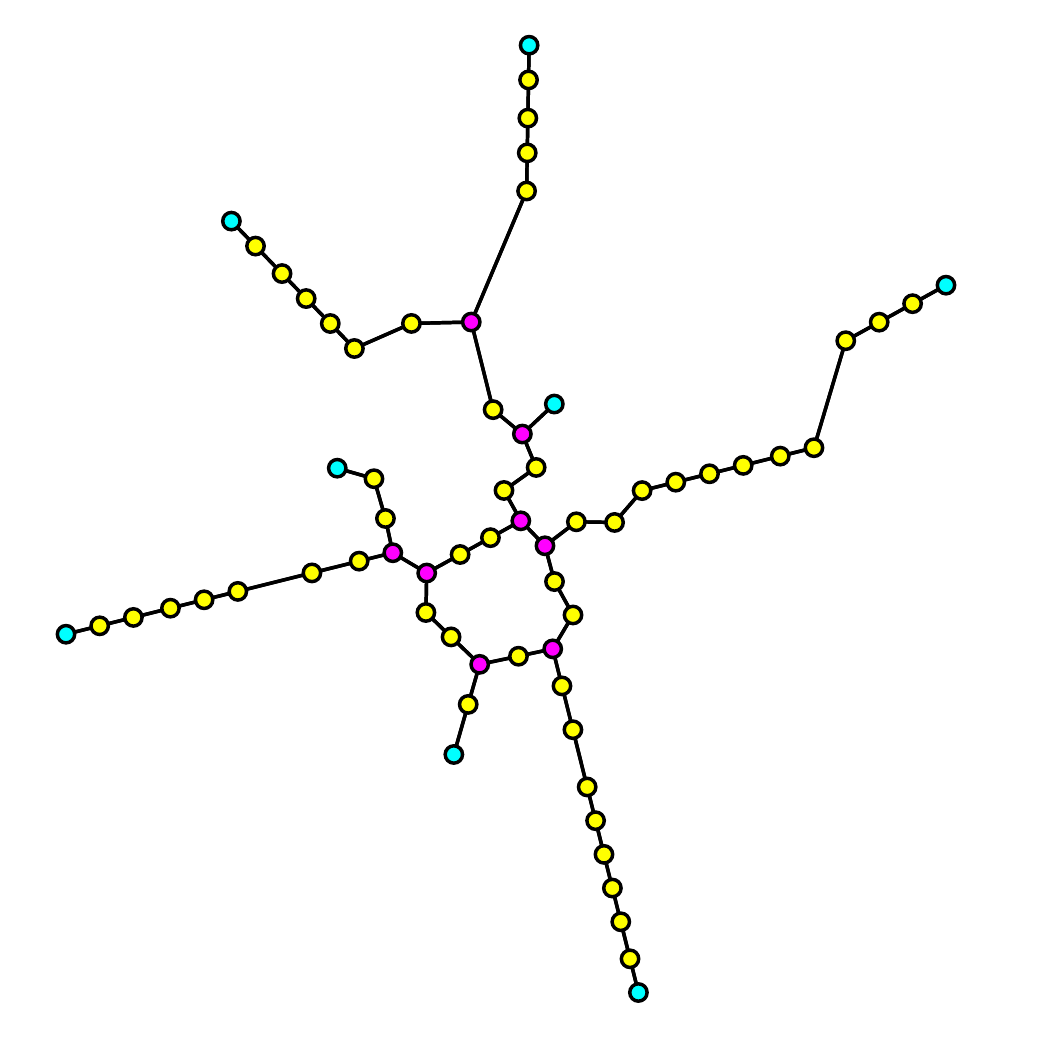}}
\subfigure[$\Delta \theta=30$, $\beta=1$]{\includegraphics[width=0.24\textwidth]{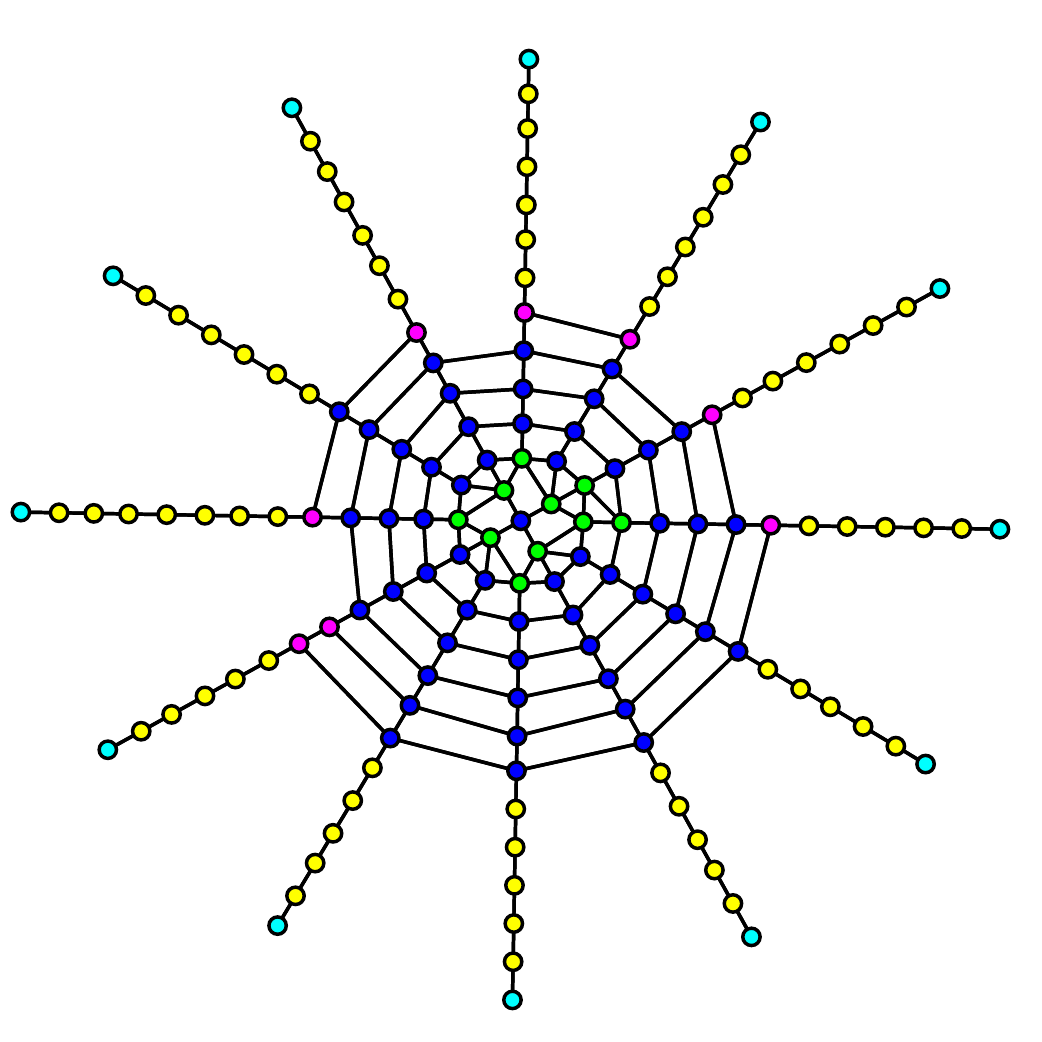}}
\subfigure[$\Delta \theta=30$, $\beta=2$]{\includegraphics[width=0.24\textwidth]{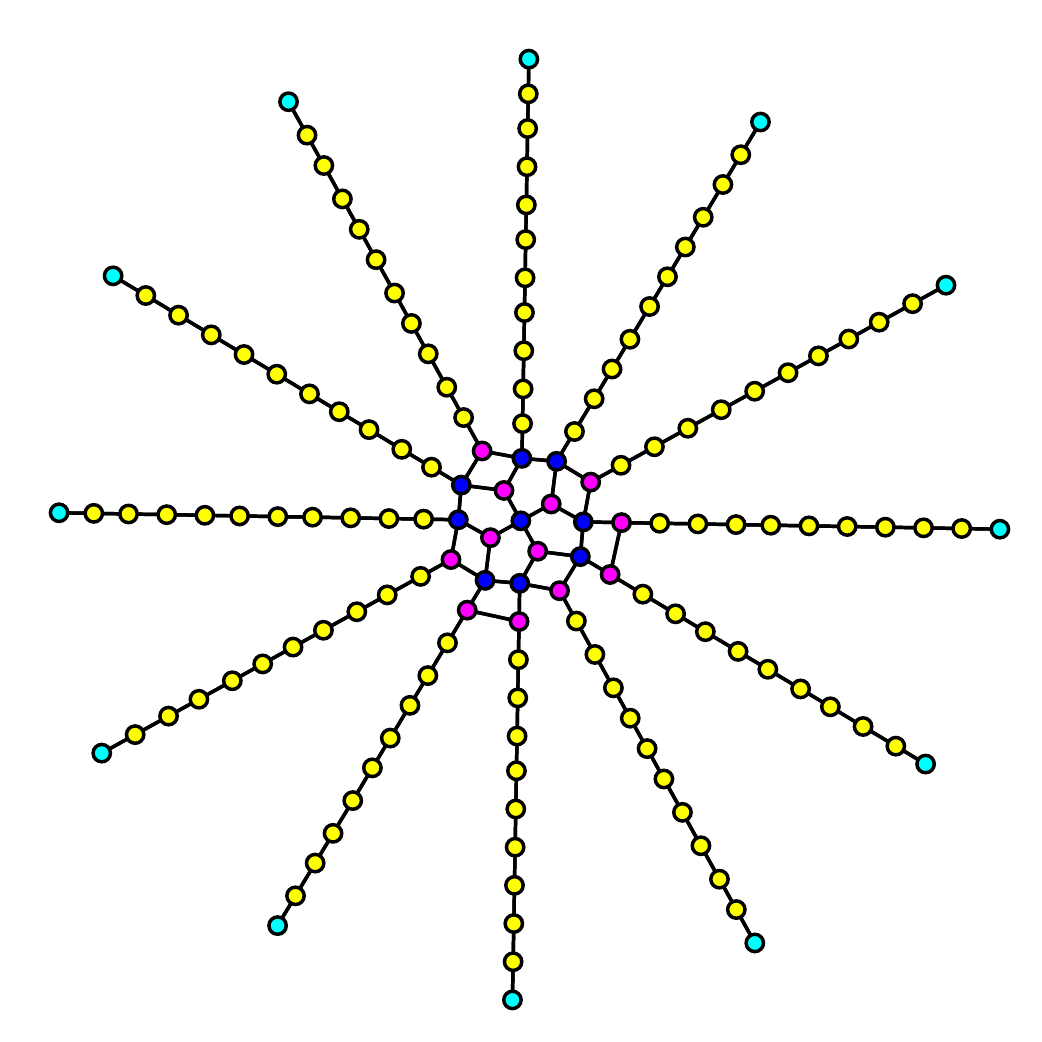}}
\subfigure[$\Delta \theta=30$, $\beta=10$]{\includegraphics[width=0.25\textwidth]{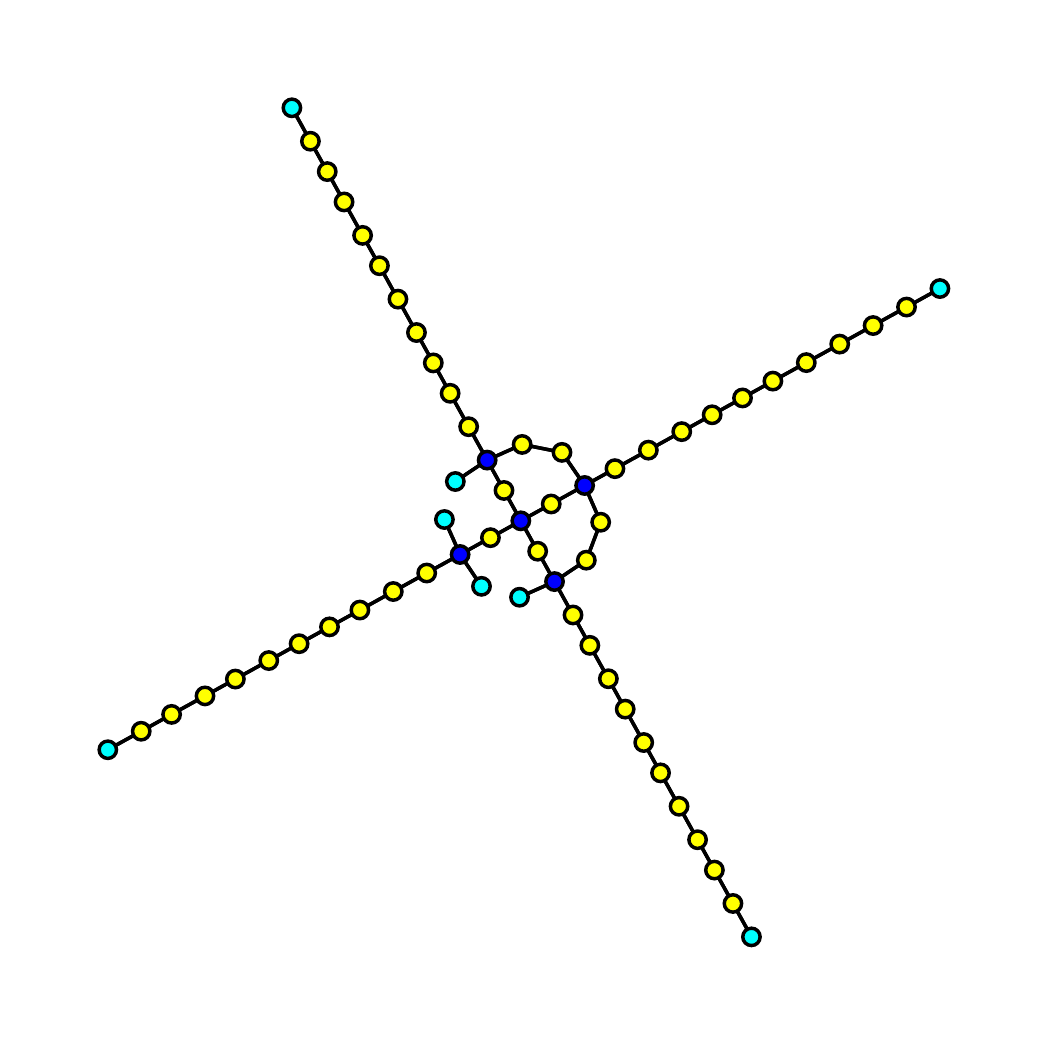}} 
\subfigure[$\Delta \theta=30$, $\beta \geq 26$]{\includegraphics[width=0.24\textwidth]{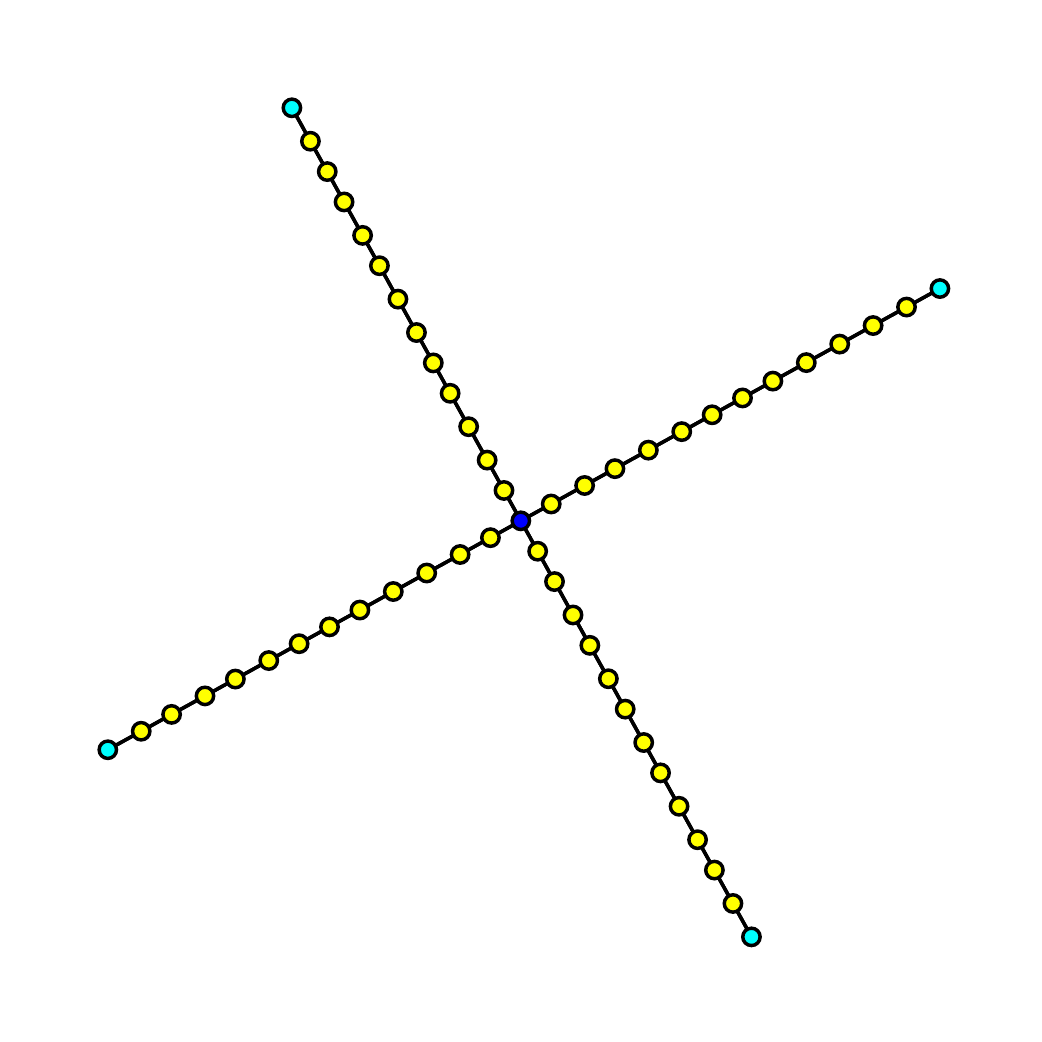}}
\subfigure[$\Delta \theta=40$, $\beta=1$]{\includegraphics[width=0.24\textwidth]{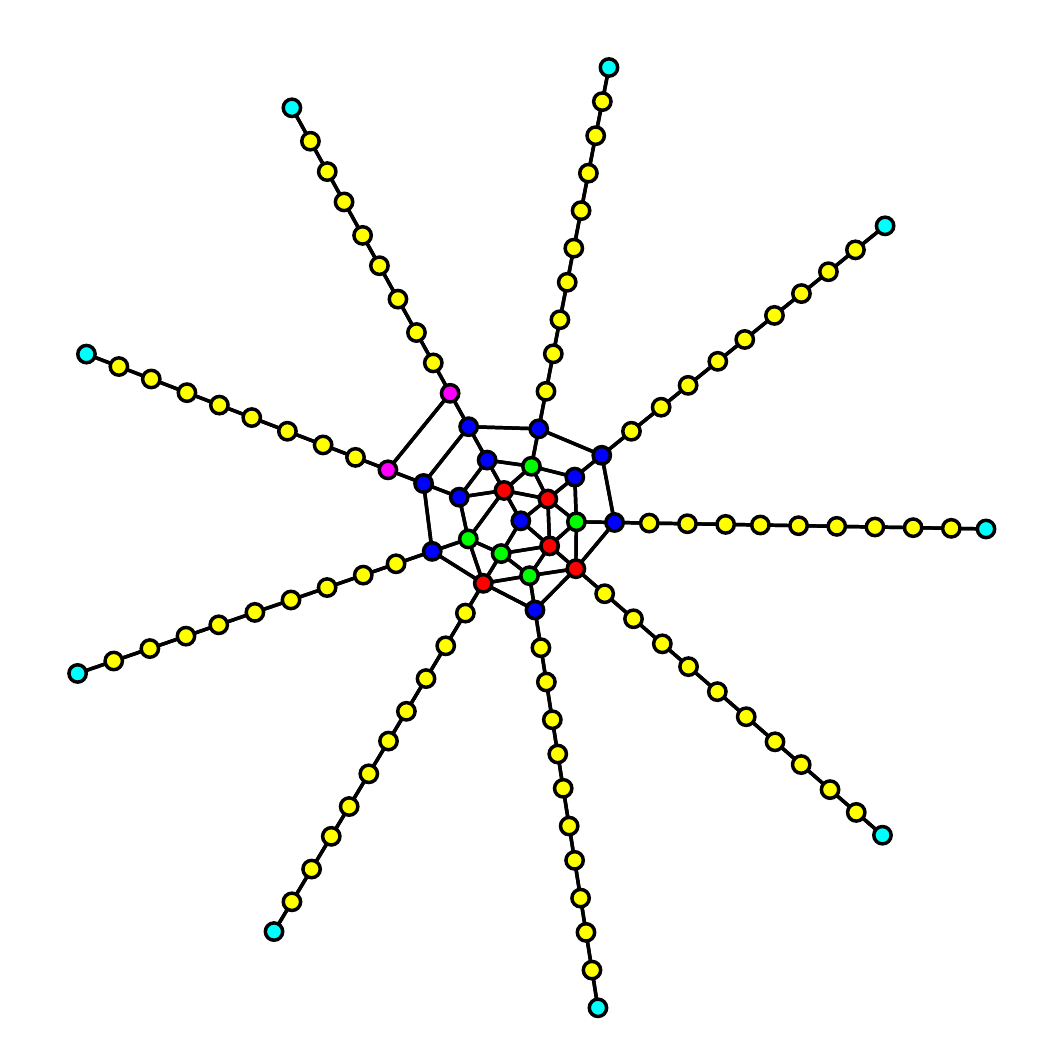}}
\subfigure[$\Delta \theta=40$, $\beta=2$]{\includegraphics[width=0.24\textwidth]{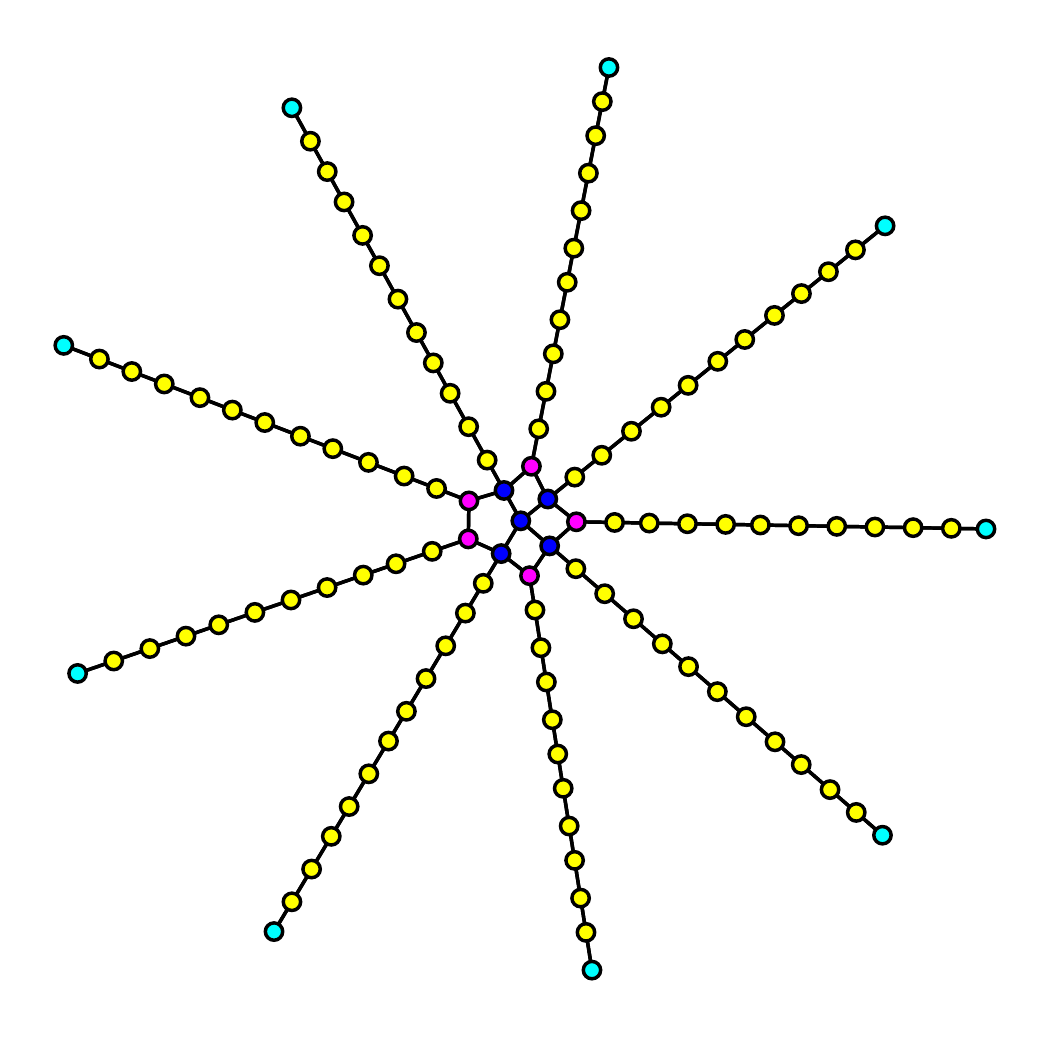}}
\subfigure[$\Delta \theta=40$, $\beta=6$]{\includegraphics[width=0.24\textwidth]{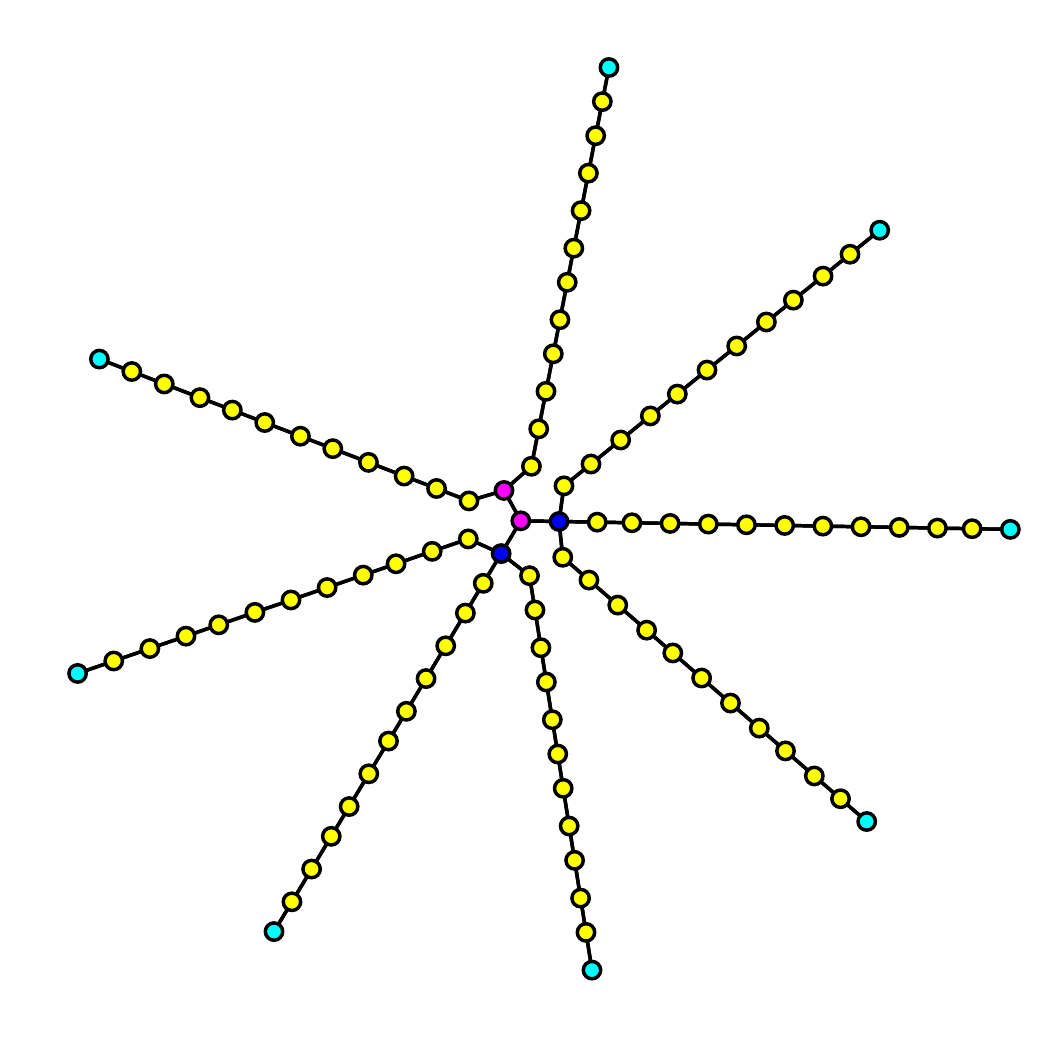}}
\subfigure[$\Delta \theta=40$, $\beta=9$]{\includegraphics[width=0.24\textwidth]{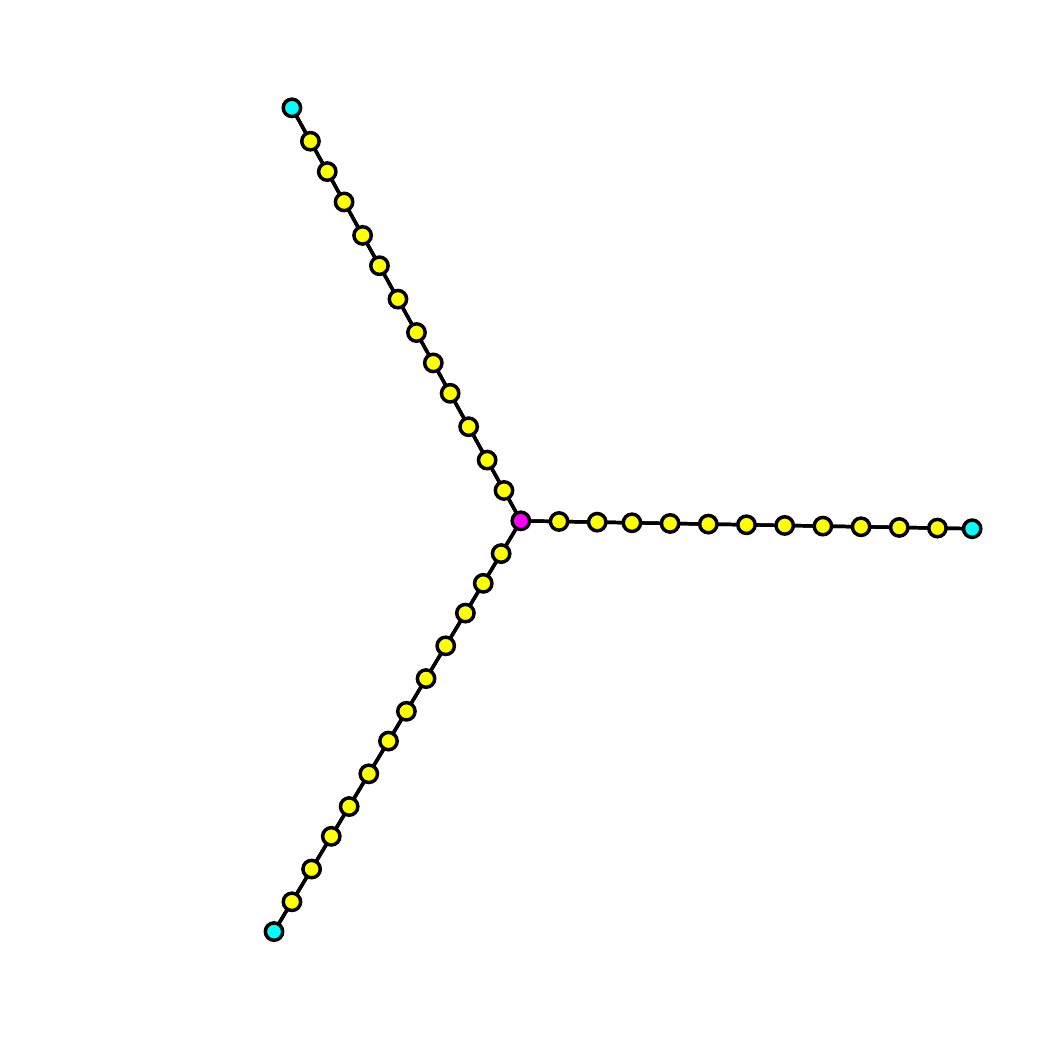}}
\caption{ $\beta$-skeletons grown with  $\Delta\theta=15$~(a--h), $\Delta\theta=30$~(i--l) and 
$\Delta\theta=40$~(m--p).}
\label{theta15}
\end{figure}

\begin{figure}[!tbp]
\centering
\subfigure[$\Delta \theta=50$, $\beta=1$]{\includegraphics[width=0.24\textwidth]{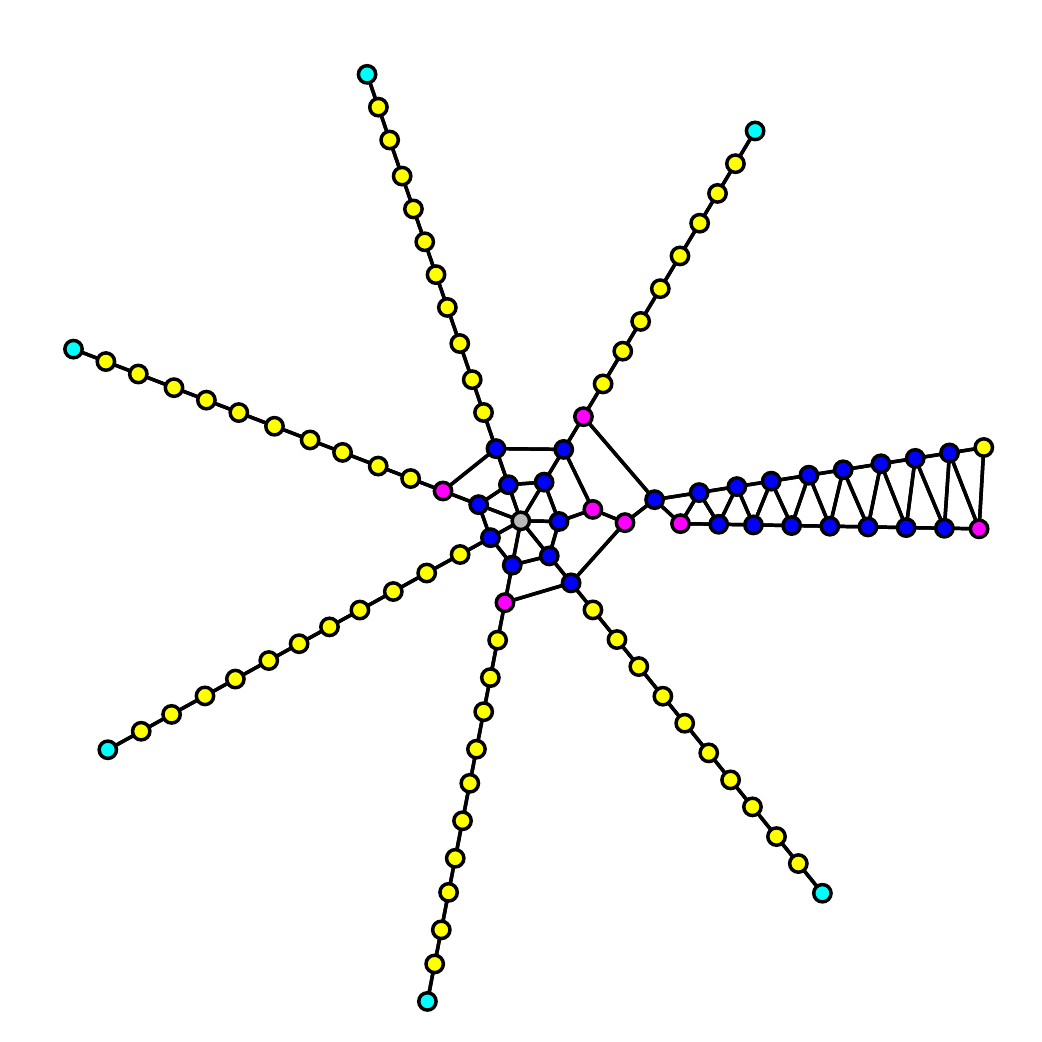}}
\subfigure[$\Delta \theta=50$, $\beta=2$]{\includegraphics[width=0.24\textwidth]{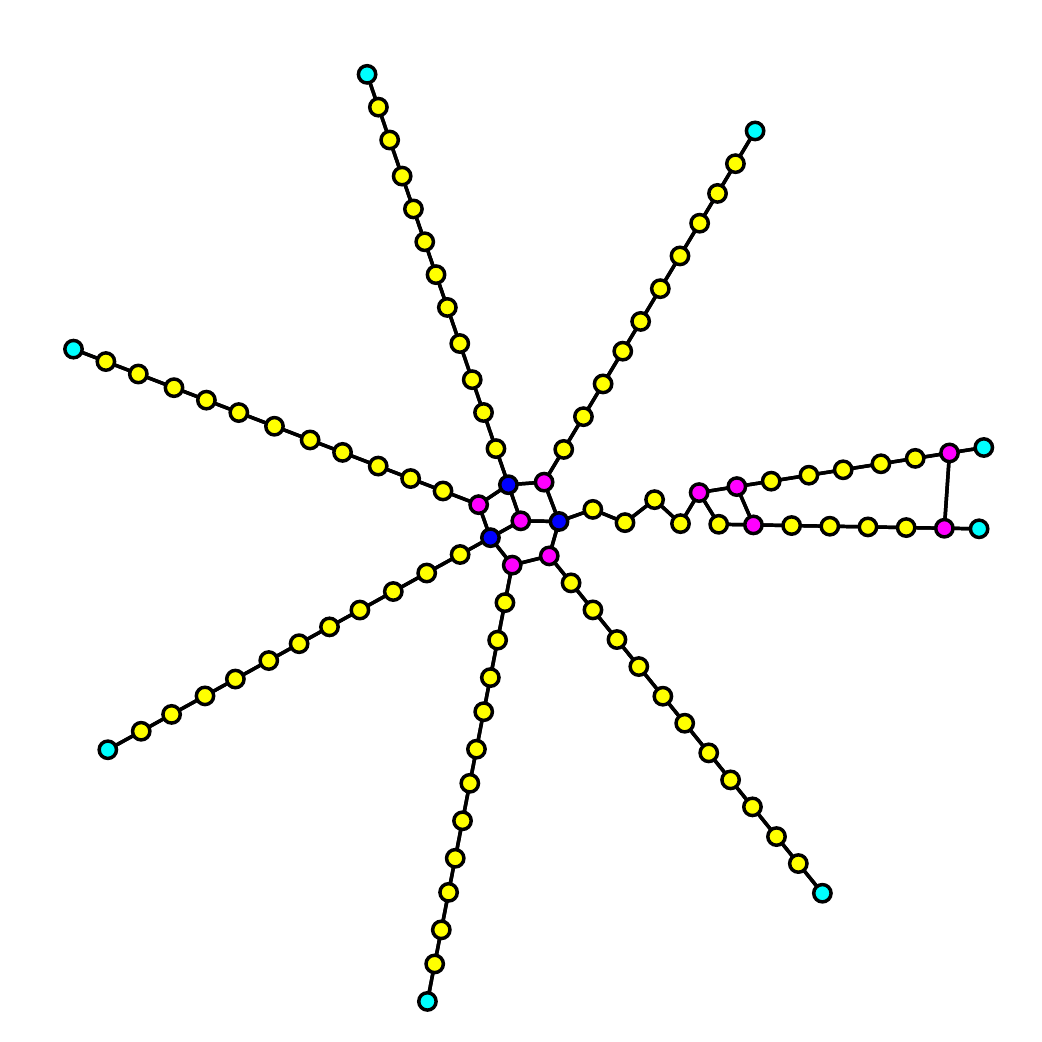}}
\subfigure[$\Delta \theta=50$, $\beta=10$]{\includegraphics[width=0.24\textwidth]{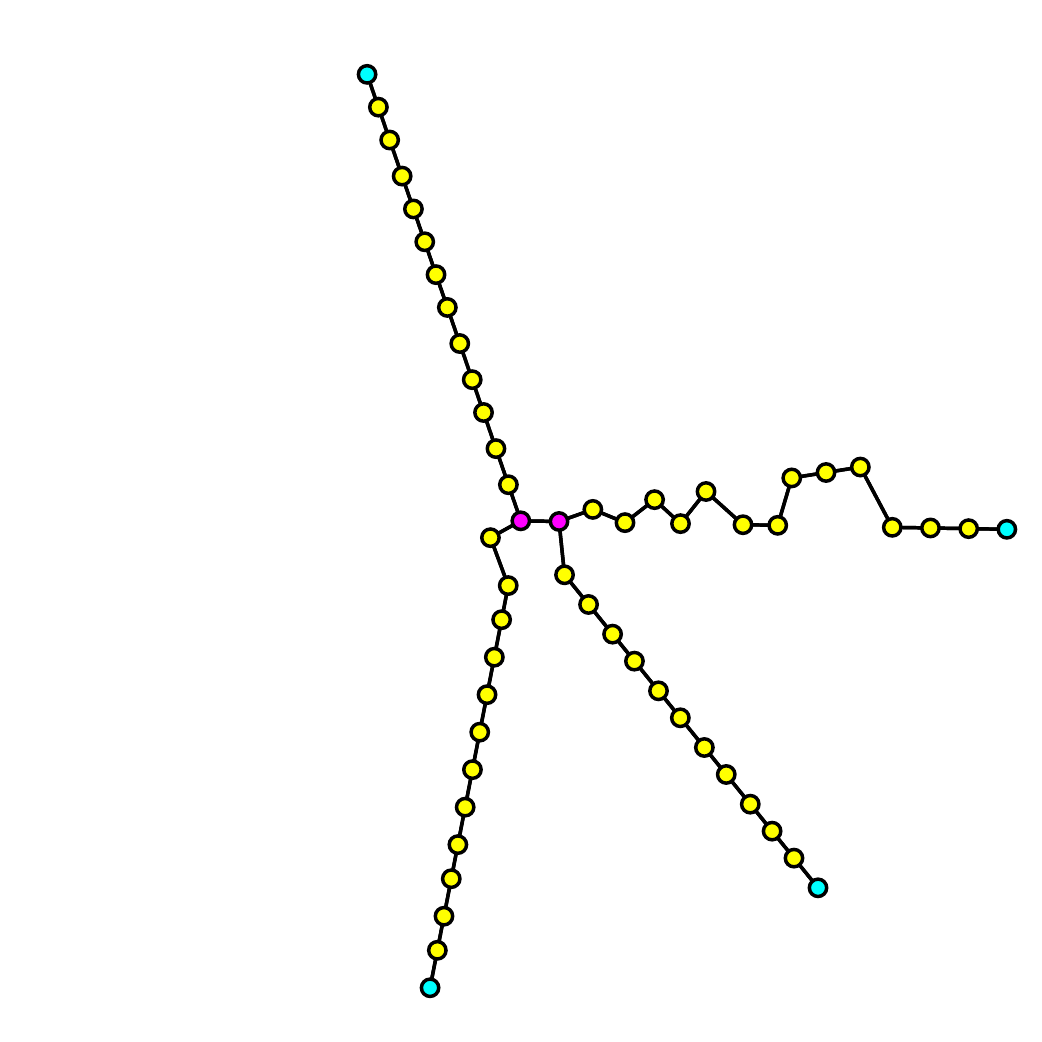}}
\subfigure[$\Delta \theta=50$, $\beta \geq 30$]{\includegraphics[width=0.25\textwidth]{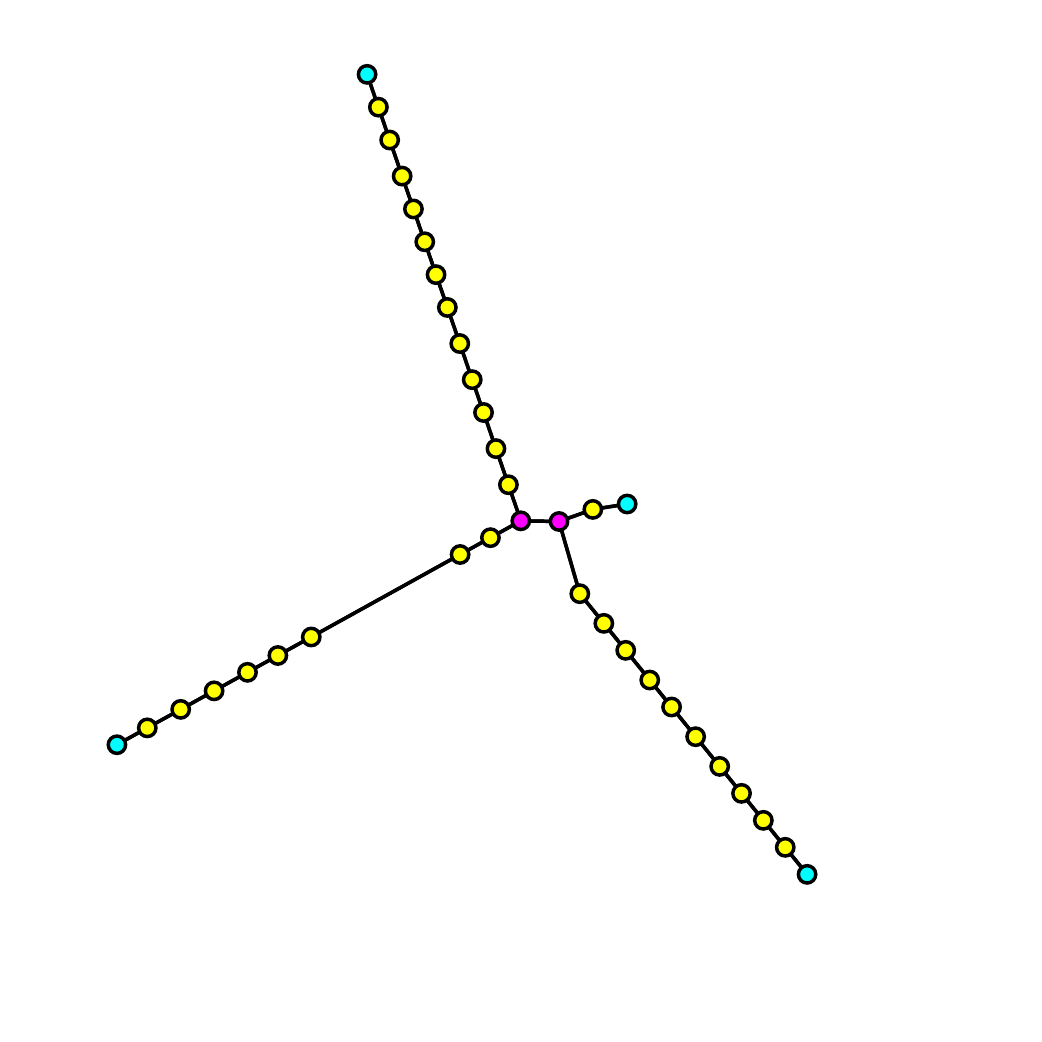}}\\
\subfigure[$\Delta \theta=60$, $\beta=1$]{\includegraphics[width=0.24\textwidth]{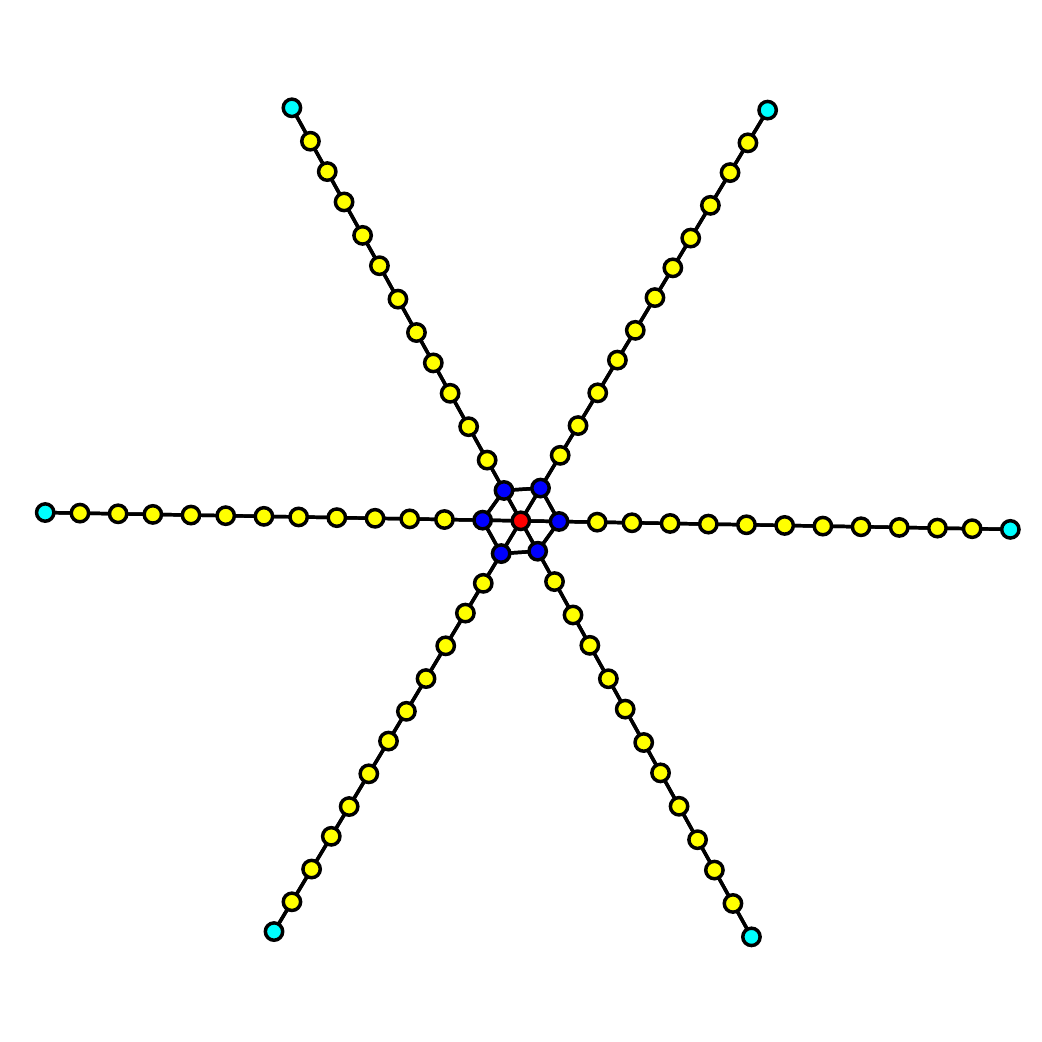}}
\subfigure[$\Delta \theta=60$, $\beta=2$]{\includegraphics[width=0.24\textwidth]{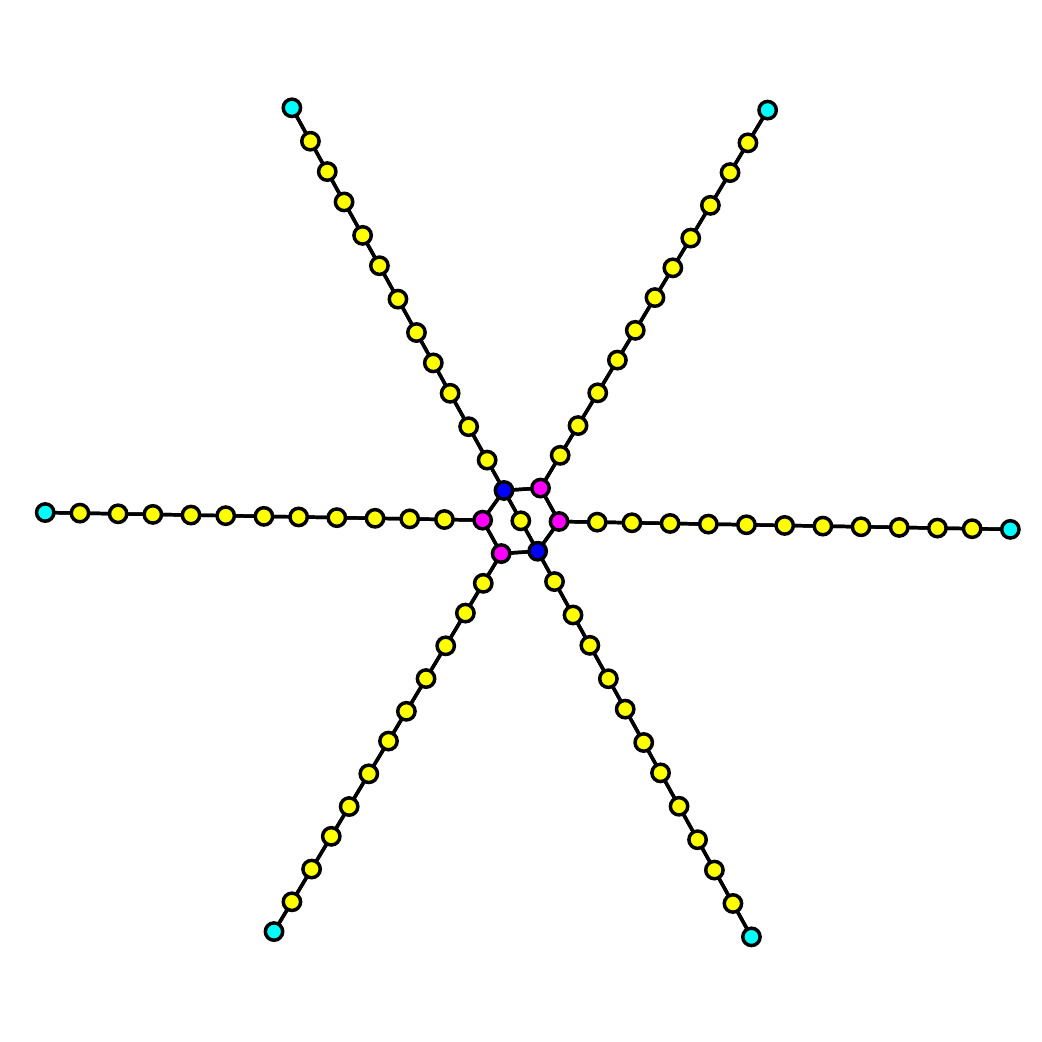}}
\subfigure[$\Delta \theta=60$, $\beta \geq 3$]{\includegraphics[width=0.24\textwidth]{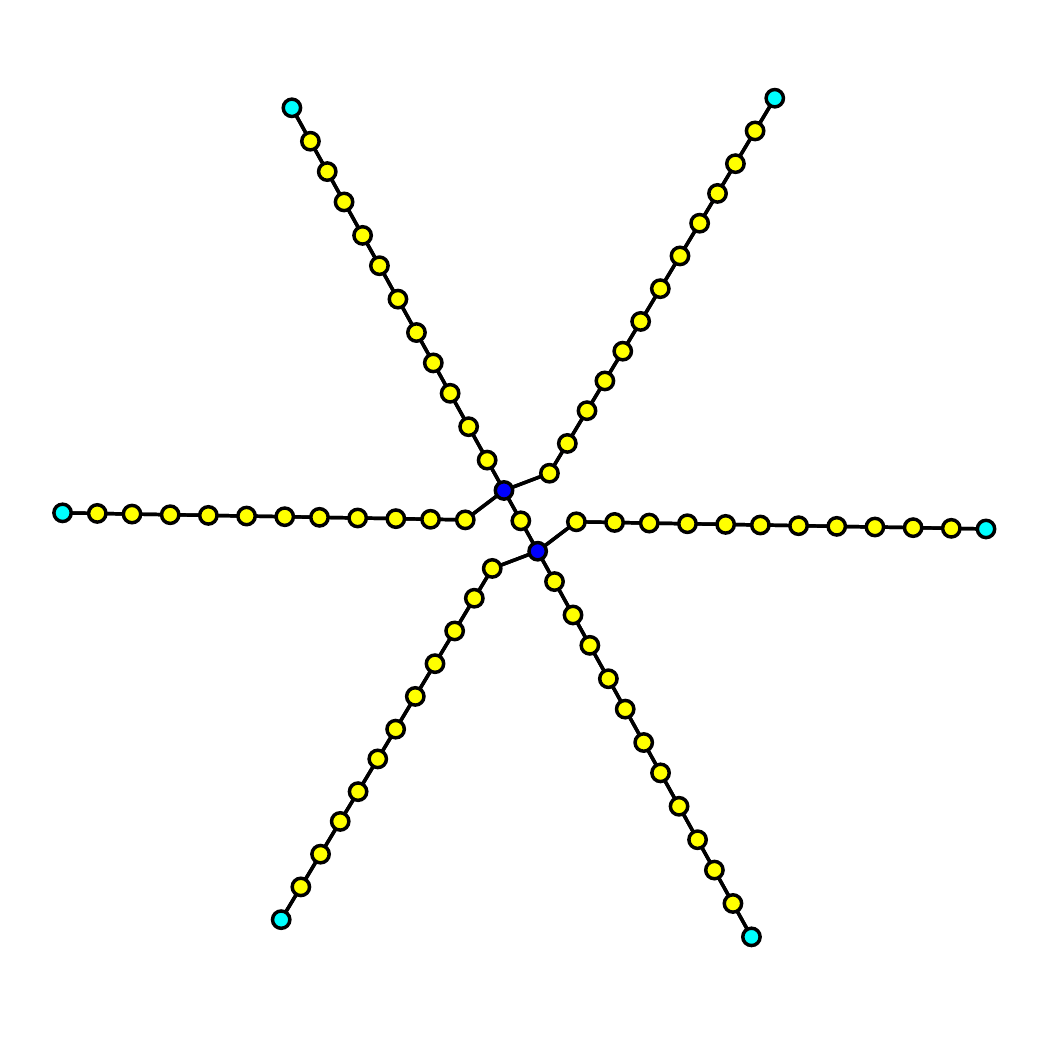}}\\
\subfigure[$\Delta \theta=70$, $\beta=1$]{\includegraphics[width=0.24\textwidth]{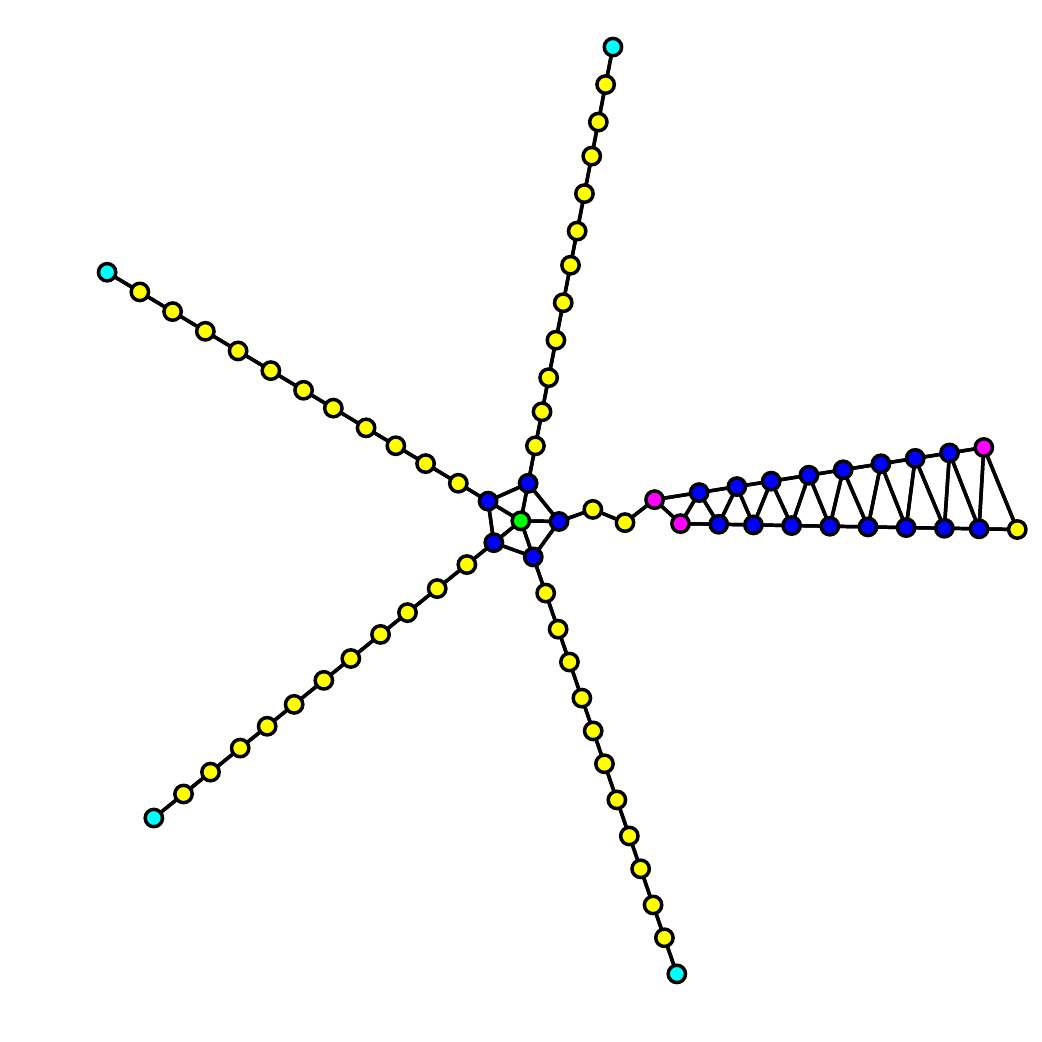}}
\subfigure[$\Delta \theta=70$, $\beta=2$]{\includegraphics[width=0.24\textwidth]{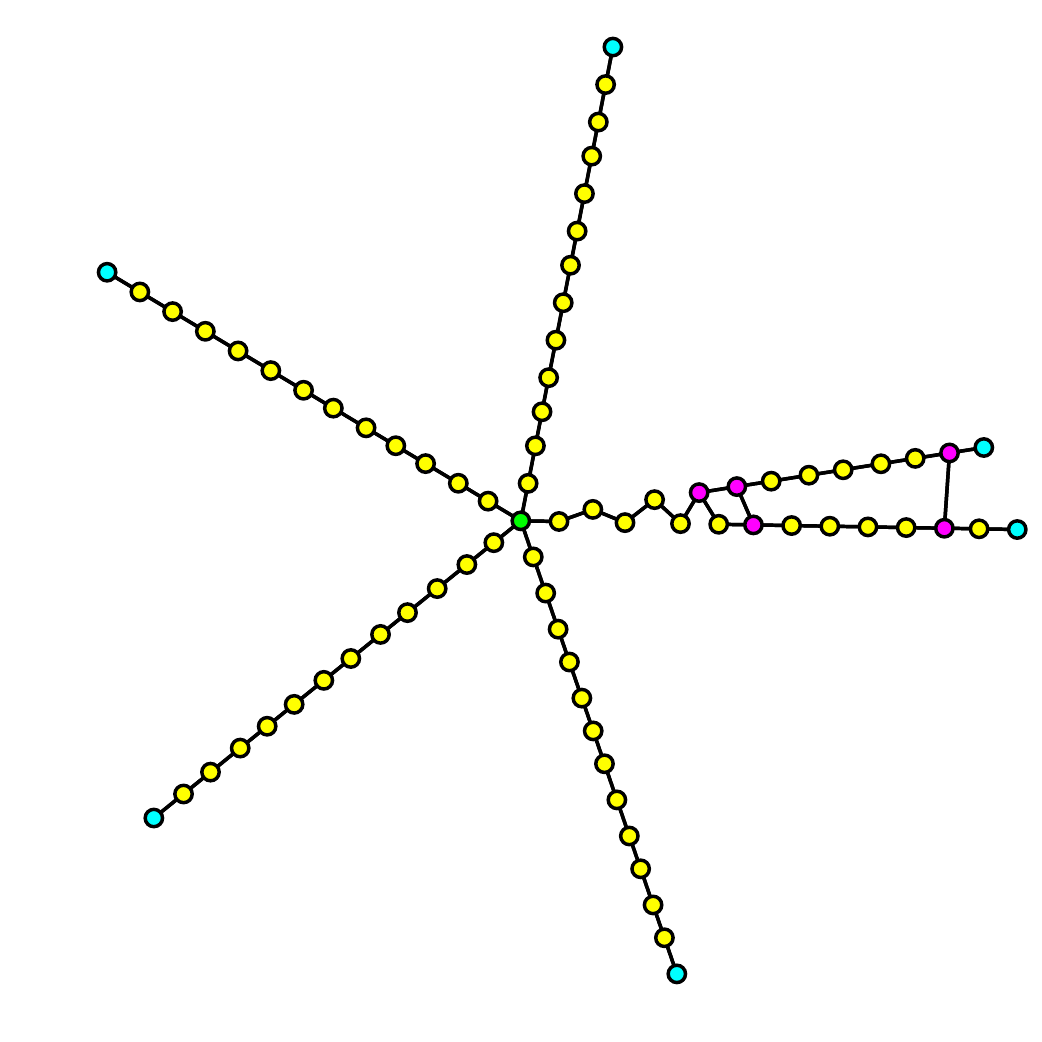}}
\subfigure[$\Delta \theta=70$, $\beta=3$]{\includegraphics[width=0.24\textwidth]{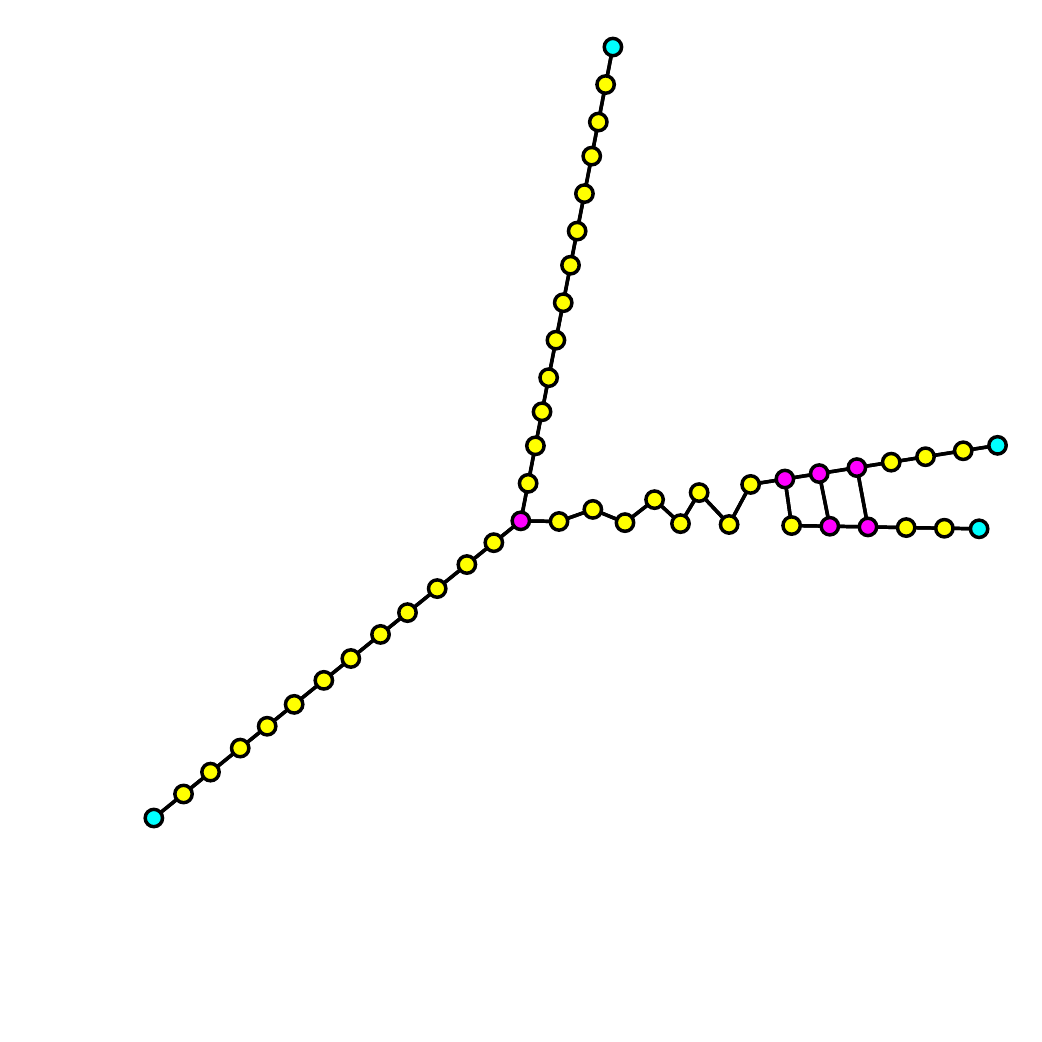}}
\subfigure[$\Delta \theta=70$, $\beta \geq 8$]{\includegraphics[width=0.24\textwidth]{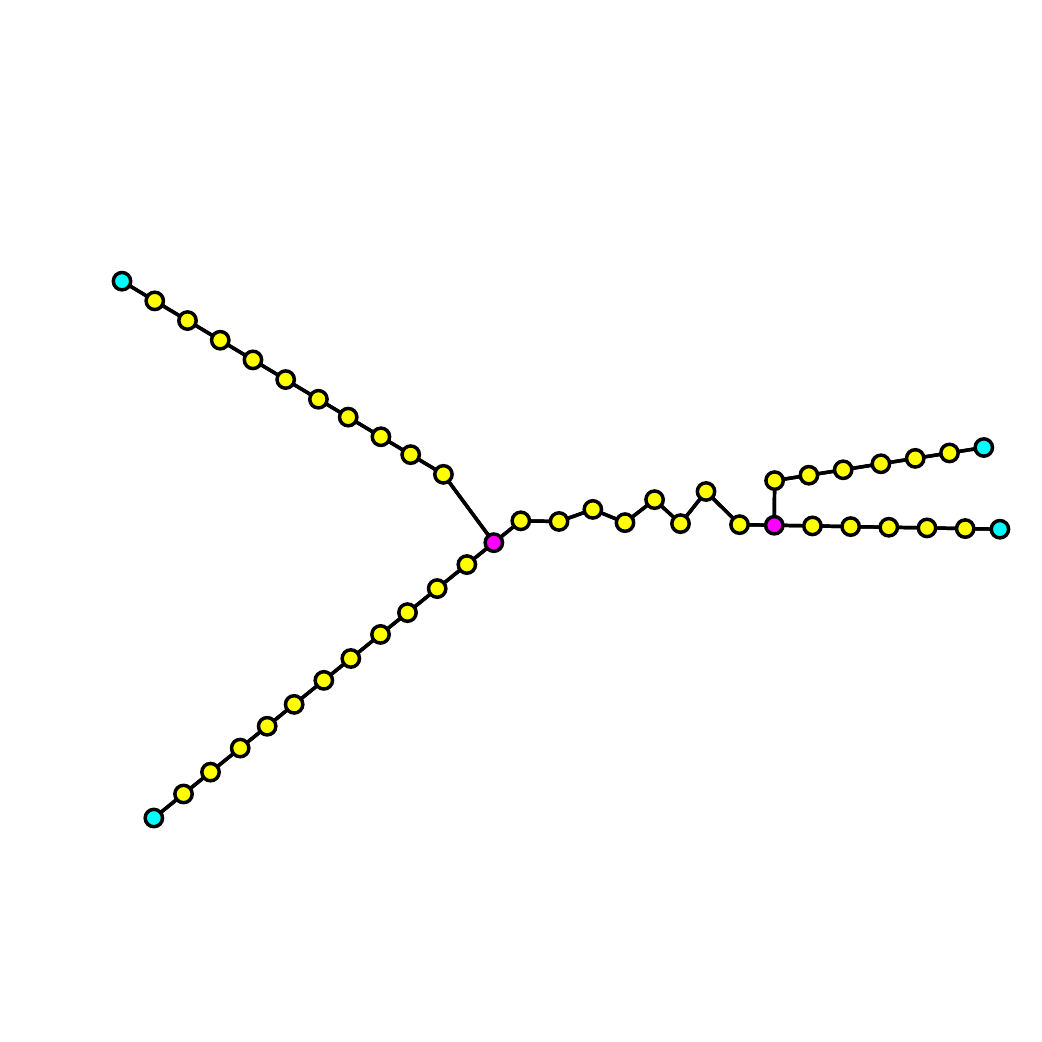}}
\subfigure[$\Delta \theta=80$, $\beta=1$]{\includegraphics[width=0.24\textwidth]{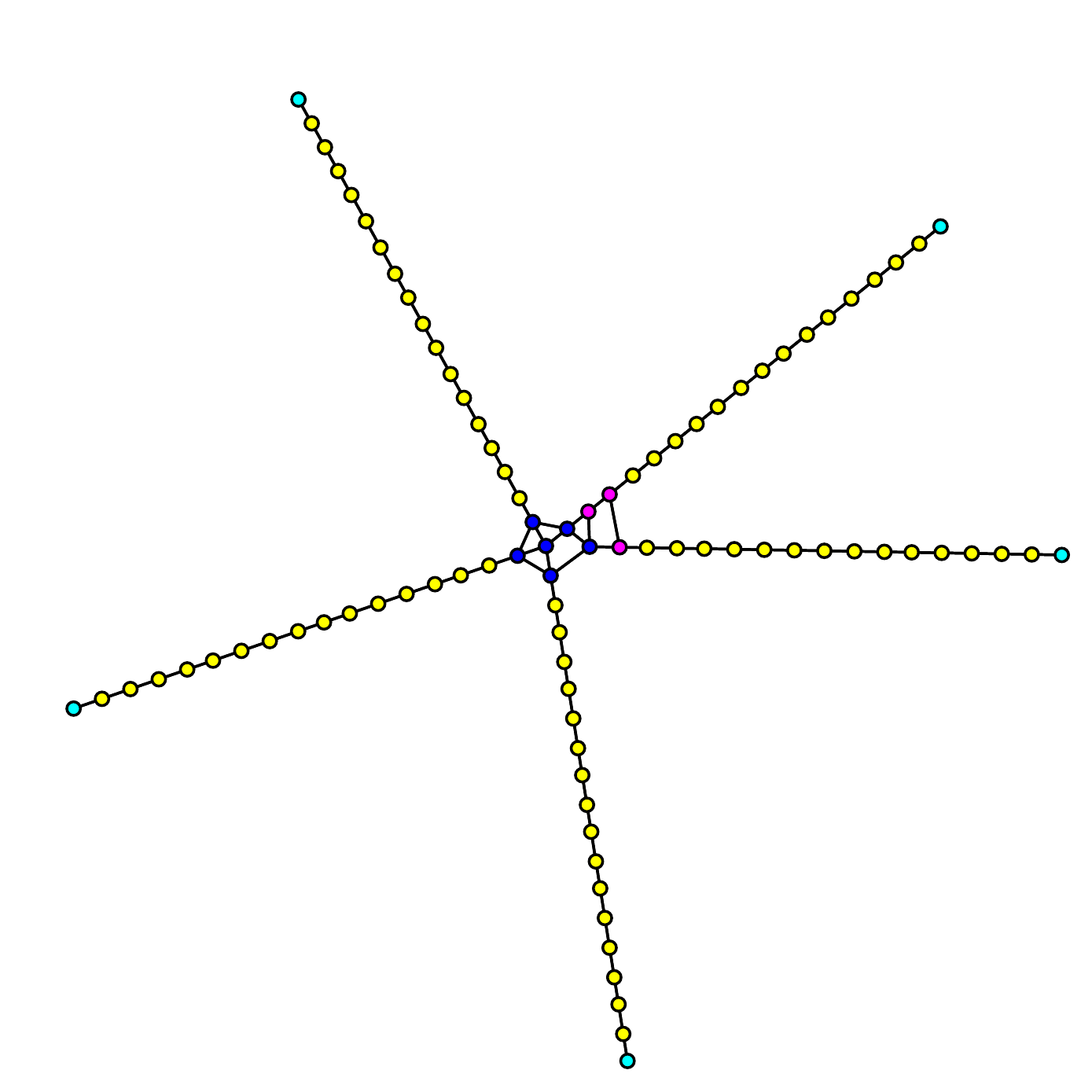}}
\subfigure[$\Delta \theta=80$, $\beta=2$]{\includegraphics[width=0.24\textwidth]{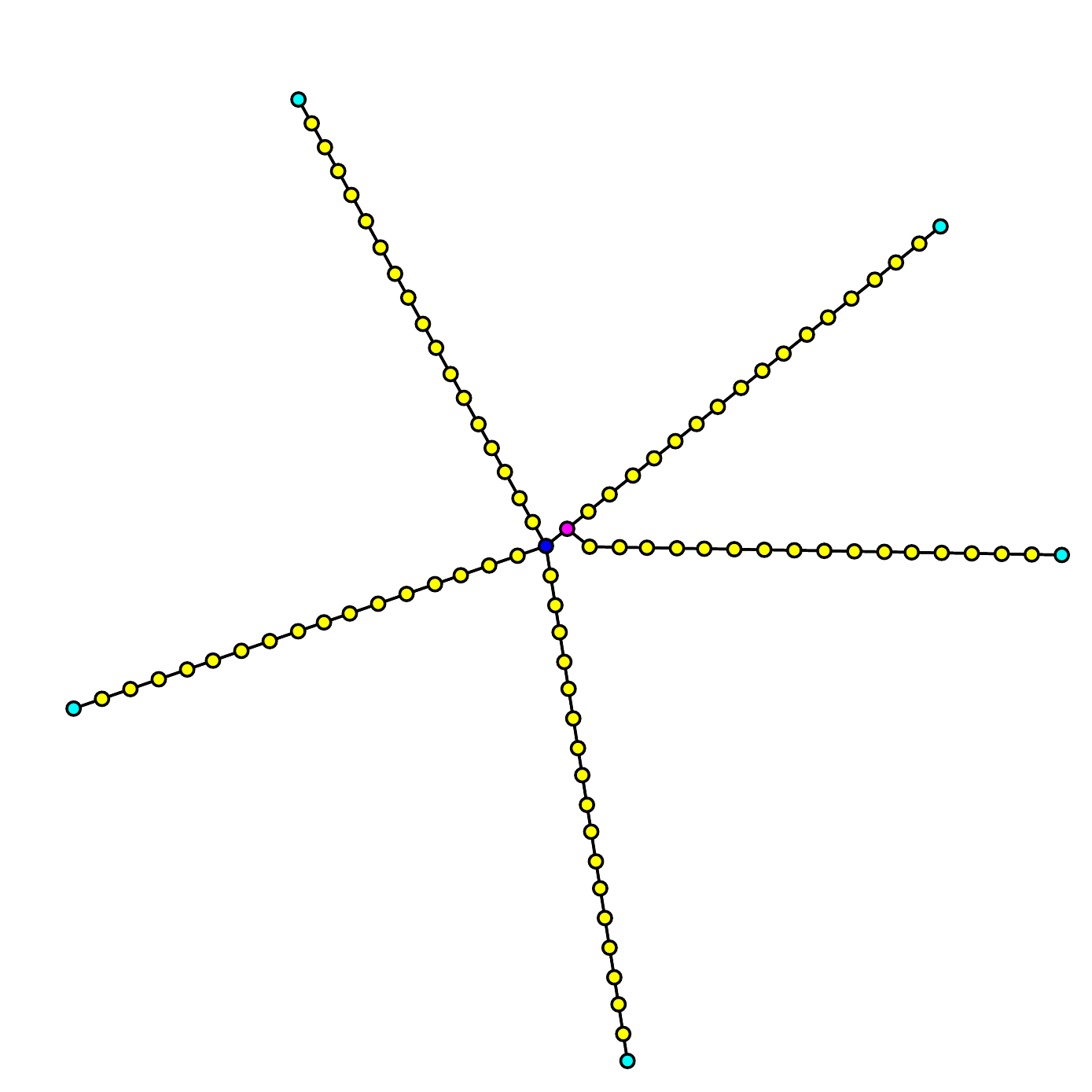}}
\subfigure[$\Delta \theta=80$, $\beta=6$]{\includegraphics[width=0.24\textwidth]{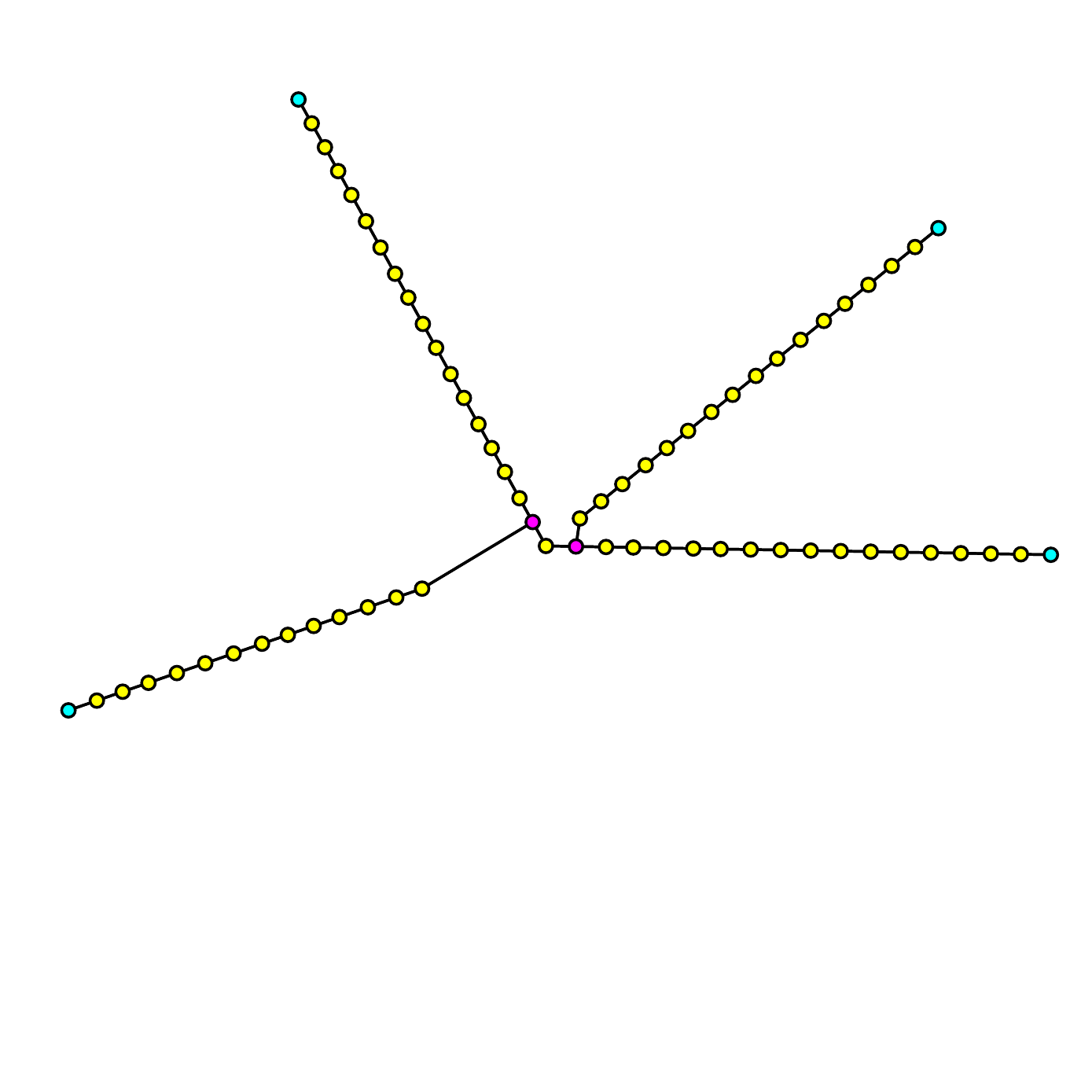}}
\subfigure[$\Delta \theta=80$, $\beta \geq 9$]{\includegraphics[width=0.24\textwidth]{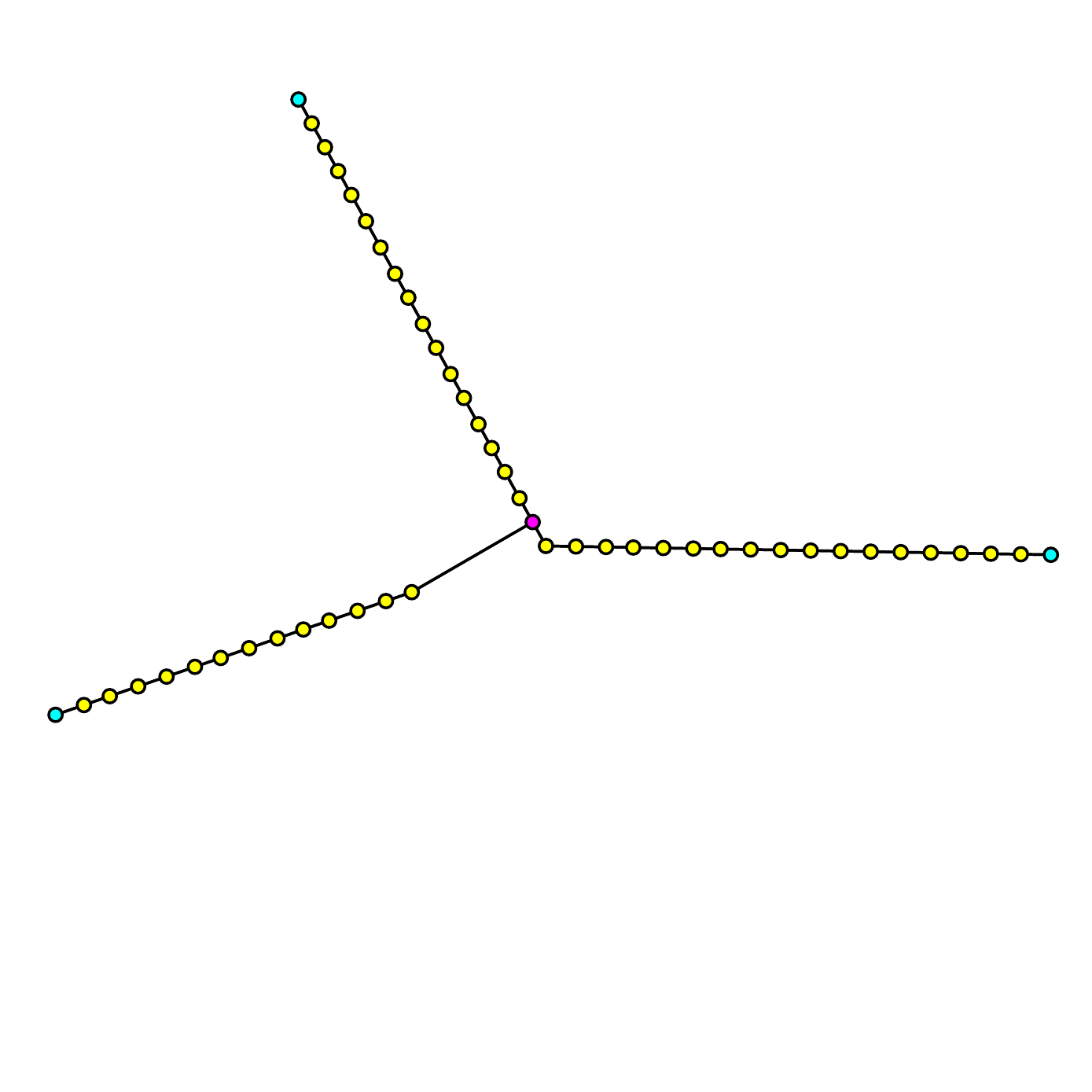}}
\subfigure[$\Delta  \theta=90$, $\beta=1$]{\includegraphics[width=0.24\textwidth]{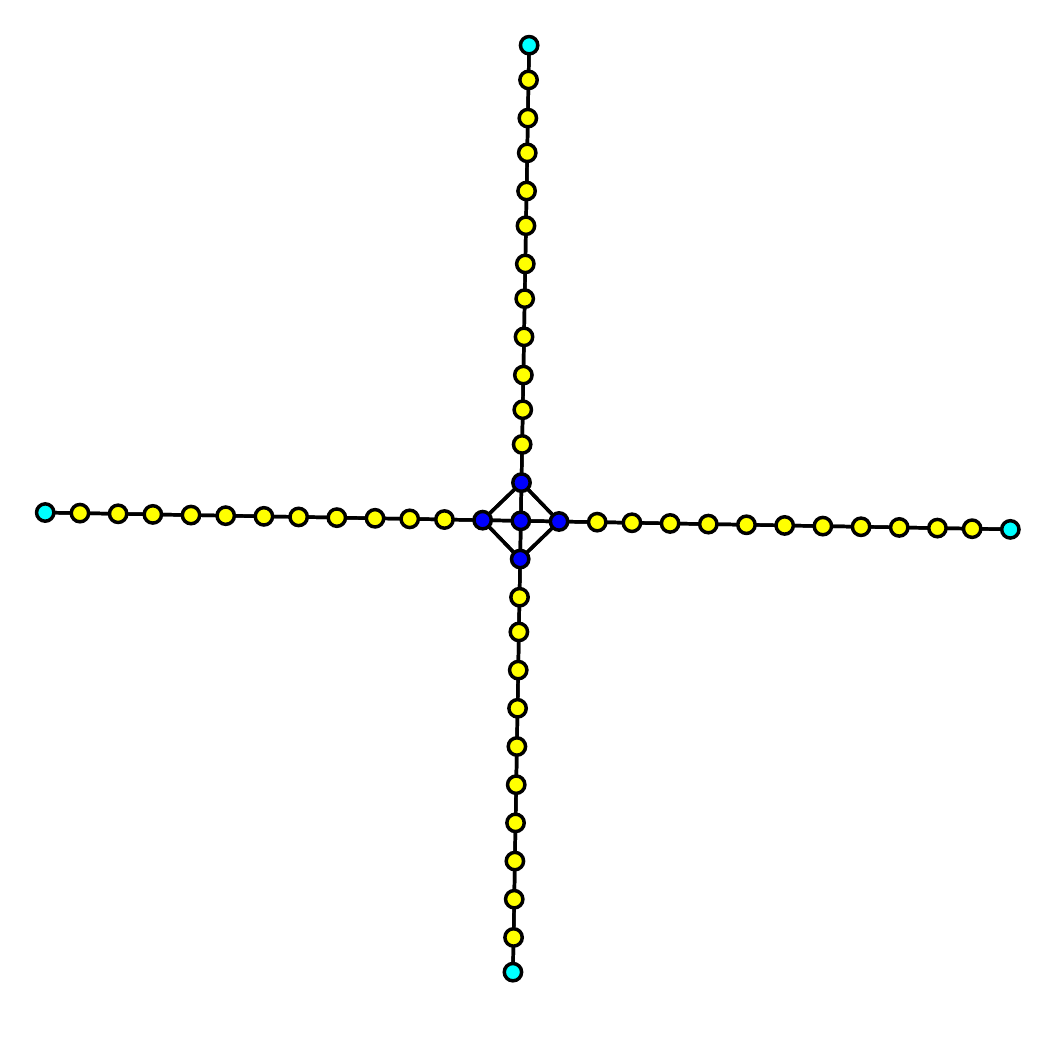}}
\subfigure[$\Delta  \theta=90$, $\beta>1$]{\includegraphics[width=0.24\textwidth]{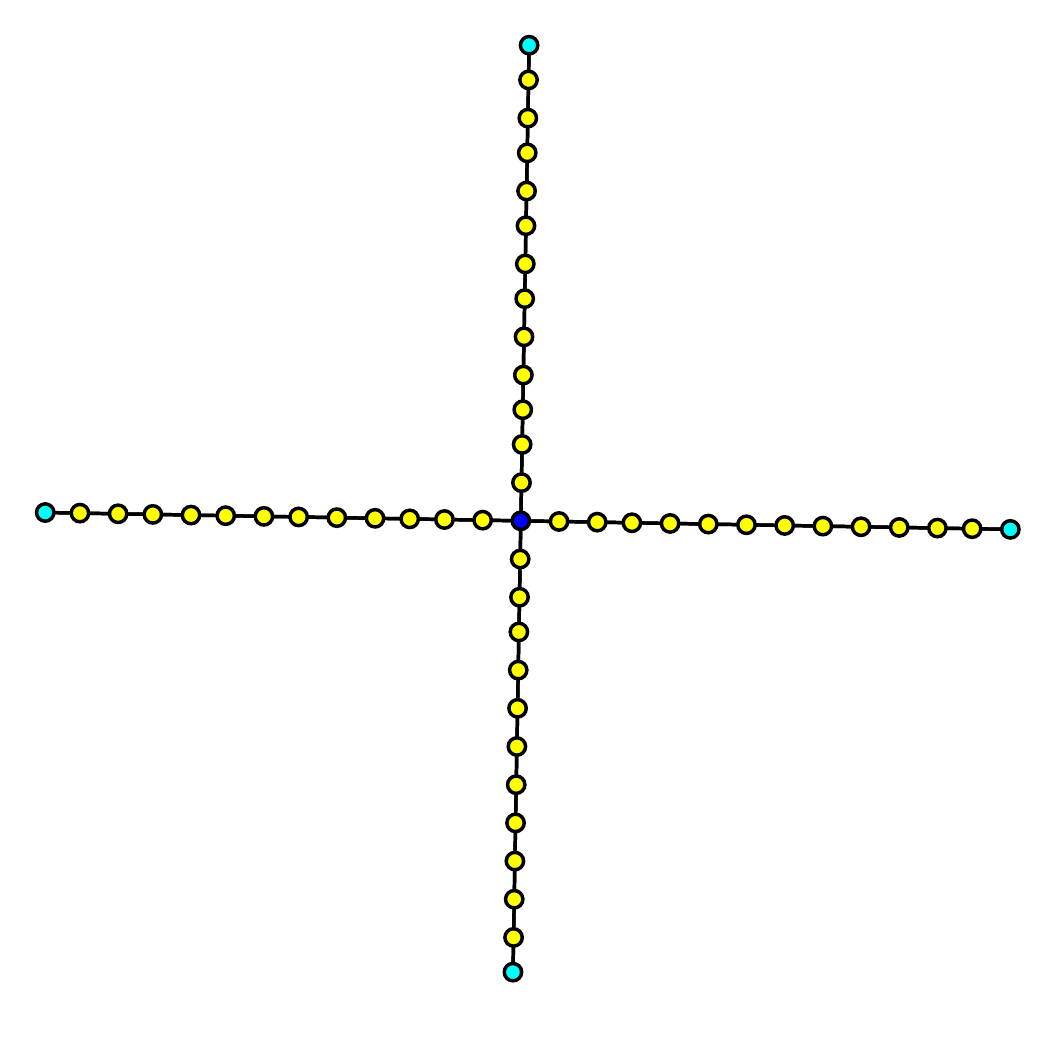}}
\caption{ $\beta$-skeletons grown with $\Delta\theta=50$~(a--d),  $\Delta\theta=60$~(e--g),
$\Delta\theta=70$~(h--k), $\Delta\theta=80$~(l--o) and $\Delta\theta=90$~(pq).}
\label{theta50}
\end{figure}

\section{$\Delta \theta$ influences morphologies}

Morphologies of $\beta$-skeletons are affected not only by values of $\beta$ but also  $\Delta \theta$. We can expect 
that with increase of  $\Delta \theta$ the skeletons converge from almost regular lattices to trees for $\beta < 4$. For example, compare skeletons grown for   $\Delta \theta=5$ (Fig.~\ref{theta05}a--l) and
$\Delta \theta=10$ (Fig.~\ref{theta05}m--t). Increase of $\Delta \theta$ causes predominant decrease of lateral (aligned along
concentric cycles centred in $p_0$) edges, see e.g.  Fig.~\ref{theta05}abc and Fig.~\ref{theta05}mno, and decrease
in branching of trees, see e.g.  Fig.~\ref{theta05}jkl and Fig.~\ref{theta05}rst.  New nodes are added 
to $\beta$-skeletons in a cycle of iterations, change of $\theta$ from 0 to 360, thus branches of the graphs are leaned towards circular arrangements (Fig.~\ref{theta05}klrs).

For $\beta$ up to 3, $\beta$-skeletons consist of two morphologically distinctive components: internal core of a quasi-regular network and spider-web like halo of radial rays connected by lateral links. The pronounced examples of this  morphological division are shown in Fig.~\ref{theta05}mno and Fig.~\ref{theta15}ab. 

Increase of $\Delta \theta$ leads to a disappearance of lateral links and shrinking of the quasi-regular network core, see transition  from Fig.~\ref{theta05}m to Fig.~\ref{theta15}a to Fig.~\ref{theta15}i.  Skeletons grown with large angle increments $\Delta \theta$ are characterised by a prevalence of radial edges, or rays. Thus, for $\Delta \theta = 30 $ we observe transition from a small spider web when $\beta=1$ (Fig.~\ref{theta15}i) to twelve rays, $\beta=2$  (Fig.~\ref{theta15}j), and 
four rays,  $\beta \geq 26$  (Fig.~\ref{theta15}j), structures. Number of rays is changed from nine, $\beta=2$, to 
seven, $\beta=6$, to three, $\beta=9$ (Fig.~\ref{theta15}nop). Skeletons grown with large angular increments, $\Delta \theta=80$ and $90$, Fig.~\ref{theta50}l--q consist of three, four or five rays, for any value of $\beta$.

In situations when  $360 \mod \Delta \theta >0$ lateral edges are formed between rays. Examples are $\beta$-skeletons grown with $\Delta \theta =50$   (Fig.~\ref{theta50}ab) and $\Delta \theta =50$   (Fig.~\ref{theta50}hij.)

\section{Possible applications in sciences}

\begin{figure}[!tbp]
\centering
\subfigure[]{\includegraphics[width=0.4\textwidth]{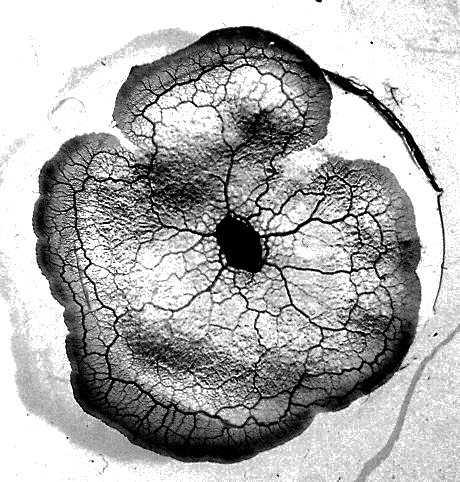}}
\subfigure[]{\includegraphics[width=0.4\textwidth]{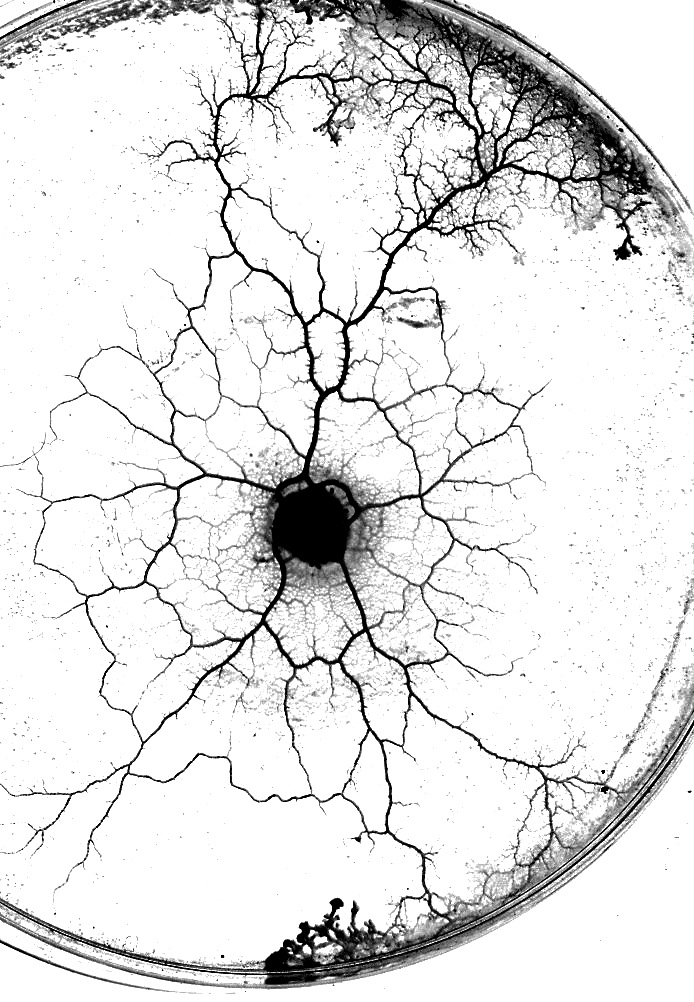}}
\subfigure[]{\includegraphics[width=0.6\textwidth]{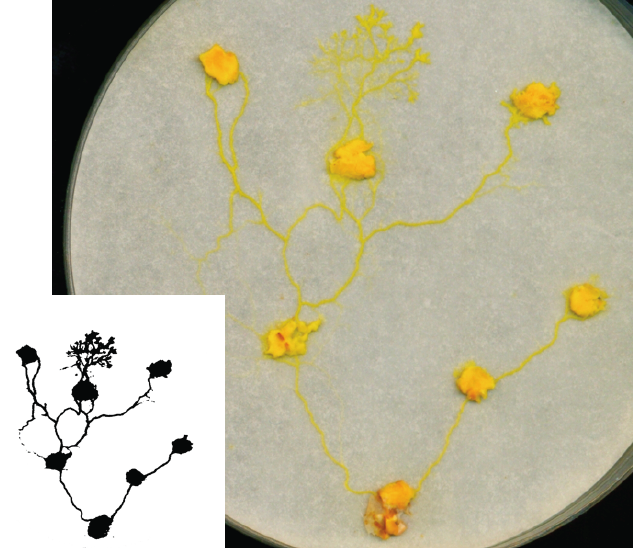}}
\caption{Slime mould \emph{P. polycephalum} growing on a corn meal 2\% agar gel~(a),
non-nutrient 2\% agar gel~(b), and a slightly wet filter paper~(c). 
See details in \cite{adamatzky_PhysarumMachines}. }
\label{physarum}
\end{figure}

Some aspects of a morphological dynamic of $\beta$-skeletons, controlled by $\beta$ and $\Delta \theta$, 
resemble substrate-induced morphological transformations in bacterial~\cite{golding_1998} and 
myxomycetes~\cite{adamatzky_PhysarumMachines} colonies.
Typically, a high concentration of nutrients in a growth substrate leads to formation of dense quasi-uniform 
omni-directionally propagating patterns. A low concentration of nutrients in a substrate leads to formation of 
branching tree-like structures. Two examples are shown in Fig.~\ref{physarum}. 

When plasmodium of  \emph{P. polycephalum} is inoculated on an agar plate with high concentration of nutrients (2\% corn meal agar) the plasmodium's growth-front propagates similarly to a circular wave. A dense network of protoplasmic tubes is formed inside the plasmodium's body (Fig.~\ref{physarum}a). Such growing pattern might be matched well by $\beta$-skeletons grown by the Node Addition Procedure, given in Sect.~\ref{procedure}, with $\Delta \theta = 0.5$, $1 \leq \beta \leq 3.5$ (Fig.~\ref{examplesofgrown}a--f), and  $\Delta \theta = 5$,    $1 \leq \beta \leq 3$ (Fig.~\ref{theta05}abc) and $\Delta \theta = 10$,  $1 \leq \beta \leq 3$ (Fig.~\ref{theta05}mno).  

On a non-nutrient  2\% agar plasmodium forms a tree like structure (Fig.~\ref{physarum}b). The plasmodium trees are alike $\beta$-skeletons generated with 
$\Delta \theta = 0.5$, $30 \leq \beta \leq 50$ (Fig.~\ref{examplesofgrown}jkl),
$\Delta \theta = 5$, $30 \leq \beta \leq 50$ (Fig.~\ref{theta05}h--l),
$\Delta \theta = 10$, $20 \leq \beta \leq 40$ (Fig.~\ref{theta05}qrs). 
A degree of branching, or 'bushiness', of protoplasmic trees decreases with increase of harshness of a growth substrate. For example, plasmodium cultivated 
on a filter paper, instead of agar gel, produces protoplasmic trees with very low degree of branching (Fig.~\ref{physarum}c). Protoplasmic trees grown in a harsh conditions resemble $\beta$-skeletons grown with $\Delta \theta = 0.5$, $600 \leq \beta \leq 800$ (Fig.~\ref{examplesofgrown}rst),
$\Delta \theta = 15$, $10 \leq \beta \leq 50$ (Fig.~\ref{theta05}d-h).

\begin{figure}[!tbp]
\centering
\subfigure[]{\includegraphics[width=0.15\textwidth]{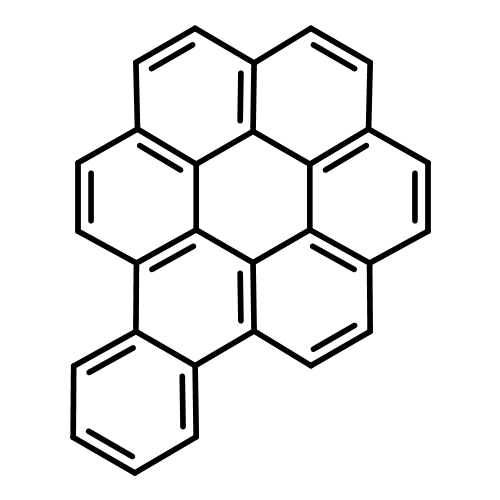}}
\subfigure[]{\includegraphics[width=0.59\textwidth]{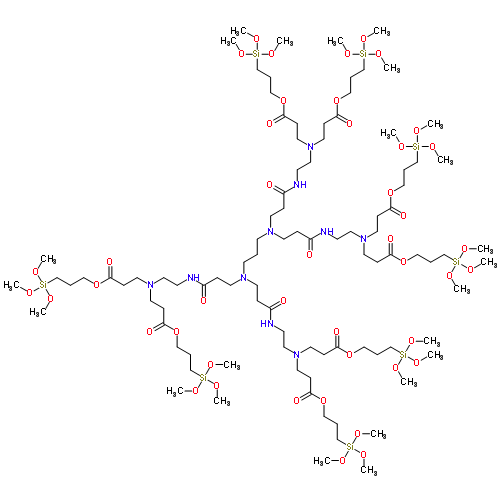}}
\subfigure[]{\includegraphics[width=0.59\textwidth]{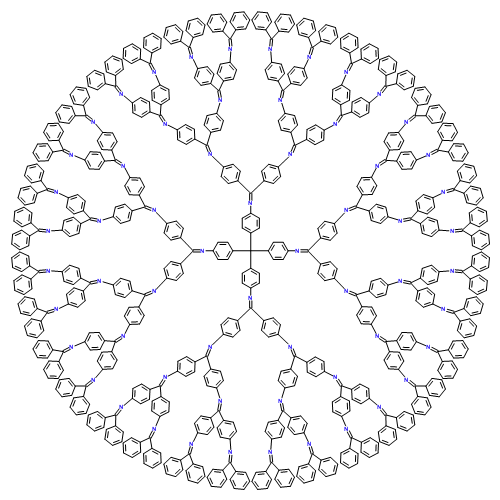}}
\caption{(a)~Benzocoronene, (b)~PAMAM-OS-trimethoxysilyl dendrimer. 
(c)~Tetrakis(4-poly(nitriromethylene-bis(p-phenylene))phenyl)methane dendrimer generation 4.
Chemical structures are drawn with www.chemspider.com.}
\label{dendrimers}
\end{figure}

\begin{figure}[!tbp]
\centering
\includegraphics[width=0.8\textwidth]{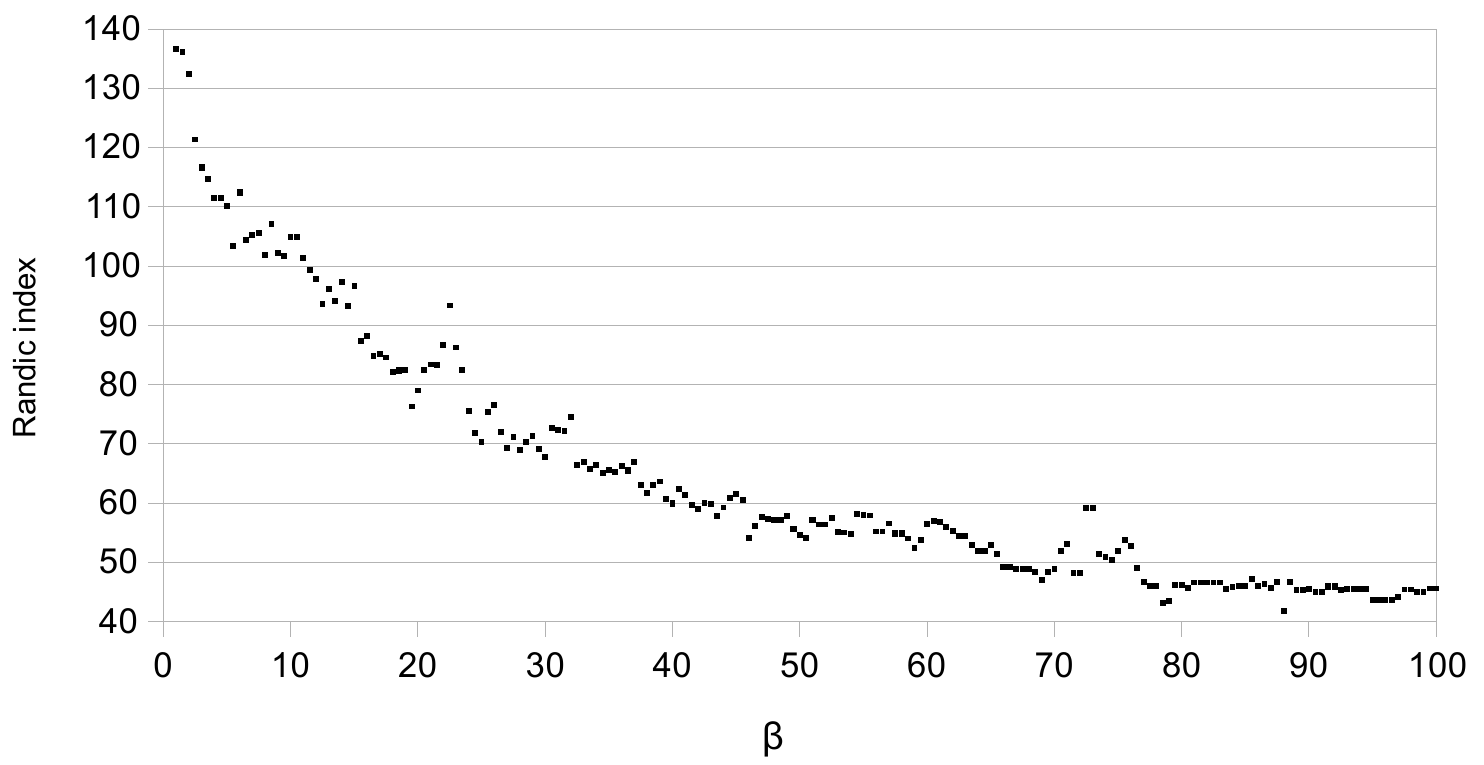}
\caption{Randi\'{c} index of $\beta$-skeletons grown with
$r=5$, $\Delta r=0.5$, $\Delta \theta=0.5$, $\delta = 2.5$}
\label{randic}
\end{figure}

The procedure of growing $\beta$-skeletons may distantly pass for, at least at a very abstract level, 
a synthesis of cyclic (Fig.~\ref{dendrimers}a) and dendrimer molecules (Fig.~\ref{dendrimers}bc).  Skeletons 
grown with low values of $\Delta \theta$ and $\beta$ may be considered analogous to cyclic molecules and 
skeletons produced with high values of $\Delta \theta$ and $\beta$ are alike dendrimer molecules.  

Whilst no direct matching between $\beta$-skeletons and cyclic or dendrimer molecules can be demonstrated we 
calculated Randi\'{c} index~\cite{Randic_1975} of the $\beta$-skeletons grown for $\beta$ up to 100 (Fig.~\ref{randic}). The Randi\'{c} index~\cite{Randic_1975} is calculated as  $R=\sum_{ij} C_{ij}*(\frac{1}{\sqrt{(d_i*d_j)}})$,  where $C_{ij}$ is an adjacency matrix of a graph. 

The Randi\'{c} index $R$ (originally called by Milan Randi\'{c} as molecular branching index)~\cite{Randic_1975}  characterises relationships between structure, property and activity of molecular components~\cite{estrada_2001}.  There are proven linear relations between the  Randi\'{c} index and molecular polarisability, enthalpy of formation, molar refraction, van der Waals areas and volumes, chromatographic retention index~\cite{kier_1976}, cavity surface areas calculated for water solubility of alcohols and hydrocarbons,  biological potencies of  anaesthetics~\cite{kier_1975}, water solubility and boiling point~\cite{hall_1975} and even bio-concentration factor of hazardous chemicals~\cite{sablijc}. Estrada~\cite{estrada_2002} suggested the following structural interpretation: the Randi\'{c} index is proportional to an area of molecular accessibility, i.e. area 'exposed' to outside environment. The  Randi\'{c} index  decreases with increase of $\beta$. Exposure of $\beta$-skeletons is proportional to $\beta$.  This is how properties of the molecules imitated by $\beta$-skeletons will change when a molecule is transformed from, e.g. aromatic to dendritic.

\section{Summary}

\begin{figure}[!tbp]
\centering
\includegraphics[width=0.8\textwidth]{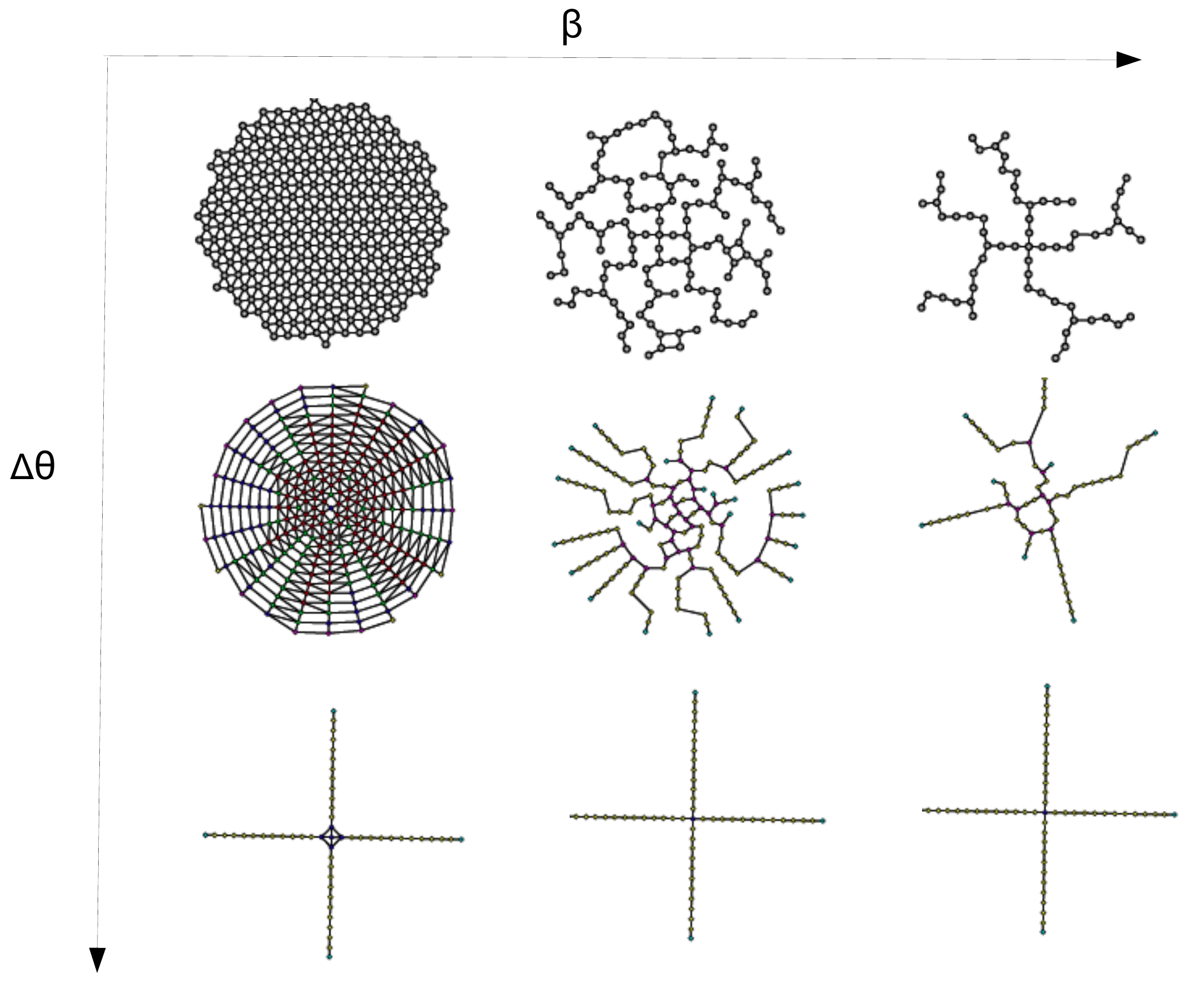}
\caption{Evolution of grown $\beta$-skeletons.}
\label{evolution}
\end{figure}

Given a random planar set, a $\beta$-skeleton of the set is, in general, a disconnected graph for $\beta>2$. We 
presented a procedure for growing the $\beta$-skeletons which remain connected for any, yet specified during the growth, value of $\beta$ however large it is. In computational experiments we demonstrated that with increase of $\beta$ and/or decrease of approximation accuracy, $\Delta \theta$, the skeletons undergo a transformation from almost regular lattices or networks to branching trees to cross-like graphs (Fig.~\ref{evolution}). We speculate that such evolution of the $\beta$-skeletons somewhat imitates morphological transformations of myxomycetes and bacterial colonies governed by concentration of nutrients in their growth substrates and transformation of molecules from aromatic to dendritic forms.

 \end{document}